# Lattice Boltzmann methods for single-phase and solid-liquid phase-change heat transfer in porous media: A review

Ya-Ling He[a,*], Qing Liu[a], Qing Li[b]

[a]Key Laboratory of Thermo-Fluid Science and Engineering of Ministry of Education, School of Energy and Power Engineering, Xi'an Jiaotong University, Xi'an, Shaanxi, 710049, China

[b]School of Energy Science and Engineering, Central South University, Changsha 410083, China

(*Corresponding author. Email: yalinghe@mail.xjtu.edu.cn)

## Abstract

Since its introduction 30 years ago, the lattice Boltzmann (LB) method has achieved great success in simulating fluid flows and modeling physics in fluids. Owing to its kinetic nature, the LB method has the capability to incorporate the essential microscopic or mesoscopic physics, and it is particularly successful in modeling transport phenomena involving complex boundaries and interfacial dynamics. The LB method can be considered to be an efficient numerical tool for fluid flow and heat transfer in porous media. Moreover, since the LB method is inherently transient, it is especially useful for investigating transient solid-liquid phase-change processes wherein the interfacial behaviors are very important. In this article, a comprehensive review of the LB methods for single-phase and solid-liquid phase-change heat transfer in porous media at both the pore scale and representative elementary volume (REV) scale. The review first introduces the fundamentals of the LB method for fluid flow and heat transfer. Then the REV-scale LB method for fluid flow and single-phase heat transfer in porous media, and the LB method for solid-liquid phase-change heat transfer, are described. Some applications of the LB methods for single-phase and solid-liquid phase-change heat transfer in porous media are provided. In addition, applications of the LB method to predict effective thermal conductivity of porous materials are also provided. Finally, further developments of the LB method in the related areas are discussed.

1. Introduction

Fluid flow and heat transfer in porous media are common phenomena in nature, and are also frequently encountered in many areas of applied science and engineering technologies, including energy and environmental science, geosciences (e.g., hydrogeology, geophysics, and petroleum geology), material science, soil science, construction engineering, chemical and petroleum engineering, and so on. Over the past half-century, many researchers have made a great number of experimental, theoretical, and numerical investigations on these areas and a lot of achievements in both theory and engineering applications have been made [1-10]. With new experimental techniques and advanced instruments, a wide variety of physical (or thermophysical) properties of the porous media can be measured, and experimental investigations on fluid flow and heat transfer therein are becoming more accessible. Theoretical and numerical analyses have also allowed researchers to investigate fluid flow and heat transfer in porous media. However, theoretical analysis is limited to very simple situations with low Reynolds numbers, and therefore, numerical analysis is widely used to investigate fluid flow and heat transfer problems in porous media systems, especially to those that have not yet been experimentally studied or are difficult to study by direct experiments with high cost and long duration.

With the rapid development of supercomputing facilities and advanced computational algorithms, it becomes more and more feasible to simulate fluid flow and heat transfer in porous media numerically.

To model fluid flow and heat transfer in porous media at three different process scales: the region scale, the representative elementary volume (REV) scale, and the pore scale (see Fig. 1), different computational approaches should be employed. The macroscopic quantities (i.e., porosity, permeability) are commonly defined by the volume average of microscopic quantities over the REV. The choice of REV can be quite a complicated process: an REV of a porous medium is defined as the minimum volume over which a measurement of the quantity of interest can be made that will yield a value representative of the whole. The characteristic size of the REV is believed to be much larger than the characteristic size of the pores (or solid matrix structures), but considerably smaller than that of the macroscopic flow region. Governing equations and dominant processes may be different at different process scales. To model the processes at the region scale that are of practical interest in engineering applications, system models should be employed [8,9]. It should be stressed that the process of up-scaling from the REV scale to the region scale heavily relies on the accurate description of flow and transport phenomena at the REV scale. To accurately model the flow and transport phenomena at the REV and pore scales, different numerical methods are required. The REV scale utilizes numerical methods based on continuum models [5,6], such as the finite-volume method (FVM), finite-difference method (FDM), and finite-element method (FDM). The pore scale requires numerical methods based on discrete models that represent the porous system by a discrete set of elements. In particular, pore-network modeling method [10] on the one hand, and some advanced numerical methods, such as the lattice Boltzmann (LB) method [11-17], on the other hand, have made it possible to simulate various transport phenomena in porous media at the pore scale.

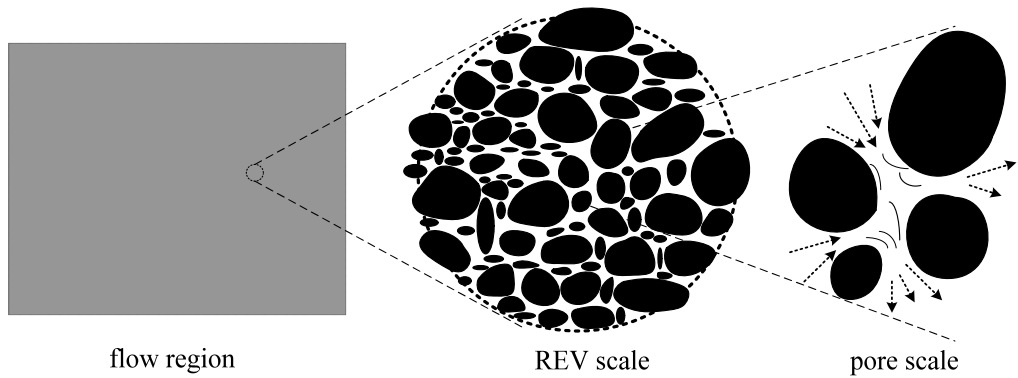

flow region      REV scale      pore scale

**Fig. 1.** Schematic of the multiscale aspects of fluid flow and heat transfer in porous media.

Historically, the LB method [11-17] originated from the lattice gas automata (LGA) method [18], a simplified, fictitious version of the molecular dynamics (MD) method in which the time, space, and particle velocities are all discrete. In 1997, He and Luo [19,20] demonstrated that the LB equation can be rigorously obtained from the linearized continuous Boltzmann equation of the single-particle distribution function. With its roots in gas kinetic theory, the LB method has been developed into a powerful and promising numerical tool for computational fluid dynamics and beyond over the last 30 years. Unlike the MD method which takes into account the movements and collisions of all the individual molecules, the LB method considers the behaviors of a collection of pseudo-particles (a pseudo-particle is comprised of a large number of molecules) moving on a regular lattice with particles residing on the nodes. Different from the conventional numerical methods based on a direct discretization of the macroscopic continuum equations, the LB method is based on minimal lattice formulations of the continuous Boltzmann equation for single-particle distribution function, and macroscopic properties can be obtained from the distribution function through moment integrations. Therefore, the LB method can be viewed as a mesoscopic numerical methodology sitting in the intermediate region between microscopic MD method and macroscopic continuum-based methods. As highlighted by Succi [21], the LB method should most appropriately be considered not just as a smart Navier-Stokes (N-S) solver "in kinetic disguise", but rather like a fully-fledged modeling strategy for a wide range of complex phenomena and processes across scales.

Since the introduction of the LB method in 1988 [11], its application in fluid flow and heat transfer in porous media has always been one of the most important themes of the method. The kinetic nature of the LB method makes it very suitable for simulating flows with fluid-fluid and/or fluid-solid interactions. Moreover, the bounce-back scheme [22-28] in the LB method is a simple and computationally efficient scheme to impose no-slip walls with irregular geometries, which supports the idea that the LB method is ideal for simulating fluid flows in porous media. Generally speaking, two approaches have been developed for the simulations of fluid flow and heat transfer in porous media: the pore-scale approach and the REV-scale approach. In the late 1980s, the LGA method has already been used to simulate flows in porous media and very Darcy's law in complicated geometries [29]. In 1989, Succi et al. [30] first utilized the LB method to measure the permeability of low Reynolds number flow in a 3D random medium, and the Darcy's law was confirmed. Since then, fluid flow [28,31-46] and heat transfer [47-58] in porous media at the pore scale have been studied numerically by many researchers using LB method. In the pore-scale approach, fluid flow and heat transfer in the pores of the medium are directly modeled by the standard LB method. For the flow field, the interaction at the fluid-solid interface can be easily handled by using the bounce-back scheme, which mimics the phenomenon that a particle reflects its momentum when colliding with the solid surface of the porous matrix. Strictly speaking, heat transfer through porous media should be referred to as conjugate heat transfer, and care must be taken in the implementation of thermal boundary conditions at the fluid-solid interface. One of the main goals of pore-scale LB simulations is to obtain values of constitutive parameters needed in the continuum models via up-scaling of the results at the pore-scale level. For instance, permeability (or relative permeability) of a porous medium can be calculated from the pore-scale flow field, which is of crucial importance for continuum modeling.

An alternative approach is to simulate fluid flow and heat transfer in porous media at the REV scale. With the development in the past two decades, many REV-scale LB models have been proposed for simulating fluid flow [59-64] and heat transfer [65-81] in porous media. In the REV-scale approach, the porous medium is viewed as a continuous medium, i.e., the detailed geometric information of the medium has been ignored. By using certain forcing schemes, an additional term is added to the evolution equation (of the flow field) to account for the presence of a porous medium, and the porosity, permeability and other statistical properties of the medium are included into the LB model based on some semi-empirical models (e.g., Darcy model, Brinkman-extended Darcy model, and generalized non-Darcy model) [6,7]. Obviously, the accuracy of the REV-scale approach depends heavily on the semi-empirical models. As compared with the pore-scale approach, the detailed local information of the flow and heat transfer in the pores is often unavailable in the REV-scale approach. However, this approach can be used for porous flow systems with large size, and it can produce reasonable results based on appropriate semi-empirical relations.

In the LB community, several excellent books [7,82-88] and comprehensive reviews [21,89-92] have been published, covering various aspects of the theory and applications of the LB method. In addition, it is worth mentioning that there have also been some reviews on the LB method for multiphase flows [93-95], particle suspension flows [96], micro- and nano-fluidics [97-99], reacting flows [100], turbulent flows [101], and fuel cells/flow batteries [102]. Moreover, in an interesting perspective by Succi [103], some encouraging prospects on the development of the LB method have been presented. Up to now, tremendous achievements have been made in the development of LB method for fluid flow and heat transfer in porous media, and the number of papers about the related areas continues to grow rapidly. With the increasing of fundamental research and practical applications, it is rather necessary to review the theoretical aspects and particular applications of the LB method for fluid flow and heat transfer in porous media.

The LB method for fluid flow and heat transfer in porous media is so diverse and interdisciplinary that it is impossible to include all the interesting topics. Hence, we limit this review to the theory and applications of the LB methods for single-phase and solid-liquid phase-change heat transfer in porous media, which are widely involved in energy/environmental science and technologies. The rest of the present review is organized as follows. Section 2 gives a brief introduction to the fundamentals of the LB method for fluid flow and heat transfer. The REV-scale LB method is comprehensively reviewed in Section 3. In Section 4, the LB method for solid-liquid phase-change heat transfer is reviewed. Section 5 reviews some applications of the LB methods for single-phase and solid-liquid phase-change heat transfer in porous media. Finally, Section 6 summarizes the key points of the present review and provides a brief discussion about the further developments of the LB method in the related areas.

## 2. The LB method for fluid flow and heat transfer

The LB method was born in 1988 [11], which can be viewed as a successor of the LGA method [18]. After many academics' tireless efforts, the LB method overcomes the shortcomings of the LGA method such as large statistical noise, non-Galilean invariance, and exponential complexity of the collision operator, while retains the merits of the LGA method (e.g., local nature of the operations, simplicity of formulation and implementation). McNamara and Zanetti [11] proposed the earliest LB model by replacing the Boolean particle variable of the LGA method by its ensemble average, i.e., the so-called single-particle distribution function. Accompanying such replacement, individual particle motion as well as particle-particle correlations in the kinetic equation have been neglected. From this respect, the LB method can be viewed as a successor of the LGA method. McNamara and Zanetti's model eliminates the statistical noise, but left all other issues of the LGA method unresolved. Subsequently, Higuera and Jiménez [12] introduced the relaxation lattice Boltzmann equation (RLBE) scheme, in which the collision operator was linearized by assuming that the particle distribution function is close to its local equilibrium state. However, the equilibrium distribution function and collision operator of the RLBE scheme were still obtained from the underlying LGA model. It was soon realized [13] that the collision matrix of the collision operator could be constructed independently regardless of the collision rules of the LGA method. According to this concept, Higuera et al. [13] proposed the RLBE model with an enhanced collision operator, which was shown to be linearly stable.

Based on the Bhatnagar-Gross-Krook (BGK) collision operator [104], the LB model using a single relaxation time, referred to as the BGK-LB model, has been independently proposed by Chen et al. [14], Koelman [15], and Qian et al. [16]. In the BGK-LB model, the equilibrium distribution function is chosen to recover the incompressible N-S equations in the incompressible limit. Due to its high efficiency and extreme simplicity, the BGK-LB model has become the most popular LB model, in spite of some well-known deficiencies (e.g., numerical instability, viscosity dependence of boundary locations). Almost at the same time when the BGK-LB method was developed, d'Humières [17] proposed the multiple-relaxation-time (MRT) LB method. The MRT-LB equation (also referred to as generalized lattice Boltzmann equation (GLBE)) is an important extension of the RLBE proposed by Higuera et al. [12,13]. This work is significant because, in addition to it overcomes some deficiencies of the BGK-LB model, it facilitates the extension of the LB method, expanding the applications to a much wider range of phenomena and processes.

In 1997, a direct connection between the LB equation and the continuous Boltzmann equation in kinetic theory was rigorously established by He and Luo [19,20]. In particular, it has been demonstrated that the LB equation can be viewed as a special finite difference scheme of the linearized continuous Boltzmann equation. The establishment of such connection not only makes the LB method more amenable to numerical analysis, but also puts the LB method on the solid theoretical foundation of kinetic theory.

*2.1. The isothermal LB method*

*2.1.1. The BGK-LB scheme*

The LB equation can be viewed as either a derivative of the LGA algorithm or an approximation to the continuous Boltzmann equation. In this review, we start from the continuous Boltzmann equation of the single-particle distribution function. An exhaustive overview of the fundamentals of the LB equation which are relevant to the LGA method can be found in Guo and Zheng [7], Wolf-Gladrow [82], and Succi [83]. The continuous Boltzmann equation can be written as [19,20,105,106]

$$\frac{\partial f}{\partial t} + \boldsymbol{\xi} \cdot \nabla_{\mathbf{x}} f + \mathbf{a} \cdot \nabla_{\boldsymbol{\xi}} f = \Omega(f), \qquad (1)$$

where $f = f(\mathbf{x}, \boldsymbol{\xi}, t)$ is the single-particle distribution function in phase space $\Gamma = (\mathbf{x}, \boldsymbol{\xi})$ (denotes the probability of finding a fluid particle at position $\mathbf{x}$ and time $t$ with microscopic velocity $\boldsymbol{\xi}$), $\mathbf{a}$ is the acceleration due to a body force acting on the particle, $\Omega(f)$ is the collision term representing the effect of inter-particle collisions.

Using the well-known single-relaxation-time approximation, the collision term $\Omega(f)$ can be expressed in the form known as the "BGK collision operator" [104]

$$\Omega(f) = -\frac{1}{\tau}(f - f^{eq}), \tag{2}$$

where $\tau$ is the relaxation time, and $f^{eq} = f^{eq}(\mathbf{x}, \boldsymbol{\xi}, t)$ is the continuous Maxwell-Boltzmann distribution function

$$f^{eq} = \frac{\rho}{(2\pi R_g T)^{D/2}} \exp\left[-\frac{(\boldsymbol{\xi} - \mathbf{u})^2}{2R_g T}\right], \tag{3}$$

where $R_g$, $D$, $T$, $\rho$, and $\mathbf{u}$ are the gas constant, spatial dimension, temperature, macroscopic density and velocity, respectively. Note that $f$ is also called as the density distribution function (the original single-particle distribution function is multiplied by the mass of particle without loss of generality). The macroscopic fluid variables are the (microscopic velocity) moments of the distribution function

$$\int f \, d\boldsymbol{\xi} = \int f^{eq} \, d\boldsymbol{\xi} = \rho, \tag{4}$$

$$\int f \boldsymbol{\xi} \, d\boldsymbol{\xi} = \int f^{eq} \boldsymbol{\xi} \, d\boldsymbol{\xi} = \rho \mathbf{u}, \tag{5}$$

$$\frac{1}{2}\int f(\boldsymbol{\xi} - \mathbf{u})^2 \, d\boldsymbol{\xi} = \frac{1}{2}\int f^{eq}(\boldsymbol{\xi} - \mathbf{u})^2 \, d\boldsymbol{\xi} = \rho e, \tag{6}$$

where $e$ is the internal energy given by $e = D_0 R_g T/2 = D_0 N_A k_B T/2$, in which $D_0$ is the number of degrees of freedom of a particle ($D_0 = 3$ and $5$ for monoatomic and diatomic gases, respectively), and $N_A$ and $k_B$ are the Avogadro's number and the Boltzmann constant, respectively.

The Boltzmann equation with the BGK collision operator (Boltzmann-BGK equation) has the

following form

$$\frac{\partial f}{\partial t}+\boldsymbol{\xi}\cdot\nabla_{\mathbf{x}}f+\mathbf{a}\cdot\nabla_{\boldsymbol{\xi}}f=-\frac{1}{\tau}\left(f-f^{eq}\right). \tag{7}$$

To obtain the LB equation, the Boltzmann-BGK equation (7) should be discretized in time and phase space. Integrating Eq. (7) over a time step $\delta_t$ (without considering the body force term), one can obtain [19,20]

$$f\left(\mathbf{x}+\boldsymbol{\xi}\delta_t,\boldsymbol{\xi},t+\delta_t\right)-f\left(\mathbf{x},\boldsymbol{\xi},t\right)=-\int_{t}^{t+\delta_t}\frac{f-f^{eq}}{\tau}dt. \tag{8}$$

Assuming that the integrand on the right-hand side of Eq. (8) is constant over a time step $\delta_t$, the first-order time discretization leads to the following time-discretized Boltzmann-BGK equation

$$f\left(\mathbf{x}+\boldsymbol{\xi}\delta_t,\boldsymbol{\xi},t+\delta_t\right)-f\left(\mathbf{x},\boldsymbol{\xi},t\right)=-\frac{1}{\tau_v}\left[f\left(\mathbf{x},t\right)-f^{eq}\left(\mathbf{x},t\right)\right], \tag{9}$$

where $\tau_v=\tau/\delta_t$ is the dimensionless relaxation time. Eq. (9) is the evolution equation of the distribution function $f$ with discrete time, which has first-order convergence in time (the leading terms neglected in the approximation are of order $O(\delta_t^2)$).

Discretizing Eq. (9) in phase space $\Gamma$, the following equation can be obtained [19,20]

$$f_i\left(\mathbf{x}+\mathbf{e}_i\delta_t,t+\delta_t\right)-f_i\left(\mathbf{x},t\right)=-\frac{1}{\tau_v}\left[f_i\left(\mathbf{x},t\right)-f_i^{eq}\left(\mathbf{x},t\right)\right], \tag{10}$$

where $f_i(\mathbf{x},t)=W_i f(\mathbf{x},\mathbf{e}_i,t)$ and $f_i^{eq}(\mathbf{x},t)=W_i f^{eq}(\mathbf{x},\mathbf{e}_i,t)$ are the discrete density distribution function and the corresponding equilibrium distribution function, respectively, in which $\{\mathbf{e}_i\}$ and $\{W_i\}$ are the abscissas (or discrete velocities) and the weights of the quadrature, respectively (see Refs. [19,20] for more details). The above equation is just the "standard" or "classical" LB equation. Through the phase space discretization step, the lattice structure and the form of the equilibrium distribution function have been constructed.

The set of the discrete velocities $\{\mathbf{e}_i\}$ is often referred to as the D$n$Q$b$ ($n$-dimensional $b$-velocity)

lattice [16]. The most widely used lattices are the D2Q9, D3Q15, and D3Q19 lattices (see Fig. 2), and the equilibrium distribution function $f_i^{eq}$ is given by [16]

$$f_i^{eq} = w_i \rho \left[ 1 + \frac{\mathbf{e}_i \cdot \mathbf{u}}{c_s^2} + \frac{\mathbf{uu} : (\mathbf{e}_i \mathbf{e}_i - c_s^2 \mathbf{I})}{2c_s^4} \right], \tag{11}$$

where $\mathbf{I}$ represents the unit vector (or unit tensor), $\{w_i\}$ are the weight coefficients, and $c_s = c/\sqrt{3}$ is the sound speed of the lattice ($c_s^2 = \sum_i w_i \mathbf{e}_i^2 / D$ [21]), in which $c = \delta_x / \delta_t$ is the lattice speed with $\delta_x$ being the lattice spacing. The equilibrium distribution function $f_i^{eq}$ is typically given by a second-order expansion in the Mach number ($Ma = |\mathbf{u}|/c_s$) of the continuous Maxwell-Boltzmann distribution function (3), it depends on the local macroscopic fluid variables ($\rho$ and $\mathbf{u}$) and satisfies the following constraints

$$\sum_i f_i^{eq} = \rho, \tag{12}$$

$$\sum_i \mathbf{e}_i f_i^{eq} = \rho \mathbf{u}, \tag{13}$$

$$\sum_i e_{i\alpha} e_{i\beta} f_i^{eq} = \rho u_\alpha u_\beta + p \delta_{\alpha\beta}, \tag{14}$$

$$\sum_i e_{i\alpha} e_{i\beta} e_{i\gamma} f_i^{eq} = c_s^2 \rho \left( u_\alpha \delta_{\beta\gamma} + u_\beta \delta_{\alpha\gamma} + u_\gamma \delta_{\alpha\beta} \right), \tag{15}$$

where Greek indices denote the Cartesian directions, and $p = \rho c_s^2$ is the pressure. The macroscopic density $\rho$ and velocity $\mathbf{u}$ are defined by

$$\rho = \sum_i f_i, \quad \rho \mathbf{u} = \sum_i \mathbf{e}_i f_i. \tag{16}$$

For the D2Q9 lattice model, the discrete velocities are given by

$$\mathbf{e} = c \begin{bmatrix} 0 & 1 & 0 & -1 & 0 & 1 & -1 & -1 & 1 \\ 0 & 0 & 1 & 0 & -1 & 1 & 1 & -1 & -1 \end{bmatrix}, \tag{17}$$

and the weight coefficients are $w_0 = 4/9$, $w_{1\sim4} = 1/9$, and $w_{5\sim8} = 1/36$.

For the D3Q15 lattice model, the discrete velocities are given by

$$\mathbf{e} = c \begin{bmatrix} 0 & 1 & -1 & 0 & 0 & 0 & 0 & 1 & -1 & 1 & -1 & 1 & -1 & 1 & -1 \\ 0 & 0 & 0 & 1 & -1 & 0 & 0 & 1 & 1 & -1 & -1 & 1 & 1 & -1 & -1 \\ 0 & 0 & 0 & 0 & 0 & 1 & -1 & 1 & 1 & 1 & 1 & -1 & -1 & -1 & -1 \end{bmatrix}, \quad (18)$$

and the weight coefficients are $w_0 = 2/9$, $w_{1\sim 6} = 1/9$, and $w_{7\sim 14} = 1/72$.

For the D3Q19 lattice model, the discrete velocities are given by

$$\mathbf{e} = c \begin{bmatrix} 0 & 1 & -1 & 0 & 0 & 0 & 0 & 1 & -1 & 1 & -1 & 1 & -1 & 1 & -1 & 0 & 0 & 0 & 0 \\ 0 & 0 & 0 & 1 & -1 & 0 & 0 & 1 & 1 & -1 & -1 & 0 & 0 & 0 & 0 & 1 & -1 & 1 & -1 \\ 0 & 0 & 0 & 0 & 0 & 1 & -1 & 0 & 0 & 0 & 0 & 1 & 1 & -1 & -1 & 1 & 1 & -1 & -1 \end{bmatrix}, \quad (19)$$

and the weight coefficients are $w_0 = 1/3$, $w_{1\sim 6} = 1/18$, and $w_{7\sim 18} = 1/36$.

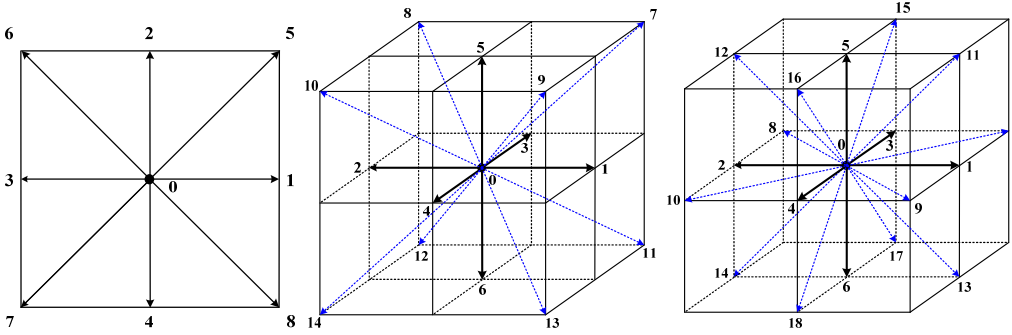

**Fig. 2.** Lattice structures of the D2Q9 (left), D3Q15 (middle), and D3Q19 (right) lattices.

Through the Chapman-Enskog analysis [91,107], the following macroscopic equations can be derived from the BGK-LB equation (10)

$$\frac{\partial \rho}{\partial t} + \nabla \cdot (\rho \mathbf{u}) = 0, \quad (20)$$

$$\frac{\partial (\rho \mathbf{u})}{\partial t} + \nabla \cdot (\rho \mathbf{u}\mathbf{u}) = -\nabla p + \nabla \cdot \mathbf{\Pi}, \quad (21)$$

where $\mathbf{\Pi} = \rho v \left[ \nabla \mathbf{u} + (\nabla \mathbf{u})^\mathrm{T} \right]$ is the shear stress tensor, and $v = c_s^2 (\tau_v - 0.5)\delta_t$ is the kinematic viscosity. $\mathbf{\Pi}$ can be rewritten as $\mathbf{\Pi} = 2\rho v \mathbf{\phi}$, where $\mathbf{\phi} = \left[ \nabla \mathbf{u} + (\nabla \mathbf{u})^\mathrm{T} \right]/2$ is the strain rate tensor. According to the Chapman-Enskog analysis of the BGK-LB equation (10), one can find that the strain rate tensor can be computed locally by [93]

$$\mathbf{\phi} \approx -\frac{1}{2\rho c_s^2 \tau_v \delta_t} \sum_i \mathbf{e}_i \mathbf{e}_i \left( f_i - f_i^{eq} \right). \quad (22)$$

In the phase space discretization step, the spatial discretization and (microscopic) velocity discretization are coupled together because the lattice spacing is chosen to be $\delta_x = c\delta_t$ with $c = \sqrt{3R_g T}$, which means that, the space is discretized in a way that is consistent with the kinetic equation (the coordinate of the nearest neighboring lattice node around $\mathbf{x}$ is $\mathbf{x} + \mathbf{e}_i \delta_t$). Therefore, the evolution process of the LB equation (10) can be divided into two processes: the collision process

$$f_i^*(\mathbf{x},t) = f_i(\mathbf{x},t) - \frac{1}{\tau_v}\left[f_i(\mathbf{x},t) - f_i^{eq}(\mathbf{x},t)\right], \tag{23}$$

and the streaming process

$$f_i(\mathbf{x} + \mathbf{e}_i \delta_t, t + \delta_t) = f_i^*(\mathbf{x},t). \tag{24}$$

It can be seen that the non-linear collision process (23) is completely local in time and space, and the non-local streaming process (24) is completely linear. The conceptual and practical simplicity of this numerical scheme lie at the heart of the computational efficiency and parallel implementation of the LB method, and form a solid basis for its success as a N-S solver "in kinetic disguise" [21].

To obtain the LB equation (10) from the Boltzmann-BGK equation (7), first-order time and phase space discretizations have been employed. Fortunately, the numerical viscosity due to discretization errors has a special form, which can be absorbed into the fluid viscosity, resulting in that the LB equation (10) has second-order accuracy both in time and space [7,91]. The macroscopic equations (20) and (21) reduce to the incompressible N-S equations in the low Mach number limit. From this perspective, the standard LB method can be viewed as a "pseudocompressible" numerical scheme for simulating incompressible flows. In order to recover the incompressible N-S equations, a couple of incompressible LB schemes have been developed [108,109].

*2.1.2. The MRT-LB scheme*

The MRT method was proposed by d'Humières [17] in 1992, which can be viewed as an important extension of the RLBE method developed by Higuera et al. [12,13]. In general, the standard MRT-LB equation can be written as (without a forcing term) [17,110,111]

$$f_i(\mathbf{x}+\mathbf{e}_i\delta_t, t+\delta_t) - f_i(\mathbf{x},t) = -\tilde{\Lambda}_{ij}\left[f_j(\mathbf{x},t) - f_j^{eq}(\mathbf{x},t)\right], \tag{25}$$

where $\tilde{\Lambda} = \mathbf{M}^{-1}\Lambda\mathbf{M}$ is the collision matrix, in which $\mathbf{M}$ is the transformation matrix, and $\Lambda$ is the relaxation matrix. Through the transformation matrix $\mathbf{M}$, the collision process of the MRT-LB equation (25) can be executed in the moment space $\mathbb{M} = \mathbb{R}^b$, i.e.,

$$\mathbf{m}^*(\mathbf{x},t) = \mathbf{m}(\mathbf{x},t) - \Lambda\left[\mathbf{m}(\mathbf{x},t) - \mathbf{m}^{eq}(\mathbf{x},t)\right], \tag{26}$$

where $\mathbf{m}^* = |m_i^*\rangle$, while the streaming process is still carried out in the velocity space $\mathbb{V} = \mathbb{R}^b$ with $\mathbf{f}^* = |f^*\rangle = \mathbf{M}^{-1}\mathbf{m}^*$ (see Eq. (24)).

The velocity space $\mathbb{V} = \mathbb{R}^b$ (spanned by $\mathbf{f} = |f\rangle$) and the moment space $\mathbb{M} = \mathbb{R}^b$ (spanned by $\mathbf{m} = |m\rangle$) are related through the following linear mapping

$$\mathbf{m} = \mathbf{M}\mathbf{f}, \quad \mathbf{f} = \mathbf{M}^{-1}\mathbf{m}. \tag{27}$$

Here, $\mathbf{f}$ represents the vector of the distribution functions $\{f_i\}$, and $\mathbf{m}$ represents the vector of the (velocity) moments $\{m_i\}$. For brevity, the Dirac notation $|\cdot\rangle$ is used to denote a b-dimensional column vector, e.g., $|m\rangle = (m_0, m_1, \ldots, m_{b-1})^\mathrm{T}$. For the D2Q9 model, $\mathbf{m}$ is given by [110]

$$\mathbf{m} = (\rho, e, \varepsilon, j_x, q_x, j_y, q_y, p_{xx}, p_{xy})^\mathrm{T}. \tag{28}$$

Each moment is associated with one physical quantity [110]: $m_0 = \rho$ is the density, $m_1 = e$ is related to energy, $m_2 = \varepsilon$ is related to energy square, $m_{3,5} = j_{x,y}$ are components of the momentum $\rho\mathbf{u} = \rho(u_x, u_y)$ (mass flux), $m_{4,6} = q_{x,y}$ are related to energy flux, and $m_{7,8} = p_{xx,xy}$ correspond to diagonal and off-diagonal components of the strain rate tensor. The transformation matrix $\mathbf{M}$ can be found in Ref. [110]. The relaxation matrix $\Lambda$ is given by

$$\Lambda=\text{diag}\left(s_\rho, s_e, s_\varepsilon, s_j, s_q, s_j, s_q, s_\nu, s_\nu\right). \tag{29}$$

The vector of the equilibrium moments $\mathbf{m}^{eq}$ corresponding to $\mathbf{m}$ is defined as ( $c=1$ ) [110]

$$\mathbf{m}^{eq} = \rho\left(1, -2+3|\mathbf{u}|^2, \tilde{\alpha}+\tilde{\beta}|\mathbf{u}|^2, u_x, -u_x, u_y, -u_y, u_x^2-u_y^2, u_x u_y\right)^\mathrm{T}, \tag{30}$$

where $\tilde{\alpha}$ and $\tilde{\beta}$ are free parameters. When $\tilde{\alpha}=1$ and $\tilde{\beta}=-3$, the equilibrium distribution function $f_i^{eq}$ in velocity space (see Eq. (11)) can be obtained via $\mathbf{f}^{eq} = \mathbf{M}^{-1}\mathbf{m}^{eq}$. From here on, $c$ is set to be 1 unless otherwise specified.

Through the Chapman-Enskog analysis of the MRT-LB equation (25), the macroscopic equations (20) and (21) can be derived with the shear stress tensor $\Pi$ is given by [7]

$$\Pi = \rho\nu\left[\nabla\mathbf{u}+(\nabla\mathbf{u})^\mathrm{T}\right] + \rho(\nu_\mathrm{B}-\nu)(\nabla\cdot\mathbf{u})\mathbf{I}. \tag{31}$$

The kinematic viscosity $\nu$ and bulk viscosity $\nu_\mathrm{B}$ are given by $\nu = c_s^2\left(s_\nu^{-1}-0.5\right)\delta_t$ and $\nu_\mathrm{B} = c_s^2\left(s_e^{-1}-0.5\right)\delta_t$, respectively. Obviously, the BGK-LB model is a special case of the MRT-LB model in which the relaxation rates are all equal to $1/\tau_\nu$, i.e., the relaxation matrix is given by $\Lambda=(1/\tau_\nu)\mathbf{I}$ with $s_\nu=1/\tau_\nu$.

In the MRT-LB method, the strain rate tensor $\boldsymbol{\phi}=\left[\nabla\mathbf{u}+(\nabla\mathbf{u})^\mathrm{T}\right]/2$ can be calculated from the non-equilibrium part of the moments, i.e., $m_i - m_i^{eq}$. According to the Chapman-Enskog analysis, the following relations can be obtained [7]

$$e^{(1)} \approx -\frac{2}{s_e}\rho\delta_t\left(\partial_x u_x + \partial_y u_y\right), \tag{32}$$

$$p_{xx}^{(1)} \approx -\frac{2}{3s_\nu}\rho\delta_t\left(\partial_x u_x - \partial_y u_y\right), \tag{33}$$

$$p_{xy}^{(1)} \approx -\frac{1}{3s_\nu}\rho\delta_t\left(\partial_x u_y + \partial_y u_x\right). \tag{34}$$

Obviously, there is no need to restrict $c=1$ in the LB model, and in general, the vector of the equilibrium moments $\mathbf{m}^{eq}$ can be expressed as

$$\mathbf{m}^{eq} = \rho \left(1, -2 + \frac{3|\mathbf{u}|^2}{c^2}, \tilde{\alpha} + \frac{\tilde{\beta}|\mathbf{u}|^2}{c^2}, \frac{u_x}{c}, -\frac{u_x}{c}, \frac{u_y}{c}, -\frac{u_y}{c}, \frac{u_x^2 - u_y^2}{c^2}, \frac{u_x u_y}{c^2}\right)^T. \qquad (35)$$

The basic theory of the MRT method has been briefly introduced. For the sake of brevity and readability, only the D2Q9-MRT model is introduced. The transformation matrix $\mathbf{M}$, the relaxation matrix $\mathbf{\Lambda}$, and the equilibrium moments $\mathbf{m}^{eq}$ of the D3Q15-MRT or D3Q19-MRT models can be found in Ref. [111]. The transformation matrix $\mathbf{M}$ of the MRT method is usually constructed based on orthogonal basis vectors obtained from the combinations of the lattice velocity components, i.e., it is an orthogonal matrix. It should be stressed that the transformation matrix is not necessary to be an orthogonal one. The transformation matrix can be constructed based on non-orthogonal basis vectors, which leads to the so-called non-orthogonal MRT-LB method [112,113]. The non-orthogonal MRT-LB model for incompressible isothermal flows provides solutions that are asymptotically consistent with the N-S equations to second-order in diffusive scaling [112,113].

Needless to say, the BGK method of Qian et al. [16] is still the most popular one in the LB community, and since its introduction 26 years ago, there have been significant developments in the theory and applications of this method. The MRT method [17], which is superior to its BGK counterpart in terms of accuracy and numerical stability, becomes more and more popular in the LB community since a detailed theoretical analysis was made by Lallemand and Luo [110] in 2000. In addition, efforts have also been made to develop efficient and accurate LB methods from different points of view. In the past two decades, several alternative LB methods, such as the entropic LB method [114,115], the two-relaxation-time LB method [116,117], and the cascaded (or central moment) LB method [112,118-122], have been proposed. These methods have also attracted significant attention.

*2.1.3. The forcing schemes*

For fluid systems in which a body force (external or internal force) is involved, the body force should be treated appropriately in order to obtain the correct hydrodynamics from the LB equation. Therefore, the forcing scheme, which is used to include the body force into the LB equation, plays an important role in the LB method. Many forcing schemes have been proposed to design LB models for fluid systems with body forces [7,123]. In what follows, the forcing schemes [106,123,124] with the discrete lattice effect being considered will be briefly reviewed.

*2.1.3.1. He et al.'s forcing scheme.* Starting from the continuous Boltzmann-BGK equation (7), He et al. [106] devised a forcing scheme to consider the effect of the body force. In order to derive the forcing term, the derivative $\nabla_{\xi} f$ has to be explicitly given. However, because the dependence of the distribution function $f$ on the microscopic velocity $\xi$ is unknown, the derivative $\nabla_{\xi} f$ cannot be calculated directly. Therefore, He et al. [106] made the following assumption

$$\nabla_{\xi} f \approx \nabla_{\xi} f^{eq} = -\frac{\xi - \mathbf{u}}{R_g T} f^{eq}. \tag{36}$$

The above assumption is based on the fact that $f^{eq}$ is the leading part of the distribution function $f$, and the gradient of $f^{eq}$ has the most important contribution to the gradient of $f$. Consequently, the following simplified continuous Boltzmann-BGK equation can be obtained

$$\frac{\partial f}{\partial t} + \xi \cdot \nabla_x f = -\frac{1}{\tau}\left(f - f^{eq}\right) + \frac{\mathbf{F} \cdot (\xi - \mathbf{u})}{\rho R_g T} f^{eq}, \tag{37}$$

where $\mathbf{F} = \rho \mathbf{a}$ is the body force. Integrating Eq. (37) over a time step $\delta_t$ leads to

$$f\left(\mathbf{x} + \xi \delta_t, \xi, t + \delta_t\right) - f\left(\mathbf{x}, \xi, t\right) = -\int_{t}^{t+\delta_t} \frac{f - f^{eq}}{\tau} dt + \int_{t}^{t+\delta_t} \frac{\mathbf{F} \cdot (\xi - \mathbf{u})}{\rho R_g T} f^{eq} dt. \tag{38}$$

As mentioned before, the integrand of the first integral on the right-hand side of Eq. (38) is assumed to be constant over a time step $\delta_t$. The first integral can be treated explicitly via the first-order approximation, while a trapezoidal rule is necessary for the second one so as to achieve second-order accuracy in time [106]. With the above considerations, Eq. (38) becomes

$$f\left(\mathbf{x}+\boldsymbol{\xi}\delta_t,\boldsymbol{\xi},t+\delta_t\right)-f\left(\mathbf{x},\boldsymbol{\xi},t\right)=-\frac{f-f^{eq}}{\tau_v}\bigg|_t+\frac{\delta_t}{2}\left[\frac{\mathbf{F}\cdot\left(\boldsymbol{\xi}-\mathbf{u}\right)}{\rho R_g T}f^{eq}\bigg|_{t+\delta_t}+\frac{\mathbf{F}\cdot\left(\boldsymbol{\xi}-\mathbf{u}\right)}{\rho R_g T}f^{eq}\bigg|_t\right]. \quad (39)$$

To eliminate the implicitness of Eq. (39), the following variable transformation is introduced

$$\bar{f}=f-\frac{\delta_t}{2}\frac{\mathbf{F}\cdot\left(\boldsymbol{\xi}-\mathbf{u}\right)}{\rho R_g T}f^{eq}. \quad (40)$$

Using $\bar{f}$, Eq. (39) becomes

$$\bar{f}\left(\mathbf{x}+\boldsymbol{\xi}\delta_t,\boldsymbol{\xi},t+\delta_t\right)-\bar{f}\left(\mathbf{x},\boldsymbol{\xi},t\right)=-\frac{\bar{f}-f^{eq}}{\tau_v}+\delta_t\left(1-\frac{1}{2\tau_v}\right)\frac{\mathbf{F}\cdot\left(\boldsymbol{\xi}-\mathbf{u}\right)}{\rho R_g T}f^{eq}. \quad (41)$$

Through the phase space discretization step, the following BGK-LB equation can be obtained

$$\bar{f}_i\left(\mathbf{x}+\mathbf{e}_i\delta_t,t+\delta_t\right)-\bar{f}_i\left(\mathbf{x},t\right)=-\frac{1}{\tau_v}\left[\bar{f}_i\left(\mathbf{x},t\right)-f_i^{eq}\left(\mathbf{x},t\right)\right]+\delta_t F_i, \quad (42)$$

where $F_i$ is He et al.'s forcing term

$$F_i=\left(1-\frac{1}{2\tau_v}\right)\frac{\mathbf{F}\cdot\left(\mathbf{e}_i-\mathbf{u}\right)}{\rho c_s^2}f_i^{eq}, \quad (43)$$

in which $c_s^2=R_g T$. According to Eq. (40), the macroscopic density $\rho$ and velocity $\mathbf{u}$ should be defined by

$$\rho=\sum_i \bar{f}_i, \quad \rho\mathbf{u}=\sum_i \mathbf{e}_i\bar{f}_i+\frac{\delta_t}{2}\mathbf{F}. \quad (44)$$

Eqs. (42)-(44) constitute He et al.'s forcing scheme. Note that Eqs. (42) and (44) are self-consistent, the hat "-" of $\bar{f}_i$ in Eqs. (42) and (44) can be dropped in practical applications. Nevertheless, Li et al. [93] stressed that the transformation given by Eq. (40) should be kept in mind in certain cases.

*2.1.3.2. Guo et al.'s forcing scheme.* From a different point of view, Guo et al. [123] proposed a forcing scheme with the discrete lattice effect as well as the contributions of the body force to the momentum flux being considered. In the presence of a body force, Guo et al. pointed out that the LB equation (10) must be modified to account for the force. By adding an additional term to the LB equation (10), one can obtain [123]

$$f_i\left(\mathbf{x}+\mathbf{e}_i\delta_t, t+\delta_t\right) - f_i(\mathbf{x},t) = -\frac{1}{\tau_v}\left[f_i(\mathbf{x},t) - f_i^{eq}(\mathbf{x},t)\right] + \delta_t F_i. \tag{45}$$

The forcing term $F_i$ can be written in a power series in the discrete velocity [96,123]

$$F_i = w_i\left[A + \frac{\mathbf{B}\cdot\mathbf{e}_i}{c_s^2} + \frac{\mathbf{C}:\left(\mathbf{e}_i\mathbf{e}_i - c_s^2\mathbf{I}\right)}{2c_s^4}\right], \tag{46}$$

where $A$, $\mathbf{B}$, and $\mathbf{C}$ are functions of $\mathbf{F}$, which can be determined by requiring that the moments of $F_i$ are consistent with the target hydrodynamic equations. The zeroth-order to second-order moments of $F_i$ are given by [123]

$$\sum_i F_i = A, \quad \sum_i \mathbf{e}_i F_i = \mathbf{B}, \quad \sum_i \mathbf{e}_i\mathbf{e}_i F_i = c_s^2 A\mathbf{I} + \frac{1}{2}\left(\mathbf{C}+\mathbf{C}^\mathrm{T}\right). \tag{47}$$

In order to recover the correct hydrodynamic equations at the N-S level, $A$, $\mathbf{B}$, and $\mathbf{C}$ should be chosen as follows [123]

$$A = 0, \quad \mathbf{B} = \left(1 - \frac{1}{2\tau_v}\right)\mathbf{F}, \quad \mathbf{C} = \left(1 - \frac{1}{2\tau_v}\right)(\mathbf{uF}+\mathbf{Fu}). \tag{48}$$

Substituting Eq. (48) into Eq. (46) gives

$$F_i = w_i\left(1 - \frac{1}{2\tau_v}\right)\left[\frac{\mathbf{e}_i\cdot\mathbf{F}}{c_s^2} + \frac{\mathbf{uF}:\left(\mathbf{e}_i\mathbf{e}_i - c_s^2\mathbf{I}\right)}{c_s^4}\right]. \tag{49}$$

Accordingly, the macroscopic density $\rho$ and velocity $\mathbf{u}$ are defined by

$$\rho = \sum_i f_i, \quad \rho\mathbf{u} = \sum_i \mathbf{e}_i f_i + \frac{\delta_t}{2}\mathbf{F}. \tag{50}$$

Eqs. (45), (49), and (50) constitute Guo et al.'s forcing scheme. Although the forms of the forcing terms are different, He et al.'s and Guo et al.'s forcing schemes are basically the same. It has been shown that the difference between He et al.'s and Guo et al.'s forcing schemes lies in the second-order moment of the forcing term (see Refs. [7,93] for more details).

*2.1.3.3. MRT forcing scheme.* Following He et al.'s work [106], McCracken and Abraham [124] proposed a forcing scheme for the MRT-LB method in 2005. The MRT-LB equation with an explicit treatment of the forcing term can be written as [124]

$$\bar{f}_i\left(\mathbf{x}+\mathbf{e}_i\delta_t,t+\delta_t\right)-\bar{f}_i\left(\mathbf{x},t\right)=-\tilde{\Lambda}_{ij}\left(\bar{f}_j-f_j^{eq}\right)\Big|_{(\mathbf{x},t)}+\delta_t\left(\tilde{S}_i-0.5\tilde{\Lambda}_{ij}\tilde{S}_j\right)\Big|_{(\mathbf{x},t)}, \quad (51)$$

where

$$\bar{f}_i = f_i - \frac{\delta_t}{2}\tilde{S}_i, \quad \tilde{S}_i = \frac{\mathbf{F}\cdot(\mathbf{e}_i-\mathbf{u})}{\rho c_s^2}f_i^{eq}. \quad (52)$$

Through the transformation matrix $\mathbf{M}$, the collision process of the MRT-LB equation (51) can be executed in moment space

$$\bar{\mathbf{m}}^*(\mathbf{x},t) = \bar{\mathbf{m}}(\mathbf{x},t) - \Lambda\left(\bar{\mathbf{m}}-\mathbf{m}^{eq}\right)\Big|_{(\mathbf{x},t)} + \delta_t\left(\mathbf{I}-\frac{\Lambda}{2}\right)\mathbf{S}, \quad (53)$$

where $\bar{\mathbf{m}} = \mathbf{M}\bar{\mathbf{f}}$, and $\mathbf{S} = \mathbf{M}\tilde{\mathbf{S}}$, in which $\bar{\mathbf{f}} = |\bar{f}\rangle$, and $\tilde{\mathbf{S}} = |\tilde{S}\rangle$. The streaming process is carried out in velocity space

$$\bar{f}_i\left(\mathbf{x}+\mathbf{e}_i\delta_t,t+\delta_t\right) = \bar{f}_i^*(\mathbf{x},t), \quad (54)$$

where $\bar{\mathbf{f}}^* = \mathbf{M}^{-1}\bar{\mathbf{m}}^*$. The forcing term in the moment space is $(\mathbf{I}-0.5\Lambda)\mathbf{S}$, and for the D2Q9 model, $\mathbf{S}$ is given by [124]

$$\mathbf{S} = \left[0, 6\mathbf{u}\cdot\mathbf{F}, -6\mathbf{u}\cdot\mathbf{F}, F_x, -F_x, F_y, -F_y, 2\left(u_xF_x-u_yF_y\right), u_xF_y+u_yF_x\right]^\mathrm{T}, \quad (55)$$

where $\mathbf{F} = (F_x, F_y)$. The macroscopic density $\rho$ and velocity $\mathbf{u}$ are also given by Eq. (44). Note that the second-order and third-order velocity terms have been neglected in Eq. (55). The same $\mathbf{S}$ can be obtained based on Guo et al.'s forcing term [123]. In addition, corresponding to the equilibrium moments given by Eq. (35), **S** should be chosen as

$$\mathbf{S} = \left(0, \frac{6\mathbf{u}\cdot\mathbf{F}}{c^2}, -\frac{6\mathbf{u}\cdot\mathbf{F}}{c^2}, \frac{F_x}{c}, -\frac{F_x}{c}, \frac{F_y}{c}, -\frac{F_y}{c}, \frac{2(u_x F_x - u_y F_y)}{c^2}, \frac{u_x F_y + u_y F_x}{c^2}\right)^{\mathrm{T}}. \quad (56)$$

In 2007, Premnath and Abraham [125] extended McCracken and Abraham's MRT forcing scheme to 3D cases. For the D3Q15 model, the components of **S** are given as follows

$$S_0 = 0, \; S_1 = 2\mathbf{u}\cdot\mathbf{F}, \; S_2 = -10\mathbf{u}\cdot\mathbf{F}, \; S_3 = F_x, \; S_4 = -\frac{7}{3}F_x, \; S_5 = F_y, \; S_6 = -\frac{7}{3}F_y,$$

$$S_7 = F_z, \; S_8 = -\frac{7}{3}F_z, \; S_9 = 2(2u_x F_x - u_y F_y - u_z F_z), \; S_{10} = 2(u_y F_y - u_z F_z),$$

$$S_{11} = u_x F_y + u_y F_x, \; S_{12} = u_y F_z + u_z F_y, \; S_{13} = u_x F_z + u_z F_x, \; S_{14} = 0, \quad (57)$$

where $\mathbf{F} = (F_x, F_y, F_z)$. For the D3Q19 model, the components of **S** are given as follows

$$S_0 = 0, \; S_1 = 38\mathbf{u}\cdot\mathbf{F}, \; S_2 = -11\mathbf{u}\cdot\mathbf{F}, \; S_3 = F_x, \; S_4 = -\frac{2}{3}F_x, \; S_5 = F_y, \; S_6 = -\frac{2}{3}F_y,$$

$$S_7 = F_z, \; S_8 = -\frac{2}{3}F_z, \; S_9 = 2(2u_x F_x - u_y F_y - u_z F_z), \; S_{10} = -(2u_x F_x - u_y F_y - u_z F_z),$$

$$S_{11} = 2(u_y F_y - u_z F_z), \; S_{12} = -(u_y F_y - u_z F_z), \; S_{13} = u_x F_y + u_y F_x,$$

$$S_{14} = u_y F_z + u_z F_y, \; S_{15} = u_x F_z + u_z F_x, \; S_{16} = S_{17} = S_{18} = 0. \quad (58)$$

The MRT forcing scheme [124] is an extension of the BGK forcing scheme of He et al. [106]. Due to the freedom of the MRT model, the forcing terms or any other source terms in the MRT-LB equation can be directly constructed in the moment space [7,93]. This feature is very useful for developing LB models for simulating multiphase flows and interfacial dynamics [126-128]. In the above mentioned forcing schemes, the discrete lattice effect has been taken into account. Through the Chapman-Enskog analysis, the macroscopic equations (20) and (21) with a body force can be recovered at the N-S level. In the literature [25,96,129-132], several forcing schemes have also been proposed to treat the body force. Detailed analyses of these forcing schemes can be found in Refs. [7,123].

*2.1.4. Boundary schemes*

Boundary conditions are crucial for meaningful simulations since they select solutions which are compatible with external constraints [83]. The boundary conditions in the LB method are expressed in terms of the distribution function $f_i$ rather than the flow variables commonly controlled in conventional numerical methods. The boundary schemes are used to determine the unknown incoming distribution functions based on the known outgoing ones at the boundary nodes, such that the physical boundary conditions of interest can be satisfied. Tremendous efforts have been devoted to devise efficient and accurate boundary schemes for treating different boundary conditions. Discussions of the boundary schemes in the LB method can be found in the books and reviews [7,83,85,87,91,96]. In what follows, the bounce-back scheme and the non-equilibrium extrapolation scheme are briefly summarized.

*2.1.4.1. Bounce-back scheme*. The bounce-back scheme [22-28] is a simple and straightforward way to impose no-slip boundary conditions at the fluid-solid interfaces in LB simulations. Bounce-back scheme in the LB method was originally taken from the LGA method [91]. Here, a particle that meets a wall node is simply reflected back with a reversed velocity, i.e., an incoming distribution function is simply equal to the corresponding outgoing one with opposite momentum. In the standard bounce-back scheme, the boundary location is situated at the solid wall, and the collision process does not occur at the boundary (fluid) nodes. It was demonstrated that the standard bounce-back scheme is only first-order in numerical accuracy at boundary [22-25]. An alternative to the standard bounce-back scheme is to place the boundary node ($\mathbf{x}_b$) half-way between the wall node ($\mathbf{x}_w$) and the adjacent fluid node ($\mathbf{x}_f$), resulting in the so-called half-way bounce-back scheme [23]. This boundary scheme can be expressed as

$$f_{\bar{i}}(\mathbf{x}_\mathrm{f}, t+\delta_t) = f_i^*(\mathbf{x}_\mathrm{f}, t), \tag{59}$$

where $\bar{i}$ represents the opposite direction of $i$ (i.e., $\mathbf{e}_{\bar{i}} = -\mathbf{e}_i$). The collision process is carried out at all fluid nodes. The half-way bounce-back scheme considers that the physical boundary is located right in the middle between the wall and fluid nodes, which makes it to be second-order accuracy in space [23].

It is important to stress that the precise location where the no-slip boundary conditions are satisfied is model dependent. The actual location of the boundary is viscosity dependent in the BGK-LB model when the half-way bounce-back scheme is applied, especially in under-relaxation situations (i.e., $\tau_v > 1$) [25]. This problem can be solved by the MRT-LB model through combining the half-way bounce-back scheme with carefully constructed collision operator [28]. The boundary location depends on a polynomial of the relaxation rates $\{s_i\}$, one can choose the relaxation rate $s_q$ as $s_q = 8(2-s_v)/(8-s_v)$ [26-28]. For sophisticated bounce-back schemes such as the interpolated bounce-back scheme, readers are referred to Refs. [26-28].

For practical simulations the half-way bounce-back scheme is very attractive because it is a simple and computationally efficient scheme for imposing no-slip walls with irregular geometries. This feature supports the idea that the LB method is ideal for simulating fluid flows in porous media at the pore scale. It has been demonstrated [28,41] that with the bounce-back scheme at the solid walls, the single-relaxation-time approximation of the BGK-LB method results in viscosity-dependent permeability for single-phase flow in porous media. This deficiency can be avoided by using the MRT-LB method, which can be seen in Fig. 3. As shown in the figure, the permeability computed by the BGK model varies significantly as the viscosity changes, while the permeabilities calculated by the MRT model are constant for different viscosities. Generally speaking, the MRT-LB method is more suitable for simulating porous flows at the pore scale.

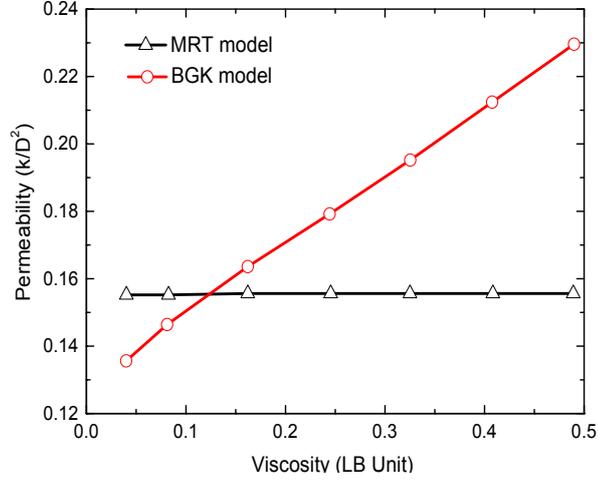

**Fig. 3.** Comparison of the permeability computed by the BGK and MRT models with various viscosities for single-phase flow in a reconstructed porous medium (carbon paper gas diffusion layer). (Adapted from Hao and Cheng [41].)

*2.1.4.2. Non-equilibrium extrapolation scheme.* The non-equilibrium extrapolation scheme was proposed by Guo et al. [133,134]. The basic idea of this scheme is to decompose the distribution function $f_i$ at a boundary node $\mathbf{x}_b$ into two parts: the equilibrium part and the non-equilibrium part, i.e.,

$$f_i(\mathbf{x}_b, t) = f_i^{eq}(\mathbf{x}_b, t) + f_i^{neq}(\mathbf{x}_b, t), \tag{60}$$

where $f_i^{neq}(\mathbf{x}_b, t)$ is the non-equilibrium part. For velocity boundary condition where $\mathbf{u}(x_b, t)$ is known but $\rho(\mathbf{x}_b, t)$ is unknown, the equilibrium distribution function $f_i^{eq}(\mathbf{x}_b, t)$ can be approximated by

$$\bar{f}_i^{eq}(\mathbf{x}_b, t) = f_i^{eq}(\rho(\mathbf{x}_f), \mathbf{u}_b, t), \tag{61}$$

where $\mathbf{x}_f$ is the nearest neighboring fluid node of $\mathbf{x}_b$ along the link $\mathbf{e}_i$, i.e., $\mathbf{x}_f = \mathbf{x}_b + \mathbf{e}_i \delta_t$. The non-equilibrium part $f_i^{neq}(\mathbf{x}_b, t)$ is approximated by

$$f_i^{neq}\left(\mathbf{x}_b,t\right) \approx f_i^{neq}\left(\mathbf{x}_f,t\right) = f_i\left(\mathbf{x}_f,t\right) - f_i^{eq}\left(\mathbf{x}_f,t\right). \tag{62}$$

Finally, the following boundary scheme for specifying the unknown distribution functions at a boundary node $\mathbf{x}_b$ can be obtained

$$f_i\left(\mathbf{x}_b,t\right) = \overline{f}_i^{eq}\left(\mathbf{x}_b,t\right) + \left[f_i\left(\mathbf{x}_f,t\right) - f_i^{eq}\left(\mathbf{x}_f,t\right)\right]. \tag{63}$$

The non-equilibrium extrapolation scheme needs no additional conditions in simulations, and can be used for both steady and unsteady flow in theory. It has been proved that this boundary scheme is of second-order accuracy in both time and space [133,134].

*2.2. The thermal LB method*

A major area of the LB applications is the simulation of a wide variety of thermal problems. In 1993, Massaioli et al. [135], Alexander et al. [136], and Qian [137] made the earliest attempts to construct thermal LB models. Since then, many thermal LB models have been proposed. In order to take the thermal effects into account, three typical approaches have been developed [7,85,87]: the multispeed approach [135-137], the double-distribution-function (DDF) approach [138-148], and the hybrid approach [149-152]. The multispeed approach was first proposed by Alexander et al. [136], in which only the density distribution function is used. As a straightforward extension of the isothermal LB method, this approach utilizes high-order lattices (i.e., additional discrete velocities are necessary) and the equilibrium distribution function contains higher-order velocity terms so that the energy equation can be recovered at the macroscopic level. However, the multispeed approach has several limitations: the numerical instability, the fixed value of the Prandtl number, and the narrow range of temperature variation. These limitations in the multispeed LB models severely restrict their applications.

The DDF approach utilizes two different distribution functions: one (density distribution function) for the flow field and the other for the energy (or temperature) field. According to the target macroscopic energy equations, the existing DDF-LB models mostly fall into the following three categories [93]: the temperature-based (or passive scalar) model [138-141], the internal-energy-based model [142-145], and the total-energy-based model [146-148]. The limitations of the multispeed approach can be partly overcome by the DDF approach. This approach has attracted much attention because of its excellent numerical stability and the adjustability of the Prandtl number. The framework of the hybrid approach was formally established by Lallemand and Luo in 2003 [149,150]. In fact, before Lallemand and Luo's work, this approach has already been used to simulate low Mach number combustion by Filippova and Hänel [151,152]. In the hybrid approach, the flow simulation is accomplished by the usual isothermal LB model, while the temperature field is solved separately by the FDM (or by other means). In what follows, we briefly introduce the internal-energy- and temperature-based DDF models without compression work and viscous heat dissipation, which are widely used in the simulations of single-phase and solid-liquid phase change heat transfer in porous media.

*2.2.1. The internal-energy- and temperature-based DDF models*

For thermal flows, the compression work and viscous heat dissipation can be neglected in many circumstances, and then the internal energy equation is given by

$$\frac{\partial}{\partial t}\left(\rho c_p T\right) + \nabla \cdot \left(\rho c_p T \mathbf{u}\right) = \nabla \cdot \left(k \nabla T\right), \tag{64}$$

where $k$ is the thermal conductivity, $c_p$ is the specific heat at constant pressure. For incompressible thermal flows ( $\rho \approx \rho_0 = \text{const}$ ), the internal energy equation (64) can be further simplified as

$\partial_t(c_p T) + \nabla \cdot (c_p T \mathbf{u}) = \nabla \cdot \left(\frac{k}{\rho_0} \nabla T\right)$. If the specific heat $c_p$ is constant, Eq. (64) reduces to the following temperature equation

$$\frac{\partial T}{\partial t} + \nabla \cdot (T\mathbf{u}) = \nabla \cdot (\alpha \nabla T), \tag{65}$$

where $\alpha = k/(\rho c_p)$ is the thermal diffusivity.

*2.2.1.1. The internal-energy-based DDF models.* Based on He et al.'s work [142], several simplified internal-energy-based DDF models have been developed, e.g., by Peng et al. [143], Shi et al. [144], and Li et al. [145], to name a few. The thermal BGK-LB equation devised for solving Eq. (64) is given by

$$g_i(\mathbf{x} + \mathbf{e}_i \delta_t, t + \delta_t) - g_i(\mathbf{x}, t) = -\frac{1}{\tau_g}\left[g_i(\mathbf{x}, t) - g_i^{eq}(\mathbf{x}, t)\right], \tag{66}$$

where $g_i$ is the internal energy distribution function, $g_i^{eq}$ is the equilibrium internal energy distribution function, and $\tau_g$ is the relaxation time related to the thermal conductivity. Generally, the equilibrium internal energy distribution function $g_i^{eq}$ can be chosen as

$$g_i^{eq} = \tilde{w}_i \rho c_p T \left[1 + \frac{\mathbf{e}_i \cdot \mathbf{u}}{c_{sT}^2} + \vartheta \frac{\mathbf{u}\mathbf{u} : (\mathbf{e}_i \mathbf{e}_i - c_{sT}^2 \mathbf{I})}{2c_{sT}^4}\right], \tag{67}$$

where $\{\tilde{w}_i\}$ are the weight coefficients, $c_{sT}$ is the sound speed of the lattice ($c_{sT}^2 = \sum_i \tilde{w}_i \mathbf{e}_i^2 / D$ [21]), and $\vartheta \in \{0,1\}$. For 2D cases, the D2Q4, D2Q5, and D2Q9 lattices can be employed (see Ref. [153] and the references therein); for 3D cases, the D3Q7, D3Q15, and D3Q19 lattices are usually employed. The parameters of the lattice models are listed in Table 1. The thermal conductivity $k$ is given by $k = \rho c_p c_{sT}^2 (\tau_g - 0.5) \delta_t$. For the D2Q4 lattice model, the discrete velocities are given by

$$\mathbf{e} = [\mathbf{e}_1, \mathbf{e}_2, \mathbf{e}_3, \mathbf{e}_4] = c\begin{bmatrix} 1 & 0 & -1 & 0 \\ 0 & 1 & 0 & -1 \end{bmatrix}. \tag{68}$$

For the D2Q5 lattice model, the discrete velocities are given by

$$\mathbf{e} = [\mathbf{e}_0, \mathbf{e}_1, \mathbf{e}_2, \mathbf{e}_3, \mathbf{e}_4] = c\begin{bmatrix} 0 & 1 & 0 & -1 & 0 \\ 0 & 0 & 1 & 0 & -1 \end{bmatrix}. \tag{69}$$

For the D3Q7 lattice model, the discrete velocities are given by

$$\mathbf{e} = [\mathbf{e}_0, \mathbf{e}_1, \mathbf{e}_2, \mathbf{e}_3, \mathbf{e}_4, \mathbf{e}_5, \mathbf{e}_6] = c \begin{bmatrix} 0 & 1 & -1 & 0 & 0 & 0 & 0 \\ 0 & 0 & 0 & 1 & -1 & 0 & 0 \\ 0 & 0 & 0 & 0 & 0 & 1 & -1 \end{bmatrix}. \tag{70}$$

**Table 1** Parameters of the lattice models for 2D and 3D ($0 < \varpi < 1$, $c = 1$).

| Lattice | $\tilde{w}_i$ | $c_{sT}$ | $\vartheta$ |
|---|---|---|---|
| D2Q4 | $\tilde{w}_{1\sim4} = 1/4$ | $1/\sqrt{2}$ | 0 |
| D2Q5 | $\tilde{w}_0 = 1-\varpi$, $\tilde{w}_{1\sim4} = \varpi/4$ | $\sqrt{\varpi/2}$ | 0 |
| D2Q9 | $\tilde{w}_0 = 4/9$, $\tilde{w}_{1\sim4} = 1/9$, $\tilde{w}_{5\sim8} = 1/36$ | $1/\sqrt{3}$ | 0,1 |
| D3Q7 | $\tilde{w}_0 = 1-\varpi$, $\tilde{w}_{1\sim6} = \varpi/6$ | $\sqrt{\varpi/3}$ | 0 |
| D3Q15 | $\tilde{w}_0 = 2/9$, $\tilde{w}_{1\sim6} = 1/9$, $\tilde{w}_{7\sim14} = 1/72$ | $1/\sqrt{3}$ | 0,1 |
| D3Q19 | $\tilde{w}_0 = 1/3$, $\tilde{w}_{1\sim6} = 1/18$, $\tilde{w}_{7\sim18} = 1/36$ | $1/\sqrt{3}$ | 0,1 |

*2.2.1.2. The temperature-based DDF models.* The temperature-based DDF models [138-141] are usually devised for solving Eq. (65). The most well-known temperature-based DDF model may be attributed to Shan [140], in which the temperature is treated as a passive scalar and the temperature field is modeled by a temperature distribution function. The thermal BGK-LB equation of the temperature distribution function $g_i$ for solving Eq. (65) is still given by Eq. (66), and the equilibrium temperature distribution function $g_i^{eq}$ can be chosen as

$$g_i^{eq} = \tilde{w}_i T \left[ 1 + \frac{\mathbf{e}_i \cdot \mathbf{u}}{c_{sT}^2} + \vartheta \frac{\mathbf{u}\mathbf{u} : (\mathbf{e}_i \mathbf{e}_i - c_{sT}^2 \mathbf{I})}{2c_{sT}^4} \right]. \tag{71}$$

The parameters of the lattice models can be found in Table 1. The thermal diffusivity $\alpha$ is given by $\alpha = c_{sT}^2 (\tau_g - 0.5) \delta_t$. Note that the velocity terms of order $O(u^2)$ can be retained in the equilibrium distribution functions (given by Eqs. (67) and (71)) in the D2Q9, D3Q15, and D3Q19 models, i.e., $\vartheta = 1$. However, because the internal energy equation (64) and temperature equation (65) do not involve second-order velocity terms, keeping the velocity terms of order $O(u)$ in the equilibrium distribution functions is sufficient ($\vartheta = 0$).

For practical applications in 2D, the D2Q4 and D2Q5 lattices are more frequently adopted because they take less computational resources, and they are more convenient in treating curved boundaries as compared with the D2Q9 lattice. The D2Q9 model has more discrete velocities and distribution function components than the D2Q4 or D2Q5 model. Therefore, one would think that the numerical accuracy and stability of the D2Q9 model may be better. However, whether the D2Q9 model is numerically better than the D2Q4 or D2Q5 model is still an open question. In a recent work by Li et al. [154], they pointed out that the D2Q5 model is more robust and accurate than the D2Q9 model when the convection effect is not very strong and the boundary effect is significant. On the contrast, when the convection effect is very strong, the D2Q9 model has better numerical accuracy. Similar conclusions can be obtained in 3D [154].

*2.2.1.3. The DDF-MRT models.* In the past years, attempts have also been made to use the MRT method to solve the convection-diffusion equation [154-157]. In the MRT model, the number of tunable parameters is sufficient to cover the anisotropic diffusion coefficient. Moreover, the error growth due to the variation of the relaxation parameter can be suppressed by using the MRT model [156]. The error $E_2$ versus the relaxation time $\tau_D$ of the Helmholtz equation is shown in Fig. 4, from which it can be observed that the rate of error increase is greatly suppressed by using the MRT model. Moreover, a number of DDF-MRT models have been developed for simulating various thermal flows since 2010 [68-70,74,76,77,113,158-164], in which the BGK-LB equations of the density and temperature distribution functions are replaced by an isothermal MRT-LB equation and a thermal MRT-LB equation, respectively. The thermal MRT-LB equation is given by

$$g_i\left(\mathbf{x}+\mathbf{e}_i\delta_t, t+\delta_t\right) - g_i\left(\mathbf{x}, t\right) = -\left(\mathbf{N}^{-1}\mathbf{\Theta}\mathbf{N}\right)_{ij}\left(g_j - g_j^{eq}\right)\Big|_{(\mathbf{x},t)}, \qquad (72)$$

where $\mathbf{N}$ is the transformation matrix, and $\mathbf{\Theta}$ is the relaxation matrix.

The velocity space $\mathbb{V} = \mathbb{R}^b$ (spanned by $\mathbf{g} = |g\rangle$) and the moment space $\mathbb{M} = \mathbb{R}^b$ (spanned by $\mathbf{n} = |n\rangle$) are related by a linear mapping: $\mathbf{n} = \mathbf{N}\mathbf{g}$ and $\mathbf{g} = \mathbf{N}^{-1}\mathbf{n}$. In thermal MRT-LB models, the most widely used lattices are D2Q5 lattice in 2D and D3Q7 lattice in 3D. The transformation matrices can be found in Refs. [156,158,160]. For the D2Q5-MRT model, the relaxation matrix is given by

$$\Theta = \text{diag}(\zeta_T, \zeta_\alpha, \zeta_\alpha, \zeta_e, \zeta_v), \tag{73}$$

and for the D3Q7-MRT model, the relaxation matrix is given by

$$\Theta = \text{diag}(\zeta_T, \zeta_\alpha, \zeta_\alpha, \zeta_\alpha, \zeta_e, \zeta_v, \zeta_v). \tag{74}$$

The equilibrium moments $\mathbf{n}^{eq}$ of the D2Q5- and D3Q7-MRT models can be obtained via $\mathbf{n}^{eq} = \mathbf{N}\mathbf{g}^{eq}$, where $\mathbf{g}^{eq} = |g^{eq}\rangle$ with $g_i^{eq}$ given by Eq. (67) or Eq. (71). For internal-energy-based MRT model, the thermal conductivity $k$ is given by $k = \rho c_p c_{sT}^2 (\zeta_\alpha^{-1} - 0.5)\delta_t$ ($\zeta_\alpha = 1/\tau_g$); For temperature-based MRT model, the thermal diffusivity $\alpha$ is given by $\alpha = c_{sT}^2 (\zeta_\alpha^{-1} - 0.5)\delta_t$. As previously mentioned, the number of tunable parameters in the MRT model is sufficient to cover the anisotropic diffusion coefficient. For details about the MRT models for convection-diffusion equation with anisotropic diffusion coefficient, readers are referred to Refs. [154-157,162].

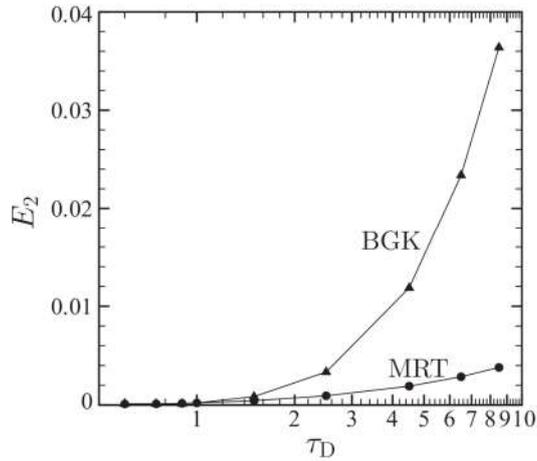

**Fig. 4.** The error $E_2$ versus the relaxation time $\tau_D$ of the Helmholtz equation. (Reprinted from Yoshida and Nagaoka [156].)

Note that the row vectors of the transformation matrix $\mathbf{N}$ [156,158,160] are orthogonal to each other, i.e., **N** is an orthogonal matrix. As mentioned in Section 2.1.2, the transformation matrix is not necessary to be an orthogonal one. Recently, Liu and He [74,113] have proposed several DDF-MRT models in 2D and 3D, in which the thermal MRT-LB models were constructed based on non-orthogonal transformation matrices. The non-orthogonal transformation matrix of the D2Q5-MRT model is given by [113]

$$\mathbf{N} = \left[|1\rangle, |e_x\rangle, |e_y\rangle, |\mathbf{e}^2\rangle, |e_x^2 - e_y^2\rangle\right]^{\mathrm{T}}$$
$$= \begin{bmatrix} 1 & 1 & 1 & 1 & 1 \\ 0 & 1 & 0 & -1 & 0 \\ 0 & 0 & 1 & 0 & -1 \\ 0 & 1 & 1 & 1 & 1 \\ 0 & 1 & -1 & 1 & -1 \end{bmatrix}, \quad (75)$$

and the non-orthogonal transformation matrix of the D3Q7-MRT model can be chosen as [74]

$$\mathbf{N} = \left[|1\rangle, |e_x\rangle, |e_y\rangle, |e_z\rangle, |\mathbf{e}^2\rangle, |e_x^2 - e_y^2\rangle, |e_x^2 - e_z^2\rangle\right]^{\mathrm{T}}$$
$$= \begin{bmatrix} 1 & 1 & 1 & 1 & 1 & 1 & 1 \\ 0 & 1 & -1 & 0 & 0 & 0 & 0 \\ 0 & 0 & 0 & 1 & -1 & 0 & 0 \\ 0 & 0 & 0 & 0 & 0 & 1 & -1 \\ 0 & 1 & 1 & 1 & 1 & 1 & 1 \\ 0 & 1 & 1 & -1 & -1 & 0 & 0 \\ 0 & 1 & 1 & 0 & 0 & -1 & -1 \end{bmatrix}. \quad (76)$$

*2.2.2. The error terms and the source term treatments*

*2.2.2.1. The error terms*. As compared with the internal energy equation (64) or temperature equation (65), there exist error terms in the macroscopic equation recovered from the thermal BGK-LB equation (66) or MRT-LB equation (72) [74,113,153,157,165-168]. For incompressible thermal flows, the error terms can be neglected in most cases. However, one should be aware of the error terms because they may result in considerable numerical errors in certain cases (e.g., if the flow field is modeled by a multiphase LB model [129], then the error terms cannot be neglected) [168]. To eliminate the error terms, several approaches have been proposed from different points of view [157,165-168]. Here we just take Chopard et al.'s approach [166] as an example to illustrate the elimination of the error terms in the internal-energy-based DDF-LB approach. Through the Chapman-Enskog analysis of the thermal BGK-LB equation (66), the following macroscopic equation can be obtained

$$\frac{\partial(\rho c_p T)}{\partial t}+\nabla\cdot(\rho c_p T\mathbf{u})=\nabla\cdot\left\{\delta_t(\tau_g-0.5)\left[c_{sT}^2\nabla(\rho c_p T)+\epsilon\partial_{t_1}(\rho c_p T\mathbf{u})+\vartheta\nabla\cdot(\rho c_p T\mathbf{uu})\right]\right\}. \quad (77)$$

Some error terms exist in the recovered macroscopic equation (77). The error terms can be neglected in most cases. If the specific heat $c_p$ is constant, the internal energy equation (64) can be recovered with $\rho\approx\rho_0$, i.e., $\partial_t(c_p T)+\nabla\cdot(c_p T\mathbf{u})=\nabla\cdot\left(\frac{k}{\rho_0}\nabla T\right)$. Here, the equilibrium distribution function $g_i^{eq}$ is given by

$$g_i^{eq}=\tilde{w}_i c_p T\left[1+\frac{\mathbf{e}_i\cdot\mathbf{u}}{c_{sT}^2}+\vartheta\frac{\mathbf{uu}:(\mathbf{e}_i\mathbf{e}_i-c_{sT}^2\mathbf{I})}{2c_{sT}^4}\right]. \quad (78)$$

For $\vartheta=0$, the approach proposed by Chopard et al. [166] can be employed to eliminate the error term $\delta_t(\tau_g-0.5)\epsilon\partial_{t_1}(\rho c_p T\mathbf{u})$ in Eq. (77). This can be done by adding a correction term into the thermal BGK-LB equation (66)

$$g_i(\mathbf{x}+\mathbf{e}_i\delta_t,t+\delta_t)-g_i(\mathbf{x},t)=-\frac{1}{\tau_g}\left[g_i(\mathbf{x},t)-g_i^{eq}(\mathbf{x},t)\right]+\delta_t Su_i, \quad (79)$$

where $Su_i$ is chosen as

$$Su_i=\tilde{w}_i\left(1-\frac{1}{2\tau_g}\right)\frac{\mathbf{e}_i\cdot\partial_t(\rho c_p T\mathbf{u})}{c_{sT}^2}. \quad (80)$$

$Su_i$ satisfies $\sum_i Su_i=0$ and $\sum_i \mathbf{e}_i Su_i=\left(1-\frac{1}{2\tau_g}\right)\partial_t(\rho c_p T\mathbf{u})$. In simulations, the explicit difference scheme $\partial_t\Delta=\left[\Delta(\mathbf{x},t)-\Delta(\mathbf{x},t-\delta_t)\right]/\delta_t$ can be used for computing $\partial_t(\rho c_p T\mathbf{u})$. To compute the time derivative term $\partial_t(\rho c_p T\mathbf{u})$ by the explicit difference scheme, a little larger memory cost is needed in simulations. Fortunately, the collision process can still be executed locally. For $\vartheta=1$, there exists another error term in Eq. (77), i.e., $\delta_t(\tau_g - 0.5)\nabla \cdot (\rho c_p T\mathbf{uu})$. Numerically, this term is a higher order term and can be neglected. To eliminate the error terms, the approaches in Refs. [157,167,168] can also be employed. Obviously, these approaches are superior to Chopard et al.'s approach. In the literature [157,168], the error terms have been eliminated by modifying the collision process of the MRT-LB equation in the moment space based on D2Q9 lattice, which is consistent with the philosophy of the LB method. It is worth mentioning that, the D2Q4 or D2Q5 lattice cannot be employed in Refs. [157,168] because the freedom of the corresponding MRT model is not enough.

*2.2.2.2. The source term treatments.* The source term treatments for the temperature-based DDF-LB approach are briefly introduced. The temperature equation with a source term can be written as

$$\frac{\partial T}{\partial t} + \nabla \cdot (T\mathbf{u}) = \nabla \cdot (\alpha \nabla T) + S(\mathbf{x},t), \tag{81}$$

where $S(\mathbf{x},t)$ is the source term. Eq. (81) is a convection-diffusion equation with a source term. Here the BGK-LB scheme is used for illustration purpose. In the presence of a source term, the thermal BGK-LB equation (66) should be modified to account for the source term. This can be done by adding an additional term into the thermal BGK-LB equation (66), as in the following

$$g_i(\mathbf{x} + \mathbf{e}_i \delta_t, t + \delta_t) - g_i(\mathbf{x},t) = -\frac{1}{\tau_g}\left[g_i(\mathbf{x},t) - g_i^{eq}(\mathbf{x},t)\right] + \delta_t S_i, \tag{82}$$

where $S_i$ is the discrete source term.

The usually used method takes $S_i = \tilde{w}_i S$, and the temperature $T$ is defined by $T = \sum_i g_i$. This source term treatment can be found elsewhere. However, through the Chapman-Enskog analysis of the thermal BGK-LB equation (82), it can be found that there exists an additional term $-\frac{\delta_t}{2}\epsilon \partial_{t_1} S$ in the recovered macroscopic equation [68,165,168]. The additional term is caused by the discrete lattice effect, and in order to match the temperature equation (81), one has to assume that $S$ varies slowly in time, namely $\partial_{t_1} S \approx 0$. Theoretically, when incorporating a source term into the thermal LB equation (66), the discrete lattice effect should also be considered. Similar to the approach in Ref. [123], the additional term can be removed by choosing

$$S_i = \tilde{w}_i \left(1 - \frac{1}{2\tau_g}\right) S. \tag{83}$$

Correspondingly, the temperature $T$ is defined by

$$T = \sum_i g_i + \frac{\delta_t}{2} S. \tag{84}$$

Note that if the source term $S$ is a function of $T$, some other methods are needed to solve the algebraic equation (84) so as to obtain the temperature $T$. When $S$ is a nonlinear function of $T$ and the temperature $T$ cannot be obtained from Eq. (84) through simple algebraic operations, the approach proposed by Shi and Guo [165] can be employed. According to Shi and Guo's approach, the discrete source term can be chosen as $S_i = \tilde{w}_i \left(1 + \frac{\delta_t}{2} \partial_t\right) S$, and the temperature $T$ is defined by $T = \sum_i g_i$. In simulations, the explicit difference scheme can be used for computing the time derivative term $\partial_t S$ (e.g., $\partial_t S = \left[S(\mathbf{x}, t) - S(\mathbf{x}, t - \delta_t)\right]/\delta_t$), which does not affect the local nature of the collision process of the thermal BGK-LB equation (82). However, since an explicit difference scheme is employed to compute the derivative term, this method can only be viewed as a compromise. The source term in the internal-energy-based DDF-LB approach can be treated in similar ways.

*2.2.3. The thermal boundary treatments*

In what follows, the thermal boundary treatments are briefly introduced. In general, the thermal boundary conditions fall into three categories: the Dirichlet (boundary value) condition, Neumann (flux) condition, and Robin (mixed) condition. There have been continuous efforts in developing thermal boundary treatments for thermal LB models. He et al. [142] developed a thermal boundary treatment for the internal-energy-based DDF-LB model. In He et al.'s work, the hydrodynamic non-equilibrium bounce-back scheme proposed by Zou and He [169] was extended to impose thermal boundary conditions. Later, D'Orazio et al. [170] proposed a counter-slip thermal boundary treatment, in which a counter-slip internal energy was introduced to specify the given temperature and/or heat flux conditions at the boundary. In the temperature-based DDF-LB model by Guo et al. [141], a thermal non-equilibrium extrapolation boundary treatment was proposed. Similar to the hydrodynamic non-equilibrium extrapolation scheme [133,134], the temperature distribution functions at boundary nodes were also decomposed into two parts, i.e., the equilibrium and non-equilibrium parts. Meanwhile, the non-equilibrium part was estimated by an extrapolation of the non-equilibrium part of the temperature distribution functions at the neighboring fluid nodes. The hydrodynamic non-equilibrium extrapolation scheme has also been extended to several DDF-LB models, e.g., by Guo and Zhao [65], Tang et al. [171], Guo et al. [146], and Li et al. [145]. In order to treat the Neumann condition by the thermal non-equilibrium extrapolation scheme, the Neumann condition (adiabatic condition) was transformed into the Dirichlet condition via some finite-difference schemes.

Furthermore, Ginzburg [172] devised a multi-reflection scheme for the LB simulations of advection and anisotropic dispersion equations with arbitrarily shaped surfaces. In the multi-reflection scheme, the equilibrium distributions were decomposed into two parts: the symmetric part was tuned to build second-order and third-order accurate Dirichlet conditions, and the anti-symmetric part was used for Neumann conditions. In 2009, Kuo and Chen [173] proposed a non-equilibrium mirror-reflection scheme for thermal boundary conditions. It was demonstrated that this scheme is of second-order accuracy and can predict the temperature and heat flux correctly. Later, Liu et al. [174] proposed a consistent thermal boundary treatment, in which the unknown energy distributions can be chosen to be functions of the known energy distributions together with appropriate corrections. Recently, based on the hydrodynamic bounce-back scheme, several bounce-back boundary treatments for thermal boundary conditions have been devised, e.g., by Zhang et al. [175], Li et al. [176], Chen et al. [177], and Cui et al. [178]. Besides the above mentioned thermal boundary treatments, several immersed-boundary-based thermal boundary treatments have also been developed [179,180]. Moreover, Huang et al. [181], Chen et al. [182], and Li et al. [176,183] have devised several thermal boundary treatments for curved boundaries.

Now attention turns to the interface treatments for conjugate heat transfer problems [184-193]. When dealing with heat transfer in heterogeneous media, one needs to consider the continuities of temperature and normal heat flux at the interface between different phases (e.g., fluid and solid, solid and liquid), which can be found elsewhere

$$T^{\text{int},+} = T^{\text{int},-}, \tag{85}$$

$$\mathbf{n} \cdot \left( k \nabla T \right)^{\text{int},+} = \mathbf{n} \cdot \left( k \nabla T \right)^{\text{int},-}, \tag{86}$$

where the superscript "int" represents the interface, the superscripts "+" and "-" denote the parameters at each side of the interface, and $\mathbf{n}$ is the unit normal vector to the interface. Eqs. (85) and (86) are usually referred to as the conjugate interface conditions (for no-slip walls with $\mathbf{u}=0$). Efficient and accurate implementation of conjugate conditions is of great importance in the study of heat transfer in multicomponent, multiphase, and porous systems.

In general, to simulate conjugate heat transfer problems, conventional numerical methods require extrapolations and iteration schemes to meet the correct conjugate conditions (see Eqs. (85) and (86)).

Such features make the conventional numerical methods particularly hard to implement and computationally demanding in complex interface geometries [184,185]. As a reliable alternative to conventional numerical methods, the LB method is a promising numerical approach for solving conjugate heat transfer problems. Wang et al. [184] proposed a "half-lattice division" treatment for fluid-solid conjugate heat transfer simulations, in which the location of any interface node was fixed half-way between two lattice nodes. No special treatment is needed in Wang et al.'s scheme and the continuity conditions at the interface can be automatically satisfied at steady scenarios. Later, the half-lattice division treatment was employed to predict effective thermal conductivity of various porous structures at the pore scale [50-52]. The continuities of temperature and heat flux at a fluid-solid interface in the LB method were theoretically analyzed by Meng et al. [186]. Meng et al. assumed that the unknown energy distribution functions of the fluid and solid phases were in equilibrium with the counter-slip internal energy, which was determined by constraints in the continuities of temperature and heat flux at the fluid-solid interface.

For conjugate heat transfer problems involving curved interface geometries, a general second-order accurate interface treatment was proposed by Li et al. [185]. In Li et al.'s scheme, the interfacial temperature and heat flux can be conveniently derived from the microscopic distribution functions without any finite-difference calculations. Based on Li et al.'s work, Guo et al. [187] proposed an interface treatment for conjugate heat transfer with discontinuities or jumps of temperature and/or heat flux (see Eqs. (2) and (3a) in Ref. [187]) at the interface. The interfacial jump conditions were intrinsically satisfied in Guo et al.'s scheme without iteration procedures. Pareschi et al. [188] presented an entropic DDF-LB model for the simulation of flows with conjugate heat transfer. In Pareschi et al.'s work, the temperature computed at the interface was used to evaluate the temperature gradients, which in turn were needed to construct non-equilibrium moments employed in Grad's approximations. Moreover, several interface schemes for conjugate transport problems involving curved interface geometries have also been proposed [189-191].

For conjugate heat transfer problems, only the following relations can be satisfied in the framework of the standard thermal LB method (see Eq . (66) or Eq. (72))

$$T^{\text{int},+} = T^{\text{int},-}, \tag{87}$$

$$\mathbf{n} \cdot \left[ \alpha \mathbf{\nabla} \left( \rho c_p T \right) \right]^{\text{int},+} = \mathbf{n} \cdot \left[ \alpha \mathbf{\nabla} \left( \rho c_p T \right) \right]^{\text{int},-}. \tag{88}$$

Obviously, there exists a difference between Eq. (86) and (88). Only when $\left( \rho c_p \right)^{\text{int},+} = \left( \rho c_p \right)^{\text{int},-}$ or at the steady state, Eq. (88) reduces to Eq. (86). Consequently, for the thermal LB method, the key to accurately model conjugate heat transfer problems is to recover the diffusion term in the internal energy equation (64) exactly [192,193]. As pointed out by Karani and Huber [192], the key to recover the diffusion term in the internal energy equation (64) exactly is to treat the spatial variation of the heat capacity $\rho c_p$ appropriately. Based on theoretical analyses, several thermal LB models for conjugate heat transfer problems have been developed by Karani and Huber [192], Chen et al. [193], Huang and Wu [194], and Gao et al. [73]. In these thermal LB models, special interface treatments of the conjugate conditions have been avoided, offering efficient strategies to deal with heat transfer in complex porous systems at the pore scale.

## 3. The REV-scale LB method for fluid flow and heat transfer in porous media

*3.1. The isothermal REV-scale LB method*

*3.1.1. The governing equations*

Owing to the complexity of the porous structure, flow in porous media is usually modeled based on some semi-empirical models. A number of models have been developed in the literature, e.g., the Darcy model, Forchheimer-extended Darcy model, Brinkman-extended Darcy model, and generalized non-Darcy (or Brinkman-Forchheimer-extended Darcy) model [6,7]. Darcy's experimental investigations [195] on steady-state unidirectional flow through a uniform medium revealed the proportionality between the flow rate and the applied pressure gradient. The mathematical formulation of the Darcy model can be expressed as

$$\nabla p = -\frac{\mu}{K}\mathbf{u}, \tag{89}$$

where $\mu$ is the dynamic viscosity of the fluid, $K$ is the permeability of the medium and it has dimensions (length)$^2$. The permeability $K$ depends on the geometry of the medium, and values of $K$ for natural materials vary widely. In 3D cases, Eq. (89) generalizes to $\mathbf{u} = -\mu^{-1}\mathbf{K}\cdot\nabla p$, here the permeability $\mathbf{K}$ is in general a second-order tensor. For an isotropic medium, the permeability is a scalar and Eq. (89) can be derived.

The Darcy equation (89) is linear in the seepage velocity $\mathbf{u}$, and it holds when $\mathbf{u}$ is sufficiently small [6]. In practice, "sufficiently small" indicates that the Reynolds number $Re_p$ (based on a typical pore or particle diameter) of the flow is of order unity or smaller. According to Joseph et al. [196], an appropriate modification to Darcy equation is to replace Eq. (89) by

$$\nabla p = -\frac{\mu}{K}\mathbf{u} - \frac{F_\phi}{\sqrt{K}}\rho|\mathbf{u}|\mathbf{u}, \tag{90}$$

where $F_\phi$ is the structure function or inertia coefficient (depends on the geometry of the medium). For simplicity, we can call Eq. (90) the Forchheimer equation and refer to the last term $-F_\phi K^{-1/2}\rho|\mathbf{u}|\mathbf{u}$ as the Forchheimer term (nonlinear inertial term).

An alternative to Darcy equation (89) is the so-called Brinkman equation. With inertial term omitted, it takes the form

$$\nabla p = -\frac{\mu}{K}\mathbf{u} + \mu_e \nabla^2 \mathbf{u}, \qquad (91)$$

where $\mu_e$ is the effective dynamic viscosity. Note that $\mu$ is not necessarily the same as $\mu_e$, and $\mu_e/\mu$ also depends on the geometry of the medium. Brinkman equation has two viscous terms: the first term is the Darcy term (linear viscous term) and the second term is the nonlinear viscous term that is analogous to the Laplacian term in the N-S equation. Brinkman equation reduces to a form of the N-S equation as $K \to \infty$ and to the Darcy equation as $K \to 0$ [6,7]. In order for Brinkman equation to be valid, the porosity should be large [6].

The generalized non-Darcy model was proposed by several groups [197-199]. The governing equation of the generalized non-Darcy model can be established by using the local volume-averaging technique. The model can be expressed by the following generalized non-Darcy equation

$$\frac{\partial \mathbf{u}}{\partial t} + (\mathbf{u} \cdot \nabla)\left(\frac{\mathbf{u}}{\phi}\right) = -\frac{1}{\rho}\nabla(\phi p) + v_e \nabla^2 \mathbf{u} - \frac{\phi v}{K}\mathbf{u} - \frac{\phi F_\phi}{\sqrt{K}}|\mathbf{u}|\mathbf{u} + \phi \mathbf{G}, \qquad (92)$$

where $v_e$ is the effective kinematic viscosity ($\mu_e = \rho v_e$), and $\mathbf{G}$ is the body force induced by external (or internal) force fields. The position of the porosity $\phi$ in relation to the spatial derivatives is important if the porous medium is heterogeneous [6]. The third and fourth terms on the right-hand side of Eq. (92) are the linear (Darcy) and nonlinear (Forchheimer) drag forces, respectively. For flow over a packed bed of particles, based on Ergun's experimental investigations [200], the structure function $F_\phi$ and the permeability $K$ can be approximated by $F_\phi = 1.75/\sqrt{150\phi^3}$, and $K = \phi^3 d_p^2/[150(1-\phi)^2]$ ($d_p$ is the solid particle diameter) [201], respectively.

The continuity equation can be expressed as [6]

$$\frac{\partial(\phi\rho)}{\partial t} + \nabla \cdot (\rho \mathbf{u}) = 0 . \tag{93}$$

For incompressible flows in porous media, Eq. (93) reduces to $\nabla \cdot \mathbf{u} = 0$.

For a given medium, the ratio between the linear and the nonlinear drag forces is about $\frac{F_\phi |\mathbf{u}|/\sqrt{K}}{\nu/K} \sim \sqrt{Da} Re$ [63], here $Da = K/L^2$ is the Darcy number, and $Re = Lu_c/\nu$ is the Reynolds number ($L$ and $u_c$ are the characteristic length and velocity, respectively). Therefore, for the cases with small Reynolds number and/or Darcy number, the nonlinear (quadratic) drag force can be neglected. On the other hand, for large Reynolds number or Darcy number, the effect of the nonlinear drag force should be considered [63]. For more details about the mechanics of fluid flow in porous media, readers are referred to the book by Nield and Bejan [6].

*3.1.2. The isothermal REV-scale LB models for Brinkman equation*

*3.1.2.1. Spaid and Phelan's model.* In 1997, Spaid and Phelan [59] developed an REV-scale LB model (referred to as SP-LB model) based on the Brinkman equation for single-component flow in heterogeneous porous media. Through a modification of the equilibrium distribution function, the Brinkman equation can be recovered from the SP-LB model. The LB equation of the SP-LB model is given by [59]

$$f_i(\mathbf{x} + \mathbf{e}_i \delta_t, t + \delta_t) - f_i(\mathbf{x}, t) = -\frac{1}{\tau_\nu} \left[ f_i - f_i^{eq}(\rho, \mathbf{u}^{eq}) \right] \Big|_{(\mathbf{x}, t)} . \tag{94}$$

The equilibrium velocity $\mathbf{u}^{eq}$ in the equilibrium distribution function $f_i^{eq}(\rho, \mathbf{u}^{eq})$ is defined as

$$\mathbf{u}^{eq} = \mathbf{u} + s(\mathbf{x}) \frac{\tau_\nu \mathbf{F}}{\rho}, \tag{95}$$

where $\mathbf{F} = -\mu \mathbf{u}/K$ is the body force accounting for the presence of a porous medium, and the variable $s(\mathbf{x})$ is either 0 or 1 depending on whether a lattice node is located in an open or porous region respectively. The macroscopic density and velocity are still defined by Eq. (16). As shown in Eq.

(95), the Shan-Chen forcing scheme [129] has been used to incorporate the body force into the SP-LB model, i.e., the body force induced by the medium is incorporated into the model by shifting the velocity in the equilibrium distribution function. Subsequently, Spaid and Phelan [60] employed the SP-LB model to simulate multicomponent fluid flow in the microstructure of a fiber preform by combining this model with Shan-Chen's multicomponent LB algorithm [129]. Based on Spaid and Phelan's work [59], Freed [202] proposed an REV-scale LB model by adding a forcing term to model flows through a resistance field. In Freed's model, a modified version of the Shan-Chen forcing scheme was employed to add the forcing term to the model. Using Freed's model, Kang et al. [203] studied the dynamic behavior of a unidirectional steady flow through homogeneous and heterogeneous porous media. With the effects of nonlinear drag force and the Brinkman viscous term being neglected, the Darcy's law can be satisfied in Kang et al.'s model.

In 2001, Martys [62] improved the SP-LB model by introducing an effective viscosity into the Brinkman equation. This model can describe the general case of $\mu_e/\mu \neq 1$, and the accuracy and stability have also been improved. In order to eliminate the second order errors of the SP-LB model, Luo's forcing scheme [131] was employed to add a forcing term $F_i$ to the LB equation (see Eq. (45)). The forcing term $F_i$ is given by

$$F_i = w_i \left[ \frac{\mathbf{e}_i - \mathbf{u}}{c_s^2} + \frac{(\mathbf{e}_i \cdot \mathbf{u})\mathbf{e}_i}{c_s^4} \right] \cdot \mathbf{F}, \tag{96}$$

where $\mathbf{F} = -\mu\mathbf{u}/K$. By using Luo's forcing scheme, some of the error terms of the SP-LB model can be eliminated. The SP-LB model and its improved versions have been demonstrated to be simple and efficient for modeling flows in porous media. The SP-LB model and its improved versions are based on the Brinkman equation. In the above mentioned REV-scale LB models, the porous media region is treated as a "resistance field". Therefore, these models can be viewed as the resistance models.

*3.1.2.2. Dardis and McCloskey's model.* In 1998, Dardis and McCloskey [61] proposed a partial-bounce-back LB model (referred to as DM-LB model) for the simulation of flow in porous media, in which the no-slip boundary condition was incorporated into the model by adding a partial-bounce-back term into the LB equation. The LB equation of the DM-LB model can be written as [61]

$$f_i(\mathbf{x}+\mathbf{e}_i\delta_t, t+\delta_t) - f_i(\mathbf{x},t) = -\frac{1}{\tau_v}(f_i - f_i^{eq})\Big|_{(\mathbf{x},t)} + \Delta f_i^{PM}(\mathbf{x}+\mathbf{e}_i\delta_t, t), \quad (97)$$

where $\Delta f_i^{PM}$ is the partial-bounce-back term

$$\Delta f_i^{PM}(\mathbf{x}+\mathbf{e}_i\delta_t, t) = n_s(\mathbf{x}+\mathbf{e}_i\delta_t)\left[f_{\bar{i}}(\mathbf{x}+\mathbf{e}_i\delta_t, t) - f_i(\mathbf{x},t)\right]. \quad (98)$$

The partial-bounce-back term $\Delta f_i^{PM}$ accounts for the influence of the porous medium on the fluid and determines the redistribution of particle momentum. $n_s(\mathbf{x})$ is a continuous variable that controls the partitioning of fluid momentum, and it is defined as the scatterer density per lattice node $\mathbf{x}$ where $0 < n_s(\mathbf{x}) < 1$. When $n_s(\mathbf{x}) = 0$, $\Delta f_i^{PM} = 0$ and Eq. (97) reduces to the case for a free fluid in an open region. For the opposite case where a lattice node is given the value of $n_s(\mathbf{x}) = 1$, the no-slip boundary condition applies and the lattice node is rendered impermeable. Through the partial-bounce-back term, an effective boundary condition between fluid and solid has been implemented. The DM-LB model corresponds to the Brinkman equation with $\mu_e = \mu$, and the permeability $K$ is related to $n_s$ as $K = \mu/(2\rho n_s)$. Based on Dardis and McCloskey's work, several partial-bounce-back LB models have been developed [204-206].

*3.1.3. The isothermal REV-scale LB models for generalized non-Darcy equation*

*3.1.3.1. Guo and Zhao's model.* Based on the generalized non-Darcy equation (92), Guo and Zhao [63]

proposed a resistance model (referred to as GZ-LB model) in 2002. The effect of the porous medium was incorporated into the model by including the porosity into the equilibrium distribution function and adding a forcing term to the LB equation to account for the linear and nonlinear drag forces of the medium. The LB equation of the GZ-LB model is given by [63]

$$f_i(\mathbf{x}+\mathbf{e}_i\delta_t, t+\delta_t) - f_i(\mathbf{x},t) = -\frac{1}{\tau_v}\left[f_i(\mathbf{x},t) - f_i^{eq}(\mathbf{x},t)\right] + \delta_t F_i, \tag{99}$$

where $f_i$ and $f_i^{eq}$ are the volume-averaged density distribution function and equilibrium distribution function at the REV scale, respectively. The equilibrium distribution function $f_i^{eq}$ and forcing term $F_i$ are defined as [63]

$$f_i^{eq} = w_i \rho \left[1 + \frac{\mathbf{e}_i \cdot \mathbf{u}}{c_s^2} + \frac{\mathbf{u}\mathbf{u}:(\mathbf{e}_i\mathbf{e}_i - c_s^2 \mathbf{I})}{2\phi c_s^4}\right], \tag{100}$$

$$F_i = w_i \rho \left(1 - \frac{1}{2\tau_v}\right)\left[\frac{\mathbf{e}_i \cdot \mathbf{F}}{c_s^2} + \frac{\mathbf{u}\mathbf{F}:(\mathbf{e}_i\mathbf{e}_i - c_s^2 \mathbf{I})}{\phi c_s^4}\right], \tag{101}$$

respectively, where $\mathbf{F}$ is the total body force

$$\mathbf{F} = -\frac{\phi v}{K}\mathbf{u} - \frac{\phi F_\phi}{\sqrt{K}}|\mathbf{u}|\mathbf{u} + \phi \mathbf{G}. \tag{102}$$

The density $\rho$ and velocity $\mathbf{u}$ are defined by $\rho = \sum_i f_i$ and $\rho \mathbf{u} = \sum_i \mathbf{e}_i f_i + 0.5\delta_t \rho \mathbf{F}$, respectively. Note that $\mathbf{F}$ also contains the velocity $\mathbf{u}$. Owing to the quadratic feature of the equation, the velocity $\mathbf{u}$ can be given explicitly by [63]

$$\mathbf{u} = \frac{\mathbf{v}}{l_0 + \sqrt{l_0^2 + l_1|\mathbf{v}|}}, \tag{103}$$

where

$$\rho \mathbf{v} = \sum_i \mathbf{e}_i f_i + \frac{\delta_t}{2}\rho\phi\mathbf{G}, \tag{104}$$

$$l_0 = \frac{1}{2}\left(1 + \phi\frac{\delta_t}{2}\frac{v}{K}\right), \quad l_1 = \phi\frac{\delta_t}{2}\frac{F_\phi}{\sqrt{K}}. \tag{105}$$

The pressure $p$ is defined as $p = \rho c_s^2/\phi$, and the effective kinematic viscosity $v_e$ is given by $v_e = c_s^2(\tau_v - 0.5)\delta_t$. It is noted that the LB equation (99) reduces to the standard LB equation for incompressible flows in the absence of porous media as $\phi \to 1$. When $F_\phi = 0$, a simplified lattice Boltzmann equation (SLBE) can be obtained for the Brinkman-extended Darcy model. Obviously, this model is applicable for simulating flows in a porous medium of variable porosity. The GZ-LB model, including its simplified version, is the most widely used REV-scale LB model. It has received significant attention since its emergence in 2002 and has been applied in many practical applications.

*3.1.3.2. REV-scale MRT-LB model.* In 2014, Liu and He [68] proposed an REV-scale MRT-LB model for generalized non-Darcy equation. According to Refs. [68,124,125], the MRT-LB equation with an explicit treatment of the forcing term can be written as

$$f_i(\mathbf{x} + \mathbf{e}_i \delta_t, t + \delta_t) - f_i(\mathbf{x}, t) = -\tilde{\Lambda}_{ij}(f_j - f_j^{eq})\big|_{(\mathbf{x},t)} + \delta_t \left(\tilde{S}_i - 0.5\tilde{\Lambda}_{ij}\tilde{S}_j\right)\big|_{(\mathbf{x},t)}, \tag{106}$$

where $\tilde{S}_i$ is given by $\tilde{S}_i = w_i \rho \left[\mathbf{e}_i \cdot \mathbf{F}/c_s^2 + \mathbf{uF} : (\mathbf{e}_i\mathbf{e}_i - c_s^2\mathbf{I})/\phi c_s^4\right]$ [63]. The collision process of the MRT-LB equation (106) is executed in moment space

$$\mathbf{m}^*(\mathbf{x}, t) = \mathbf{m}(\mathbf{x}, t) - \Lambda(\mathbf{m} - \mathbf{m}^{eq})\big|_{(\mathbf{x},t)} + \delta_t \left(\mathbf{I} - \frac{\Lambda}{2}\right)\mathbf{S}, \tag{107}$$

and the streaming process is carried out in velocity space

$$f_i(\mathbf{x} + \mathbf{e}_i \delta_t, t + \delta_t) = f_i^*(\mathbf{x}, t). \tag{108}$$

Note that the hat "-" of $\bar{\mathbf{m}}$ and $\bar{f}_i$ in Eqs. (106)-(108) have been dropped (in moment space, the transformation given by Eq. (52) can be expressed as $\bar{\mathbf{m}} = \mathbf{m} - 0.5\delta_t \mathbf{S}$). Considering the influence of the total body force $\mathbf{F}$, the vector of the moments $\mathbf{m}$ for the D2Q9 model is given by

$$\mathbf{m} = \left(\rho, e, \varepsilon, j_x - \frac{\delta_t}{2}\rho F_x, q_x, j_y - \frac{\delta_t}{2}\rho F_y, q_y, p_{xx}, p_{xy}\right)^\mathsf{T}, \tag{109}$$

The vector of the equilibrium moments $\mathbf{m}^{eq}$ corresponding to $\mathbf{m}$ is defined as [68]

$$\mathbf{m}^{eq} = \rho \left( 1, -2 + \frac{3|\mathbf{u}|^2}{\phi}, \tilde{\alpha} + \frac{\tilde{\beta}|\mathbf{u}|^2}{\phi}, u_x, -u_x, u_y, -u_y, \frac{u_x^2 - u_y^2}{\phi}, \frac{u_x u_y}{\phi} \right)^{\mathrm{T}}, \tag{110}$$

When $\tilde{\alpha} = 1$ and $\tilde{\beta} = -3$, the equilibrium distribution function $f_i^{eq}$ in velocity space (see Eq. (100)) can be obtained via $\mathbf{f}^{eq} = \mathbf{M}^{-1} \mathbf{m}^{eq}$. For the D2Q9 model, $\mathbf{S}$ can be chosen as [68]

$$\mathbf{S} = \rho \left( 0, \frac{6\mathbf{u} \cdot \mathbf{F}}{\phi}, -\frac{6\mathbf{u} \cdot \mathbf{F}}{\phi}, F_x, -F_x, F_y, -F_y, \frac{2(u_x F_x - u_y F_y)}{\phi}, \frac{u_x F_y + u_y F_x}{\phi} \right)^{\mathrm{T}}, \tag{111}$$

where $F_x$ and $F_y$ are x- and y-components of the total body force $\mathbf{F}$ (see Eq. (102)), respectively. The density $\rho$ is defined by $\rho = \sum_i f_i$, the velocity $\mathbf{u}$ is also calculated by Eq. (103), and the effective kinematic viscosity is defined as $v_e = c_s^2 (s_v^{-1} - 0.5) \delta_t$. When the relaxation rates are all equal to $1/\tau_v$ (the relaxation matrix is given by $\mathbf{\Lambda} = (1/\tau_v)\mathbf{I}$), the MRT-LB equation (106) reduces to the BGK-LB equation (99). In addition, based on the non-orthogonal MRT method [113], Liu and He [70] have proposed a non-orthogonal D2Q9-MRT model for generalized non-Darcy equation. For details about the D3Q15-MRT and D3Q19-MRT models for incompressible flows in porous media, readers are referred to a recent work by Liu and He [74].

*3.1.3.3. Incompressible REV-scale LB models.* Through the Chapman-Enskog analysis of the MRT-LB equation (106), the following macroscopic equations can be obtained

$$\frac{\partial \rho}{\partial t} + \nabla \cdot (\rho \mathbf{u}) = 0, \tag{112}$$

$$\frac{\partial (\rho \mathbf{u})}{\partial t} + \nabla \cdot \left( \frac{\rho \mathbf{u}\mathbf{u}}{\phi} \right) = -\nabla(\phi p) + \nabla \cdot \mathbf{\Pi} + \nabla \cdot \left\{ (1 - \phi^{-1}) \left[ v_e \left( \mathbf{u} \nabla \rho + (\mathbf{u} \nabla \rho)^{\mathrm{T}} \right) + (v_B - v_e)(\mathbf{u} \cdot \nabla \rho) \mathbf{I} \right] \right\} + \rho \mathbf{F}, \tag{113}$$

where $\mathbf{\Pi} = \rho v_e \left[ \nabla \mathbf{u} + (\nabla \mathbf{u})^{\mathrm{T}} \right] + \rho (v_B - v_e)(\nabla \cdot \mathbf{u}) \mathbf{I}$. The underlined term arises from the inclusion of the porosity into the equilibrium moments, and this unwanted term vanishes as $\phi \to 1$. For the GZ-LB model [63], through the Chapman-Enskog analysis, it can be found that the macroscopic momentum equation recovered from BGK-LB equation (99) is

$$\frac{\partial(\rho\mathbf{u})}{\partial t}+\nabla\cdot\left(\frac{\rho\mathbf{u}\mathbf{u}}{\phi}\right)=-\nabla(\phi p)+\nabla\cdot\mathbf{\Pi}+\underline{\nabla\cdot\left\{\left(1-\phi^{-1}\right)\left[v_{\mathrm{e}}\left(\mathbf{u}\nabla\rho+\left(\mathbf{u}\nabla\rho\right)^{\mathrm{T}}\right)\right]\right\}}+\rho\mathbf{F}, \quad (114)$$

where $\mathbf{\Pi}=\rho v_{\mathrm{e}}\left[\nabla\mathbf{u}+(\nabla\mathbf{u})^{\mathrm{T}}\right]$. The underlined term is the unwanted term, and it vanishes as $\phi\to 1$.

Without the unwanted term, Eq. (113) (or Eq. (114)) reduces to the following macroscopic equation

$$\frac{\partial(\rho\mathbf{u})}{\partial t}+\nabla\cdot\left(\frac{\rho\mathbf{u}\mathbf{u}}{\phi}\right)=-\nabla(\phi p)+\nabla\cdot\mathbf{\Pi}+\rho\mathbf{F}. \quad (115)$$

In the incompressible limit ($\rho=\rho_0+\delta\rho\approx\rho_0$, where $\rho_0$ is the mean density, and $\delta\rho$ is the density fluctuation), Eq. (115) reduces to the generalized non-Darcy equation (92). Theoretically, the unwanted term in Eq. (113) (or Eq. (114)) cannot be eliminated because of the inclusion of the porosity into the equilibrium moments (or equilibrium distribution function). It is not clear whether the unwanted term is harmful to the accuracy or numerical stability of the model in certain cases. To eliminate the influence of the unwanted term, the incompressible LB schemes [108,109] can be employed. In 2014, Liu and He [68] constructed an incompressible REV-scale MRT-LB model for incompressible porous flows, in which Guo et al.'s incompressible LB model (D2G9 model) [109] was employed. In the model, the following pressure-based equilibrium moments are introduced [68]

$$\mathbf{m}^{eq}=\begin{pmatrix}\rho_0\\-4\rho_0+6\phi p+3\rho_0|\mathbf{u}|^2/\phi\\4\rho_0-9\phi p-3\rho_0|\mathbf{u}|^2/\phi\\\rho_0 u_x\\-\rho_0 u_x\\\rho_0 u_y\\-\rho_0 u_y\\\rho_0\left(u_x^2-u_y^2\right)/\phi\\\rho_0 u_x u_y/\phi\end{pmatrix}. \quad (116)$$

The incompressibility approximation, i.e., $\rho\approx\rho_0$ and $\mathbf{J}=(j_x,j_y)\approx\rho_0\mathbf{u}$, have been used in Eq. (116). In simulations, the mean density $\rho_0$ is usually set to be 1 for simplicity. In this model, the pressure $p$ instead of the density $\rho$ becomes the independent dynamic variable. The velocity $\mathbf{u}$ is still calculated by Eq. (103) with $\rho_0 \mathbf{v} = \sum_i \mathbf{e}_i f_i + 0.5\delta_t \rho_0 \phi \mathbf{G}$, but the density $\rho$ in Eq. (111) should be replaced by $\rho_0$. The pressure $p$ is defined as [68]

$$p = \frac{c_s^2}{\phi(1-w_0)}\left[\sum_{i \neq 0} f_i + \rho_0 s_0(\mathbf{u})\right], \tag{117}$$

where $s_i(\mathbf{u})$ is given by $s_i(\mathbf{u}) = w_i\left[\mathbf{e}_i \cdot \mathbf{u}/c_s^2 + \mathbf{uu}:(\mathbf{e}_i\mathbf{e}_i - c_s^2\mathbf{I})/(2\phi c_s^4)\right]$. However, to recover the incompressible governing equation (92) exactly based on the D2G9 model, the definition of the pressure given by Eq. (117) should be modified (see Eq. (10) in Ref. [71]).

The pressure-based equilibrium distribution function $f_i^{eq}$ in velocity space is given by [68]

$$f_i^{eq} = \begin{cases} \rho_0 - (1-w_0)\dfrac{\phi p}{c_s^2} + \rho_0 s_0(\mathbf{u}), & i=0 \\ w_i \dfrac{\phi p}{c_s^2} + \rho_0 s_i(\mathbf{u}), & i=1 \sim 8 \end{cases}. \tag{118}$$

Based on He and Luo's incompressible LB scheme [108], the following equilibrium moments can be obtained ($\tilde{\alpha} = 1$, and $\tilde{\beta} = -3$) [76]

$$\mathbf{m}^{eq} = \left(\rho, -2\rho + \frac{3\rho_0|\mathbf{u}|^2}{\phi}, \rho - \frac{3\rho_0|\mathbf{u}|^2}{\phi}, \rho_0 u_x, -\rho_0 u_x, \rho_0 u_y, -\rho_0 u_y, \frac{\rho_0(u_x^2 - u_y^2)}{\phi}, \frac{\rho_0 u_x u_y}{\phi}\right)^{\mathrm{T}}. \tag{119}$$

The density $\rho$ is defined by $\rho = \sum_i f_i$, and the velocity $\mathbf{u}$ is calculated via Eq. (103) with $\rho_0 \mathbf{v} = \sum_i \mathbf{e}_i f_i + 0.5\delta_t \rho_0 \phi \mathbf{G}$. In addition, the density $\rho$ in Eq. (111) should be replaced by $\rho_0$. Note that pressure-based equilibrium moments can also be constructed based on He and Luo's incompressible LB scheme. The equilibrium distribution function $f_i^{eq}$ in velocity space is given by [76]

$$f_i^{eq} = w_i\left\{\rho + \rho_0\left[\frac{\mathbf{e}_i \cdot \mathbf{u}}{c_s^2} + \frac{(\mathbf{e}_i \cdot \mathbf{u})^2}{2\phi c_s^4} - \frac{|\mathbf{u}|^2}{2\phi c_s^2}\right]\right\}. \tag{120}$$

In the literature [64], Guo and Zhao proposed an incompressible BGK-LB model for incompressible porous flows, in which the distribution function was constructed based on the pressure instead of the fluid density. In Guo and Zhao's model, the pressure-based equilibrium distribution function $f_i^{eq}$ is defined as [64]

$$f_i^{eq} = w_i \left\{ \frac{\phi p}{c_s^2 \rho_0} + \frac{\mathbf{e}_i \cdot \mathbf{u}}{c_s^2} + \frac{(\mathbf{e}_i \cdot \mathbf{u})^2}{2\phi c_s^4} - \frac{|\mathbf{u}|^2}{2\phi c_s^2} \right\}. \qquad (121)$$

The pressure $p$ is defined as $p = \left(c_s^2 \rho_0 / \phi\right) \sum_i g_i$, and the velocity $\mathbf{u}$ is calculated via Eq. (103) with $\mathbf{v} = \sum_i \mathbf{e}_i f_i + 0.5 \delta_t \phi \mathbf{G}$. Guo and Zhao's model can be used to simulate incompressible porous flows with a large pressure gradient [64]. Moreover, Wang et al. [71] also developed an incompressible BGK-LB model for the generalized non-Darcy equation (92).

In the above mentioned improved REV-scale LB models [64,68,71,76], the fluid density is decoupled from the fluid velocity (i.e., $\mathbf{J} = (j_x, j_y) \approx \rho_0 \mathbf{u}$) so that the compressibility can be reduced effectively. Moreover, the influence of the unwanted term can be eliminated because $\nabla \rho_0 = 0$.

*3.2. The thermal REV-scale LB method*

*3.2.1. The energy equations*

For heat transfer in porous media at the REV scale, two different models have been used to describe the heat transfer process [6]: the local thermal equilibrium (LTE) model and the local thermal non-equilibrium (LTNE) model. For simplicity, we only introduce the situation where the thermal dispersion effects, viscous heat dissipation, and compression work are negligible. Taking averages over an REV of the medium, the corresponding energy equations can be obtained [6]: the energy equation for the fluid phase ($f$) is

$$\phi(\rho c_p)_f \frac{\partial T_f}{\partial t} + \nabla \cdot (\rho_f c_{pf} T_f \mathbf{u}) = \nabla \cdot (\phi k_f \nabla T_f) + \phi q_f''', \qquad (122)$$

and the energy equation for the solid matrix (*m*) is

$$(1-\phi)(\rho c_p)_m \frac{\partial T_m}{\partial t} = \nabla \cdot \left[(1-\phi)k_m \nabla T_m\right] + (1-\phi)q_m''', \qquad (123)$$

where $q'''$ is the internal heat source. Usually, it is a good approximation to assume that the fluid and solid matrix are in local thermal equilibrium, i.e., the interfacial heat transfer between the fluid and solid matrix can be neglected so that $T_f = T_m = T$. Therefore, the following energy equation can be obtained [6]

$$\frac{\partial}{\partial t}\left(\overline{\rho c_p} T\right) + \nabla \cdot \left(\rho_f c_{pf} T \mathbf{u}\right) = \nabla \cdot \left(k_e \nabla T\right) + q_e''', \qquad (124)$$

where $\overline{\rho c_p} = \phi(\rho c_p)_f + (1-\phi)(\rho c_p)_m$, $k_e = \phi k_f + (1-\phi)k_m$, and $q_e''' = \phi q_f''' + (1-\phi)q_m'''$ are the effective (or overall) heat capacity, effective thermal conductivity, and effective heat production per unit volume of the medium, respectively. If the specific heat $c_{pf}$ is constant, Eq. (124) reduces to the following temperature equation

$$\frac{\partial}{\partial t}(\sigma T) + \nabla \cdot (T\mathbf{u}) = \nabla \cdot (\alpha_e \nabla T) + Q_e''', \qquad (125)$$

where $\sigma = \overline{\rho c_p}/(\rho c_p)_f$ is the heat capacity ratio, $\alpha_e = k_e/(\rho c_p)_f$ is the effective thermal diffusivity, and $Q_e''' = q_e'''/(\rho c_p)_f$ is the internal heat source.

The assumption of LTE between the fluid and solid matrix breaks down in many practical applications, and then the interfacial heat transfer between the fluid and solid matrix cannot be neglected [6,207]. The interfacial heat transfer between the fluid and solid matrix can be described by the LTNE model, of which the energy equations can be expressed as follows [6,207]

$$\phi(\rho c_p)_f \frac{\partial T_f}{\partial t} + \nabla \cdot \left(\rho_f c_{pf} T_f \mathbf{u}\right) = \nabla \cdot \left(\phi k_f \nabla T_f\right) + h_v \left(T_m - T_f\right) + \phi q_f''', \qquad (126)$$

$$(1-\phi)(\rho c_p)_m \frac{\partial T_m}{\partial t} = \nabla \cdot \left[(1-\phi)k_m \nabla T_m\right] + h_v \left(T_f - T_m\right) + (1-\phi)q_m''', \qquad (127)$$

where $h_v$ is the volumetric heat transfer coefficient. A critical aspect of using the LTNE model lies in the determination of the appropriate value of $h_v$. For more details about the LTE and LTNE models, readers are referred to Ref. [6].

*3.2.2. The thermal REV-scale LB models under LTE condition*

In 2005, Guo and Zhao [65,66] developed a temperature-based DDF-BGK model for convection heat transfer in porous media, in which a thermal BGK-LB model was introduced to simulate the temperature field (governed by Eq. (125) without the heat source term) in addition to the GZ-LB model for the flow field. In Guo and Zhao's model, the equilibrium temperature distribution function $g_i^{eq}$ is given by [65,66]

$$g_i^{eq} = \tilde{w}_i T \left( \sigma + \frac{\mathbf{e}_i \cdot \mathbf{u}}{c_{sT}^2} \right). \tag{128}$$

The temperature is determined by $\sigma T = \sum_i g_i$, and the effective thermal diffusivity is defined as $\alpha_e = \sigma c_{sT}^2 \left( \tau_g - 0.5 \right) \delta_t$. Guo and Zhao [65] demonstrated that their model has a second-order accuracy in space. Subsequently, an internal-energy-based DDF-BGK model for convection heat transfer in porous media was developed by Seta et al. [67]. In their work, the flow field was solved by the GZ-LB model [63], while the temperature field (the heat capacity ratio $\sigma = 1$) was simulated by the simplified internal-energy-based BGK-LB model of Peng et al. [143]. Seta et al. [67] showed that the LB method is more efficient than the FDM in simulating natural convection in porous media.

To improve the numerical stability of the DDF-LB method for simulating convection heat transfer in porous media, several temperature-based DDF-MRT models have been proposed by Liu and He [68-70], in which the thermal MRT-LB models were constructed based on the D2Q5 lattice. Liu and He showed that the MRT method has better numerical stability than its BGK counterpart in simulating convection heat transfer in porous media. Following the idea of the lattice kinetic scheme, Wang et al. [71] devised a modified DDF-BGK model for simulating convection heat transfer in porous media. By using a modified equilibrium temperature distribution function and adding a source term into the thermal BGK-LB equation, the temperature equation can be correctly recovered from this model. Wang et al. showed that their modified DDF-BGK model has better numerical stability than Guo and Zhao's DDF-BGK model [65].

It can be found that the macroscopic equation recovered from Guo and Zhao's model [65,66] is given by $\partial_t(\sigma T) + \nabla \cdot (T\mathbf{u}) = \nabla \cdot \left[\delta_t c_{sT}^2 (\tau_g - 0.5) \nabla (\sigma T)\right]$ (the error term $\delta_t (\tau_g - 0.5) \epsilon \partial_{t_1}(T\mathbf{u})$ has been dropped). If $\sigma$ is a constant or varies slowly in space, the temperature equation (125) can be obtained (without the heat source term). As pointed out by Chen et al. [72], Guo and Zhao's model suffers from a drawback that it cannot address an investigated domain where the heat capacity ratio $\sigma$ varies spatially obviously (i.e., $\nabla(\sigma T) \neq \sigma \nabla T$). To remedy such shortcoming in a simple way, Chen et al. proposed a thermal BGK-LB model by including a reference heat capacity ratio $\sigma_0$ into the equilibrium temperature distribution function $g_i^{eq}$, and the corresponding effective thermal diffusivity is defined as $\alpha_e = \sigma_0 c_{sT}^2 (\tau_g - 0.5) \delta_t$.

In the literature [65,66,68-71], the effective thermal diffusivity depends on the heat capacity ratio. Obviously, the dependence of effective thermal diffusivity on heat capacity ratio is non-physical. Moreover, the introduction of a reference heat capacity ratio $\sigma_0$ into the model is not necessary. To avoid the artificial coupling, the equilibrium moments can be defined as (based on D2Q5 lattice) [208]

$$\mathbf{n}^{eq} = \left(\sigma T, u_x T, u_y T, -4\sigma T + 5\varpi T, 0\right)^\mathsf{T}, \tag{129}$$

and the equilibrium temperature distribution function $g_i^{eq}$ is given by

$$g_i^{eq} = \begin{cases} \sigma T - (1-\tilde{w}_0)T, & i = 0, \\ \tilde{w}_i T \left(1 + \dfrac{\mathbf{e}_i \cdot \mathbf{u}}{c_{sT}^2}\right), & i = 1 \sim 4. \end{cases} \tag{130}$$

The effective thermal diffusivity is defined as $\alpha_e = c_{sT}^2(\tau_g - 0.5)\delta_t$. By modifying the equilibrium moments appropriately, the artificial coupling has been avoided. This feature is very useful in practical applications.

In Ref. [73], Gao et al. proposed a modified thermal LB model for simulating conjugate heat transfer in a system simultaneously containing a porous medium and other media. In their work, a free parameter $\gamma$, which keeps unvaried in the entire domain, has been introduced into the equilibrium internal energy distribution function. Therefore, no special treatment is needed in Gao et al.'s model and the temperature and heat flux continuities at the interface can be automatically satisfied. Recently, Liu and He [74] have developed a 3D DDF-MRT model for convection heat transfer in porous media. They showed that the DDF-MRT model has a second-order accuracy in space. Moreover, Rong et al. [209], Liu and He [210], and Grissa et al. [211] have developed axisymmetric DDF-LB models for studying axisymmetric thermal flows in porous media. Besides the above models, several DDF-LB models have also been proposed for convection heat transfer in anisotropic porous media [212,213]. Seta [212] pointed out that for incompressible flows in a hydrodynamically anisotropic porous medium, the velocity $\mathbf{u}$ cannot be calculated via Eq. (103). To obtain the velocity $\mathbf{u}$ from $\rho\mathbf{u} = \sum_i \mathbf{e}_i f_i + 0.5\delta_t \rho \mathbf{F}$, one needs to use some other methods (e.g., Newton-Raphson method) [212]. In order to avoid using some complex methods, Hu et al. [213] employed an approximation method to derive the velocity $\mathbf{u}$. In Hu et al.'s method, the nonlinear term $|\mathbf{u}(t+\Delta t)|\mathbf{u}(t+\Delta t)$ is replaced by $|\mathbf{u}(t)|\mathbf{u}(t+\Delta t)$ so as to derive the velocity $\mathbf{u}$ from the nonlinear equation (see Eqs. (27) and (28) in Ref. [213]).

*3.2.3. The thermal REV-scale LB models under LTNE condition*

As previously mentioned, the assumption of LTE between the fluid and solid matrix breaks down in many practical applications, and then the LTNE model should be employed to take account of the interfacial heat transfer between the fluid and solid matrix. In 2014, Gao et al. [214] devised a thermal LB model for convection heat transfer in porous media under LTNE condition. Gao et al.'s model was devised in the framework of the triple-distribution-function (TDF) approach: two temperature-based BGK-LB equations were proposed for the temperature fields of the fluid and solid matrix in addition to the BGK-LB equation of the density distribution function for the velocity field described by the generalized non-Darcy model. In Gao et al.'s model, the source terms accounting for the interfacial heat transfer and internal heat source were included based on Shi and Guo's approach [165]. Subsequently, Liu and He [215] constructed a TDF-MRT model for convection heat transfer in porous media under LTNE condition. In Liu and He's work, the source terms were included into the model based on Eq. (83), and the the temperatures $T_f$ and $T_s$ can be obtained through some simple algebraic operations. Gao et al.'s model and Liu and He's model were developed based on the energy equations (126) and (127) without considering the compression work and viscous heat dissipation. Recently, following Guo et al.'s total-energy-based DDF-LB approach [146], Wang et al. [216] have constructed a TDF-LB model for thermal flows in porous media under LTNE condition, in which the compression work and viscous heat dissipation were both considered.

## 4. The LB method for solid-liquid phase-change heat transfer

Latent heat storage (LHS) using solid-liquid phase-change materials (PCMs) has attracted a great deal of attention due to its importance for energy saving, efficient and rational utilization of available resources, and optimum utilization of renewable energies [217-219]. Solid-liquid PCMs absorb or release thermal energy by taking advantage of their latent heat (heat of fusion) during solid to liquid or liquid to solid phase-change process. In the past two decades, the LB method has been extensively used to study solid-liquid phase-change problems. In the LB community, the first attempt to use the LB method to simulate solid-liquid phase change was made by De Fabritiis et al. [220] in 1998. Since then, many LB models for solid-liquid phase change have been developed from different points of view. Most of the existing thermal LB models for solid-liquid phase change can be generally classified into two major categories: the phase-field method [221-228] and the enthalpy-based method [54-58,75-81,194,229-246]. Besides, a couple of LB models were recently developed based on some interface-tracking methods [180,247].

In the phase-field LB method, the solid and liquid phases are distinguished by a continuous variable, namely the order parameter or the phase field. The solid-liquid interface is implicitly captured by solving an interface-capturing equation of the order parameter. The first solid-liquid phase-field LB model might be attributed to Miller et al. [221], who simplified and extended De Fabritiis et al.'s model [220] by using only one type of quasiparticles and a phase-field method. Many phase-field LB models [222-228] have been developed along this line, and the phase-field LB method can serve as a powerful and accurate numerical tool for solid-liquid phase-change problems.

The enthalpy-based LB method provides an alternative approach to study solid-liquid phase-change problems. In this method, the solid and liquid phases as well as the solid-liquid interface are distinguished through the liquid fraction. Unlike the phase-field method, the liquid fraction can be obtained when the energy equation is solved, and simultaneously the phase interface is implicitly captured by the liquid fraction. Because of its simplicity and easy implementation, the enthalpy-based method has been extensively employed in simulating solid-liquid phase-change heat transfer problems. In what follows, the enthalpy-based LB models for simulating solid-liquid phase-change heat transfer are reviewed. Moreover, several numerical results of solid-liquid phase-change heat transfer in the absence of a porous medium are provided.

*4.1. The enthalpy-based LB models for solid-liquid phase change*

*4.1.1. Governing equations*

For solid-liquid phase-change heat transfer, the energy equation can be written as [229,231,232, 236,248]

$$\frac{\partial}{\partial t}(\rho c_p T) + \nabla \cdot (\rho c_p T \mathbf{u}) = \nabla \cdot (k \nabla T) + S, \tag{131}$$

where $S$ is the latent-heat source term. In phase-change problems, the source term $S$ depends on the nature of the latent-heat evolution and should be carefully defined [248]. The enthalpy can be expressed as $H = c_p T + \Delta H$, i.e., the sum of sensible enthalpy $h = c_p T$ and latent enthalpy $\Delta H$. The latent heat contribution can be specified as a function of the temperature, i.e., $\Delta H = f(T) = L_a f_l$, where $L_a$ is the latent heat of melt, and $f_l$ is the liquid fraction ($f_l = 0$ represents the solid phase, $f_l = 1$ represents the liquid phase, and $0 < f_l < 1$ represents the interface region or mushy zone). Once the enthalpy $H$ is determined, the liquid fraction $f_l$ is then defined as

$$f_l = \begin{cases} 0, & H \leq H_s, \\ \dfrac{H - H_s}{H_l - H_s}, & H_s < H < H_l, \\ 1, & H \geq H_l, \end{cases} \tag{132}$$

where $H_s = c_{ps} T_s$ is the enthalpy at the solidus temperature $T_s$, and $H_l = c_{pl} T_l + L_a$ is the enthalpy at the liquidus temperature $T_l$; $c_{ps}$ and $c_{pl}$ are the specific heat of the solid and liquid phases, respectively. The latent-heat source term $S$ in the energy equation (131) is given by

$S = -\left[\partial_t (\rho \Delta H) + \nabla \cdot (\rho \Delta H \mathbf{u})\right]$ [232, 236, 248]. $S$ is caused by the absorption or release of thermal energy during phase-change process. For a pure material undergoing phase change, the term $\nabla \cdot (\rho \Delta H \mathbf{u})$ vanishes and the latent-heat source term reduces to $S = -\partial_t (\rho \Delta H)$. Correspondingly, the energy equation (131) becomes

$$\frac{\partial}{\partial t}(\rho c_p T) + \nabla \cdot (\rho c_p T \mathbf{u}) = \nabla \cdot (k \nabla T) - \frac{\partial}{\partial t}(\rho L_a f_l). \tag{133}$$

*4.1.2. The enthalpy-based LB models*

The first enthalpy-based LB model for solid-liquid phase change was introduced by Jiaung et al. [229] in 2001. Jiaung et al.'s model was constructed based on Eq. (133) without considering the convection term, i.e., $\partial_t T = \nabla \cdot (\alpha \nabla T) - L_a \partial_t f_l / c_p$. The latent-heat source term $S = -L_a \partial_t f_l / c_p$ was taken into account by adding a source term into the thermal BGK-LB equation (66), i.e.,

$$g_i(\mathbf{x} + \mathbf{e}_i \delta_t, t + \delta_t) - g_i(\mathbf{x}, t) = -\frac{1}{\tau_g}\left[g_i(\mathbf{x}, t) - g_i^{eq}(\mathbf{x}, t)\right] - \delta_t \tilde{w}_i \frac{L_a}{c_p} \frac{\partial f_l}{\partial t}. \tag{134}$$

To keep consistent with the thermodynamics and kinetics of phase change, Jiaung et al. employed a forward finite-difference scheme to calculate the transient term $\partial_t f_l$ in the nonlinear latent-heat source term, i.e., $\partial_t f_l = \left[f_l(\mathbf{x}, t + \delta_t) - f_l(\mathbf{x}, t)\right]/\delta_t$. Because the liquid fraction $f_l$ at the new time level $t + \delta_t$ is unknown, an iteration procedure is needed in Jiaung et al.'s model so as to obtain convergent solutions of the temperature and liquid-fraction fields at each time step. Hence, Jiaung et al.'s method can be viewed as an iterative-enthalpy-based method.

In 2005, Chatterjee and Chakraborty [230] proposed an enthalpy-based LB model for simulating conduction dominated phase-change problem. They introduced a thermodynamically consistent enthalpy updating scheme for the iteration procedure, and a suitable relaxation factor has been used so as to guarantee the convergence of the iteration procedure. Later, following the line of He et al.'s internal-energy-based DDF-LB model [142], Chatterjee and Chakraborty [231] further extended their model for solid-liquid phase change in the presence of fluid flow by coupling a modified thermal LB model with a fixed-grid enthalpy-porosity technique. In their work, the morphology of the phase-change region was treated as an equivalent porous medium that offers a resistance force towards fluid flow through the phase-change region, and consequently, one does not need to impose hydrodynamic or thermal boundary conditions at the solid-liquid interface. Shortly afterwards, Chakraborty and Chatterjee [232] proposed a hybrid thermal LB model for solid-liquid phase change in the presence of convective transport. In the hybrid thermal LB model, a control-volume-based fully implicit finite-difference scheme in conjunction with the iterative-enthalpy formulation was employed to solve the temperature field governed by Eq. (133).

Based on a modified version of Jiaung et al.'s model, Huber et al. [55] developed a DDF-LB model for simulating melting coupled with natural convection. Huber et al. used the bounce-back scheme to impose the no-slip velocity condition at the solid-liquid interface (the collision process of the LB equation for the flow filed is executed only for $f_l > 0.5$), and in order to reduce the computational costs, they artificially set the number of iterations to be one at the expense of the numerical accuracy. Using an iterative-enthalpy-based LB model, Jourabian et al. [233] have assessed the applicability of several heat transfer enhancement techniques (including the implementation of the multitubes and the insertion of the Cu nanoparticles) for accelerating the melting of ice. In Fig. 5, the streamlines and temperature contours at different times ($\theta = FoSte$ is the dimensionless time) for array (II) with $\varphi = 0.0, 0.01,$ and $0.02$ are shown ($\varphi$ is the volume fraction of nanoparticles). As shown in the figure, it can be concluded that adding nanoparticles can accelerate the melting rate more efficiently in the bottom of storage unit.

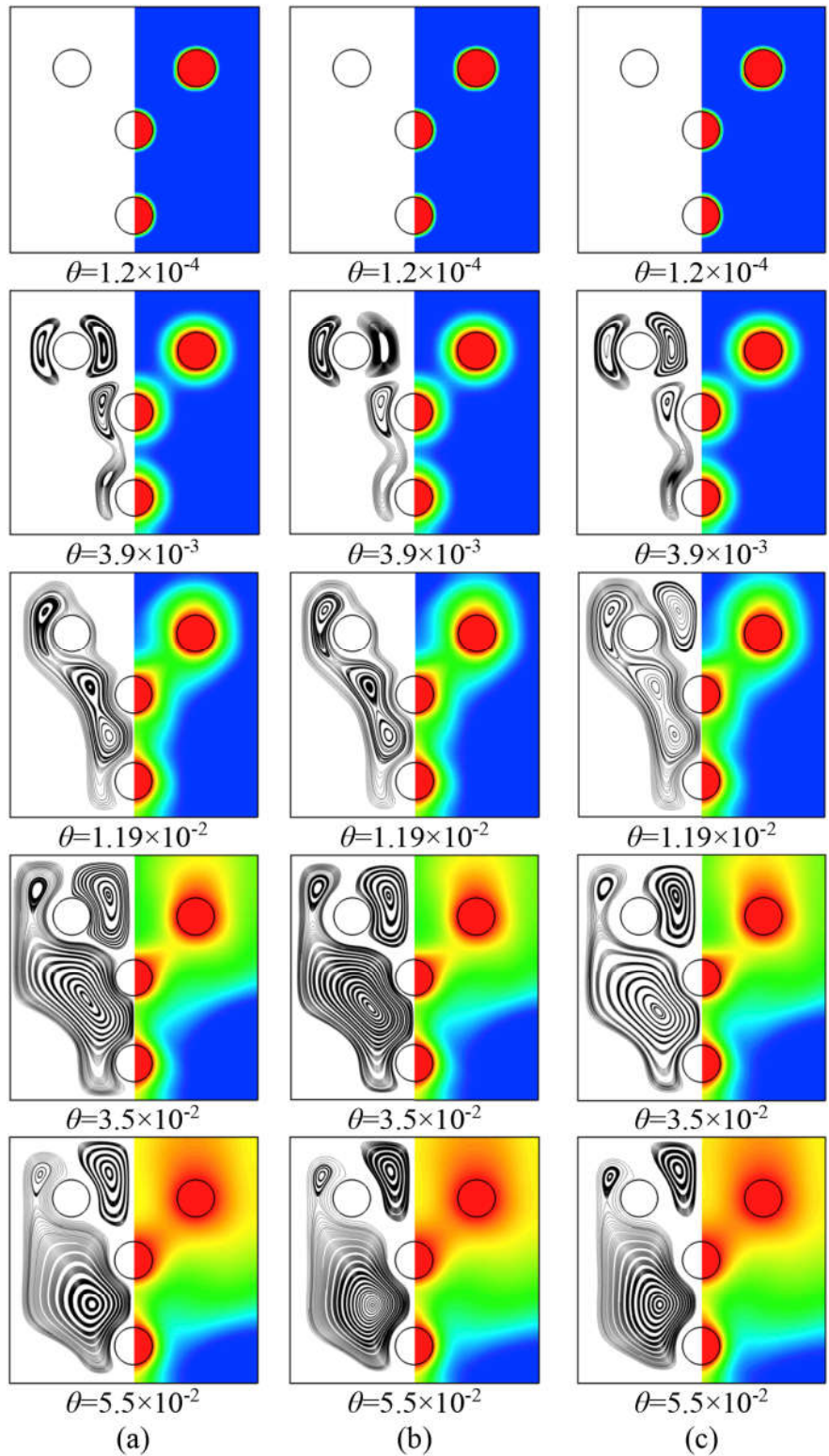

**Fig. 5.** Streamlines (left) and temperature contours (right) at different times for array (II): (a) $\varphi = 0.0$, (b) $\varphi = 0.01$, and (c) $\varphi = 0.02$. (Reprinted from Jourabian et al. [233].)

In the above solid-liquid enthalpy-based LB models (except the hybrid model in Ref. [232]), the nonlinear latent-heat source term accounting for the phase change is treated as a source term in the thermal LB equation, which causes the explicit time-matching LB equation to be implicit. Therefore, an iteration procedure is needed at each time step so as to obtain the convergent solution of the implicit LB equation. The iteration procedure severely affects the computational efficiency, and the inherent merits of the LB method have been lost. Note that in the hybrid model [232], iterative-enthalpy formulation is also needed to treat the solid-liquid phase-change process. The iteration procedure in the iterative-enthalpy-based LB models severely restricts their applications. An alternative approach was therefore proposed by Eshraghi and Felicelli [234], who used an implicit formulation to treat the latent-heat source term. By solving a group of linear equations at each lattice node, the temperature and liquid fraction can be obtained without using the iterative enthalpy-based method. Based on Eshraghi and Felicelli's conduction melting LB model, Feng et al. [235] developed a solid-liquid enthalpy-based LB model to investigate the melting behavior of nanoparticle-enhanced PCM in a 2D rectangular cavity heated from below. In Feng et al.'s work, the no-slip velocity condition in the interface and solid phase regions was treated by the immersed moving boundary scheme [236,249].

In 2013, Huang et al. [236] developed a new enthalpy-based LB model for solid-liquid phase change. By combing the transient term $\partial_t(\rho c_p T)$ with the latent-heat source term $\partial_t(\rho L_a f_l)$, the energy equation (133) becomes $\partial_t(\rho H) + \nabla \cdot (\rho c_p T \mathbf{u}) = \nabla \cdot (k \nabla T)$. To recover this enthalpy-based energy equation, Huang et al. defined the following equilibrium enthalpy distribution function

$$g_i^{eq} = \begin{cases} H - c_p T + \tilde{w}_0 c_p T \left(1 - \dfrac{|\mathbf{u}|^2}{2c_{sT}^2}\right), & i = 0, \\ \tilde{w}_i c_p T \left[1 + \dfrac{\mathbf{e}_i \cdot \mathbf{u}}{c_{sT}^2} + \dfrac{\mathbf{u}\mathbf{u}:(\mathbf{e}_i \mathbf{e}_i - c_{sT}^2 \mathbf{I})}{2c_{sT}^4}\right], & i = 1 \sim 8. \end{cases} \quad (135)$$

The enthalpy $H$ is calculated by $H = \sum_i g_i$. The relationship between the enthalpy $H$ and temperature $T$ is given by

$$T = \begin{cases} H/c_p, & H \leq H_s, \\ T_s + \dfrac{H - H_s}{H_l - H_s}(T_l - T_s), & H_s < H < H_l, \\ T_l + (H - H_l)/c_p, & H \geq H_l. \end{cases} \quad (136)$$

The liquid fraction $f_l$ is determined by Eq. (132). By introducing the enthalpy into the equilibrium distribution function, the enthalpy $H$ becomes the basic evolution variable of the thermal LB equation. Actually, the treatment of the latent-heat source term in Huang et al.'s model can be viewed as a special implicit scheme. By taking advantage of the kinetic nature of the LB method, the nonlinear latent-heat source term vanishes in the thermal LB equation, and the iteration procedure as well as solving a group of linear equations has been avoided in Huang et al.'s model. Owing to its conceptual simplicity, computational efficiency and accuracy, Huang et al.'s method has attracted a great deal of attention and has been applied to many solid-liquid phase-change problems.

Subsequently, Huang and Wu [194] further improved their enthalpy-based LB model by considering the phase interface effects. For the differences in thermophysical properties between solid and liquid phases, a reference specific heat $c_{p,\text{ref}}$ was introduced into the equilibrium enthalpy distribution function (see Eq. (135)), which makes the thermal conductivity and specific heat decoupled. Huang and Wu have shown that the numerical diffusion across the phase interface can be dramatically reduced by using the MRT collision model with $\Lambda = (s_e^{-1} - 0.5)(s_j^{-1} - 0.5) = 1/4$ (here, $\Lambda$ is the so-called "magic" parameter). Most recently, Huang and Wu [237] developed an enthalpy-based LB method with adaptive mesh refinement for accurate and efficient simulation of solid-liquid phase-change problems. To accurately realize the no-slip velocity condition on the diffusive phase interface and in the solid phase, a volumetric LB scheme based on a kinetic assumption has been proposed, which can avoid non-physical flow in the solid phase.

In a recent study carried out by Luo et al. [238], different enthalpy-based LB models for solid-liquid phase change have been tested and analyzed. They found that the HLBM-MRT scheme (i.e., MRT-LB model [110] for flow field, and enthalpy-based LB model [236] for temperature field) is more suitable for modeling convection dominated melting process. For melting with convection in a square cavity, Luo et al. showed that the temperature obtained by the TLBM (the thermal LB equation is given by Eq. (134), but the transient term $\partial_t f_l$ is calculated by $\partial_t f_l = \left[ f_l(\mathbf{x},t) - f_l(\mathbf{x},t-\delta_t) \right]/\delta_t$ so as to avoid the iteration procedure) in the solid phase near the solid-liquid interface locally decreases below the initial solid temperature, while the temperature obtained by Huang et al.'s model [236] remains precisely (see Fig. 6 in Ref. [238]).

In 2017, Huo and Rao [239] developed a new phase-change LB model based on the quasi-enthalpy method. In Huo and Rao's model, the update of temperature consists of two steps: the "prediction" and "consumption (for melting)" or "release (for solidification)". By using the quasi-enthalpy method, the temperature and liquid fraction can be obtained directly without iterations. Huo and Rao argued that their method is more accurate than Huang et al.'s enthalpy-based method [236]. For axisymmetric solid-liquid phase change, an enthalpy-based LB method has been recently proposed by Li and He [240]. They have shown that, when the parallel algorithm is applied, the LB method is superior to the traditional FVM in computational time with the same time step and lattice number. For solid-liquid phase change with natural convection in an annulus between two coaxial vertical cylinders, Li and He discussed the influence of the $Pr$ on the melting processes, which can be seen in Fig. 6.

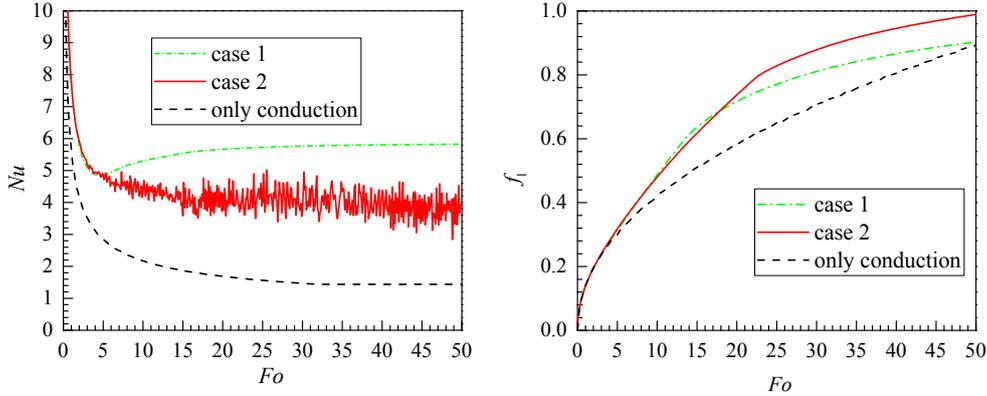

**Fig. 6.** The average Nusselt number along the left wall $Nu$ versus $Fo$ (left) and the total liquid fraction $f_l$ versus $Fo$ (right) for case 1 ( $Ra = 10^5$, $Pr = 0.7$, and $St = 0.01$ ) and case 2 ( $Ra = 10^5$, $Pr = 0.02$, and $St = 0.01$ ). (Reprinted from Li and He [240].)

Now attention turns to the 3D solid-liquid enthalpy-based LB models. In the literature [241], Zhu et al. extended Huber et al.'s enthalpy-based method [55] to 3D case using the D3Q7 lattice. In order to avoid the iteration procedure, Zhu et al. employed an explicit difference scheme to treat the transient term $\partial_t f_l$, i.e., $\partial_t f_l = \left[ f_l(\mathbf{x},t) - f_l(\mathbf{x},t-\delta_t) \right]/\delta_t$. Li and He [242] developed an enthalpy-based MRT model for 3D solid-liquid phase change. The reasonable relationship of the relaxation rates in the MRT model has been analyzed. Based on the D3Q7 lattice, Li and He found that the numerical diffusion across the solid-liquid interface can be eliminated by using the MRT collision model with $\sigma_1 + \sigma_4 = 2$. Hu et al. [243] extended Huang et al.'s enthalpy-based LB model to 3D case using the D3Q19 lattice. For the flow field, Hu et al. adopted a smoothed profile method to treat the no-slip velocity condition at the moving solid-liquid interface. By using the smoothed profile method, the force acting on the solid phase can be calculated directly, and as a result, Hu et al.'s model can be used to simulate phase-change problems when the solid phase can move freely. Some numerical results can be found in Fig. 7.

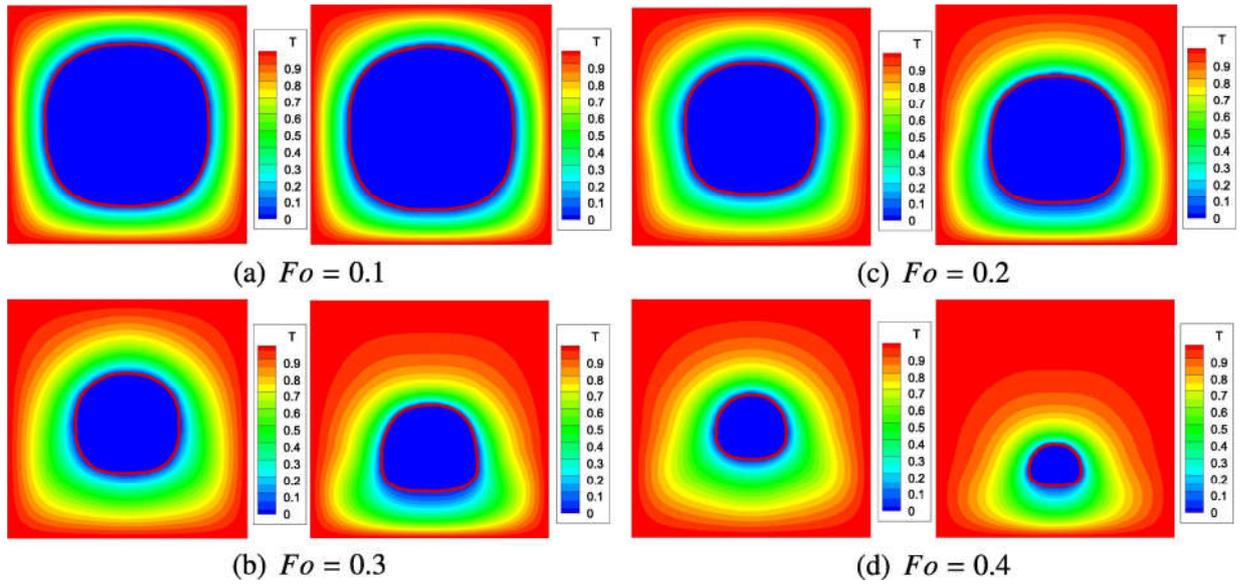

**Fig. 7.** Temperature fields and phase interfaces in *y*-plane at $y = 0.5$ for different $Fo$ (Left: the solid phase is fixed; Right: the solid phase is free). (Reprinted from Hu et al. [243].)

Zhao and Cheng [244] developed a new 3D solid-liquid phase-change LB method for simulating laser cutting of thin metal plates. Zhao and Cheng's method was constructed based on the enthalpy-based LB model of Huang et al. [236] and the pseudopotential LB model of Shan and Chen [129]. The pseudopotential LB model was employed to treat the interactions between the liquid phase of the molten materials and the gas phase existing in the computational domain (including the environmental gas and the gas jet). The performances of the laser cutting of a thin metal plate have been evaluated by Zhao and Cheng for several dimensionless parameters. Some numerical results can be found in Fig. 8.

To simulate triple phase-change heat transfer phenomena, Li and Cheng [245] proposed a triple phase-change LB model, which couples Gong-Cheng's improved liquid-vapor phase-change model [250] and the solid-liquid enthalpy-based LB model of Jiaung et al. [229]. In Li and Cheng's work, the no-slip velocity condition on the solid-liquid interface was treated by the immersed moving boundary scheme [236,249]. Using the model, Li and Cheng successfully simulated some triple phase-change heat transfer problems.

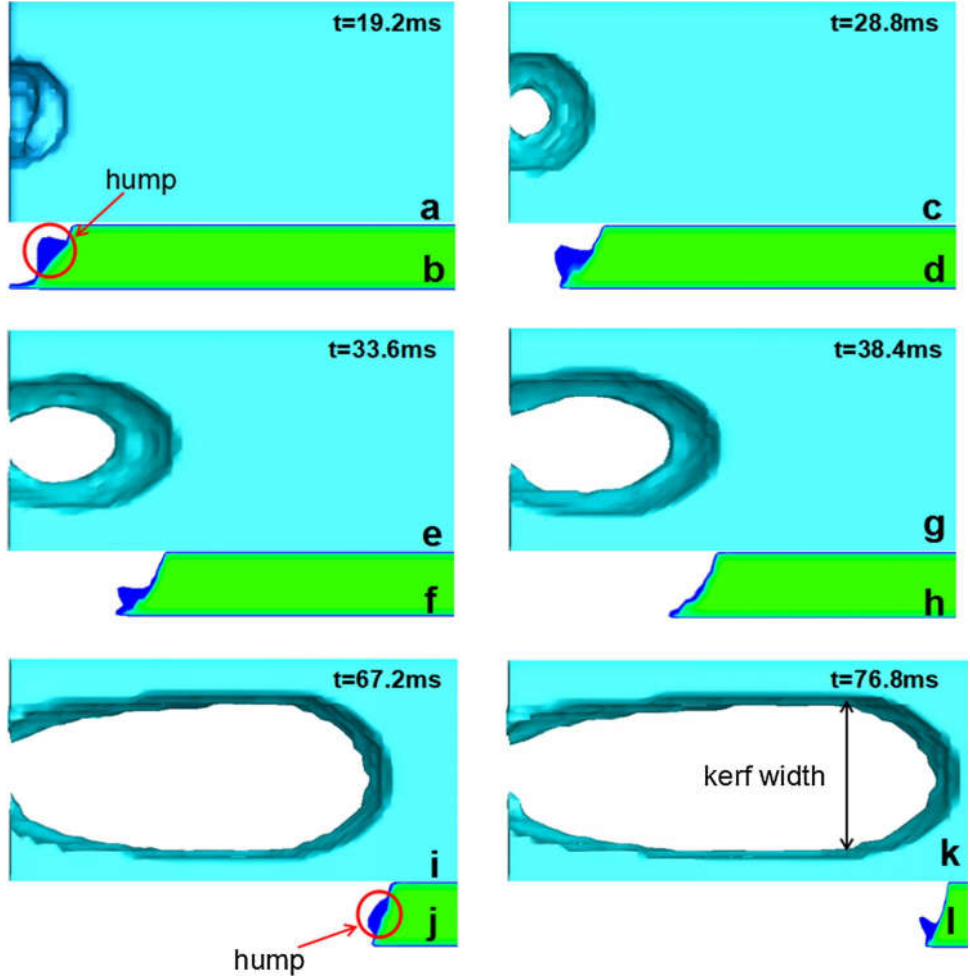

**Fig. 8.** The LB simulation results of the kerf's geometry and melt's shape under experimental condition for a 304 stainless steel work piece. (Reprinted from Zhao and Cheng [244].)

*4.2. The enthalpy-based LB models for solid-liquid phase change in porous media*

*4.2.1. Governing equations*

For solid-liquid phase-change heat transfer in porous media under LTNE condition, the volume-averaged energy equations of the PCM (including liquid phase ($l$) and solid phase ($s$)) and the solid matrix can be written as follows [80,81,214,251]

$$\frac{\partial}{\partial t}\left\{\phi\left[f_l\rho_l c_{pl}+(1-f_l)\rho_s c_{ps}\right]T_f\right\}+\nabla\cdot\left(\rho_l c_{pl}T_f\mathbf{u}\right)=\nabla\cdot\left(\phi k_f\nabla T_f\right)+h_v\left(T_m-T_f\right)-\frac{\partial}{\partial t}\left(\phi\rho_l L_a f_l\right), \quad (137)$$

$$\frac{\partial}{\partial t}\left[(1-\phi)\rho_m c_{pm}T_m\right]=\nabla\cdot\left[(1-\phi)k_m\nabla T_m\right]+h_v\left(T_f-T_m\right). \quad (138)$$

The effective thermal conductivities of the PCM and solid matrix are defined by $k_{e,f} = \phi k_f$ and $k_{e,m} = (1-\phi)k_m$, respectively; the thermal conductivity and specific heat of the PCM are given by $k_f = f_l k_l + (1-f_l)k_s$ and $c_{pf} = f_l c_{pl} + (1-f_l)c_{ps}$, respectively.

Under LTE condition, i.e., $T_f = T_m = T$, the energy equations (137) and (138) can be replaced by the following single-temperature equation [252]

$$\frac{\partial}{\partial t}\left(\overline{\rho c_p} T\right) + \nabla \cdot \left(\rho_l c_{pl} T\mathbf{u}\right) = \nabla \cdot (k_e \nabla T) - \frac{\partial}{\partial t}(\phi \rho_l L_a f_l), \tag{139}$$

where $\overline{\rho c_p} = \phi\left[f_l \rho_l c_{pl} + (1-f_l)\rho_s c_{ps}\right] + (1-\phi)\rho_m c_{pm}$, and $k_e = \phi k_f + (1-\phi)k_m$. Eq. (139) can be simplified as

$$\frac{\partial}{\partial t}(\sigma T) + \nabla \cdot (T\mathbf{u}) = \nabla \cdot (\alpha_e \nabla T) - \phi \frac{L_a}{c_{pl}} \frac{\partial f_l}{\partial t}, \tag{140}$$

where $\sigma = \overline{\rho c_p}/(\rho c_p)_l$ is the heat capacity ratio, and $\alpha_e = k_e/(\rho c_p)_l$ is the effective thermal diffusivity.

*4.2.2. The REV-scale enthalpy-based LB models*

In 2011, Gao and Chen [75] developed a temperature-based DDF-BGK model for melting with convection heat transfer in porous media under LTE condition. Gao and Chen's model was constructed based on Eq. (140), and the nonlinear latent-heat source term was treated by Jiaung et al.'s iterative-enthalpy-based method [229]. This model can be used to simulate freezing and solidification in porous media. Later, Liu and He [76] have presented a temperature-based DDF-MRT model for solid-liquid phase change with natural convection in porous media at the REV scale. However, to deal with the nonlinear latent-heat source term, Liu and He's model still needs the iteration procedure.

Following the line of Huang et al.'s model [236], Wu et al. [77] developed a new enthalpy-based LB model for solid-liquid phase change with convection heat transfer in porous media. By combing the transient term $\partial_t \left( \overline{\rho c_p} T \right)$ with the latent-heat source term $\partial_t \left( \phi \rho_l L_a f_l \right)$, the energy equation (139) becomes $\partial_t \left( \rho_l H_e \right) + \nabla \cdot \left( \rho_l c_{pl} T \mathbf{u} \right) = \nabla \cdot \left( k_e \nabla T \right)$, where $H_e = \sigma c_{pl} T + \phi L_a f_l$ is the effective enthalpy. To recover this effective-enthalpy-based energy equation, Wu et al. defined the following equilibrium enthalpy distribution function

$$g_i^{eq} = \begin{cases} H_e - \sigma_0 c_{pl} T + \tilde{w}_0 \sigma c_{pl} T \left( \dfrac{\sigma_0}{\sigma} - \dfrac{u^2}{2\sigma^2 c_{sT}^2} \right), & i = 0, \\ \tilde{w}_i \sigma c_{pl} T \left[ \dfrac{\sigma_0}{\sigma} + \dfrac{\mathbf{e}_i \cdot \mathbf{u}}{c_{sT}^2 \sigma} + \dfrac{\mathbf{uu} : \left( \mathbf{e}_i \mathbf{e}_i - c_{sT}^2 \mathbf{I} \right)}{2\sigma^2 c_{sT}^4} \right], & i = 1 \sim 8, \end{cases} \quad (141)$$

where $\sigma_0$ is the reference thermal capacity ratio. The equilibrium moments in the MRT model can be obtained via $\mathbf{n}^{eq} = \mathbf{N} \mathbf{g}^{eq}$. By introducing the effective enthalpy into the equilibrium distribution function, the effective enthalpy $H_e$ becomes a basic evolution variable of the thermal LB equation. As a result, the iteration procedure is not needed in Wu et al.'s model. As compared with previous models [75,76], Wu et al.'s model can provide a higher computational efficiency.

In a recent study carried out by Gao et al. [246], a modified enthalpy-based LB model has been developed by incorporating the enthalpy and a free parameter into the equilibrium distribution function for the PCM temperature field. In Gao et al.'s model, the equilibrium enthalpy distribution function is given by

$$g_i^{eq} = \begin{cases} h_{sys} - \gamma T + \tilde{w}_0 \gamma T, & i = 0, \\ \tilde{w}_i T \left( \gamma + \rho c_p \dfrac{\mathbf{e}_i \cdot \mathbf{u}}{c_{sT}^2} \right), & i = 1 \sim 8, \end{cases} \quad (142)$$

where $h_{sys}$ is the enthalpy of the porous matrix/solid/liquid mixture system, and $\gamma$ is a free parameter (keeps unvaried in the entire domain). Gao et al.'s model avoids the iteration procedure and can be used to simulate solid-liquid phase change in porous media with conjugate heat transfer. Moreover, Liu and He have proposed a 3D enthalpy-based DDF-MRT model for solid-liquid phase change with convection heat transfer in porous media and the equilibrium moments of the D3Q7-MRT model are defined as $\mathbf{n}^{eq} = \left(H_e, c_{pl}Tu_x, c_{pl}Tu_y, c_{pl}Tu_z, \varpi c_{p,\text{ref}}T, 0, 0\right)^\text{T}$, where $c_{p,\text{ref}}$ is the reference specific heat.

In recent years, some progress has also been achieved in simulating solid-liquid phase-change heat transfer in porous media under LTNE condition using the enthalpy-based LB method. In 2010, Gao et al. [78] proposed a thermal LB model to simulate the melting process coupled with natural convection in open-cell metal foams under LTNE condition. Their model was constructed in the framework of the TDF approach: the flow field, and the temperature fields of PCM and metal foam were solved separately by three different BGK-LB models. Subsequently, Gao et al. [214] further developed a thermal LB model for solid-liquid phase-change heat transfer in porous media under LTNE condition. By appropriately choosing the equilibrium temperature distribution functions and discrete source terms, the energy equations of the PCM and solid matrix can be correctly recovered. Recently, Tao et al. [79] also developed an enthalpy-based LB model to study the LHS performance of copper foams/paraffin composite PCM.

In these models [78,79,214], iteration procedure is needed, which severely affects the computational efficiency, and the inherent merits of the LB method have been lost. In 2017, Gao et al. [80] proposed an enthalpy-based TDF-BGK model for solid-liquid phase change in porous media under LTNE condition. By introducing the enthalpy and a free parameter into the equilibrium distribution function for the PCM temperature field, Gao et al.'s model avoids the iteration procedure and the relaxation time can be adjusted to reduce the numerical diffusion across phase interface. Most recently, Liu and He [81] have constructed an enthalpy-based TDF-MRT model for solid-liquid phase-change heat transfer in metal foams under LTNE condition. In Liu and He's model, the vector of the equilibrium moments $\mathbf{n}_g^{eq}$ is given by

$$\mathbf{n}_g^{eq} = \left( H_f, -4H_f + 2c_{f,\text{ref}}T_f, 4H_f - 3c_{f,\text{ref}}T_f, \frac{c_{pl}T_f u_x}{\phi c}, -\frac{c_{pl}T_f u_x}{\phi c}, \frac{c_{pl}T_f u_y}{\phi c}, -\frac{c_{pl}T_f u_y}{\phi c}, 0, 0 \right)^T, \quad (143)$$

where $H_f$ is the enthalpy of the PCM, and $c_{f,\text{ref}}$ is a reference specific heat.

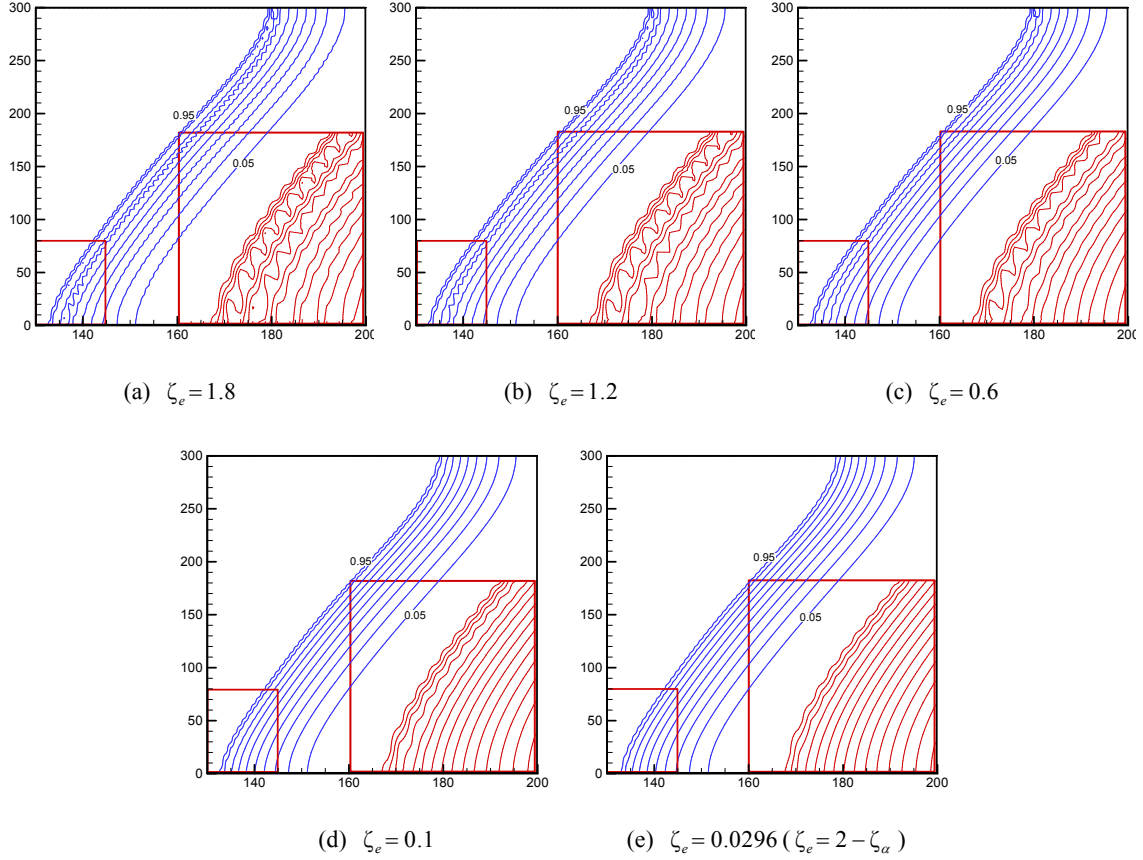

(a) $\zeta_e = 1.8$  (b) $\zeta_e = 1.2$  (c) $\zeta_e = 0.6$

(d) $\zeta_e = 0.1$  (e) $\zeta_e = 0.0296$ ($\zeta_e = 2 - \zeta_\alpha$)

**Fig. 9.** Local enlargement view of the phase interfaces for different values of $\zeta_e$ at $Fo = 0.002$. (Adapted from Liu and He [81].)

Liu and He's model has three distinctive features. First, the iteration procedure has been avoided. Second, by using the volumetric LB scheme, the no-slip velocity condition in the interface and solid phase regions can be accurately realized. Moreover, the MRT collision model is employed, and with additional degrees of freedom, it has the ability to reduce the numerical diffusion across phase interface induced by solid-liquid phase change. All the inherent merits of the standard LB method have been preserved in Liu and He's model. Comparisons and discussions have been made by Liu and He to offer some insights into the roles of the collision model, phase-interface-treatment scheme, enthalpy formulation, and relaxation rate $\zeta_e$ in the enthalpy-based MRT-LB model, which are very useful for practical applications. As shown in Fig. 9, the numerical diffusion across the phase interface can be effectively reduced by the MRT collision model with $\zeta_e = 2 - \zeta_\alpha$.

## 5. Applications

In the above sections, the fundamentals of the LB method for fluid flow and heat transfer have been briefly summarized, and the advances in the theory of the thermal LB methods for single-phase and solid-liquid phase-change heat transfer in porous media have been comprehensively reviewed. In this section, some representative applications of the thermal LB methods in these areas are reviewed. The applications of the LB method to predict effective thermal conductivity of porous materials are also briefly reviewed.

*5.1. Pore-scale heat transfer*

In pore-scale simulations, an adequate description of the pore structure of the porous medium is necessary. For a wide range of natural (e.g., soil, sandstone, limestone, coal) and man-made (composite materials, high-porosity metal foams) porous media, the pore structures can be reconstructed from images obtained by imaging techniques such as X-ray computed tomography (CT). An example is shown in Fig. 10, where the pore structure is reconstructed based on the CT images [253]. On the other hand, some artificial porous media, which are constructed according to certain rules so as to satisfy some required statistical properties, can also be used for pore-scale simulations. For instance, one can randomly place some solid objects in the domain, or randomly select a node on the lattice as a solid one with a certain probability, until a given porosity is reached [7,87]. Some more sophisticated structure generation methods have also been developed from different points of view, such as the hard-sphere Monte-Carlo method [254], random obstacle location method [255], and quartet structure generation set (QSGS) method [50]. Among these methods, the QSGS method has been employed by many researchers in pore-scale LB simulations [45,46,256,257]. More details about the structure generation methods can be found in a review paper by Wang and Pan [258]. In what follows, the pore-scale LB simulations of heat transfer in porous media are reviewed.

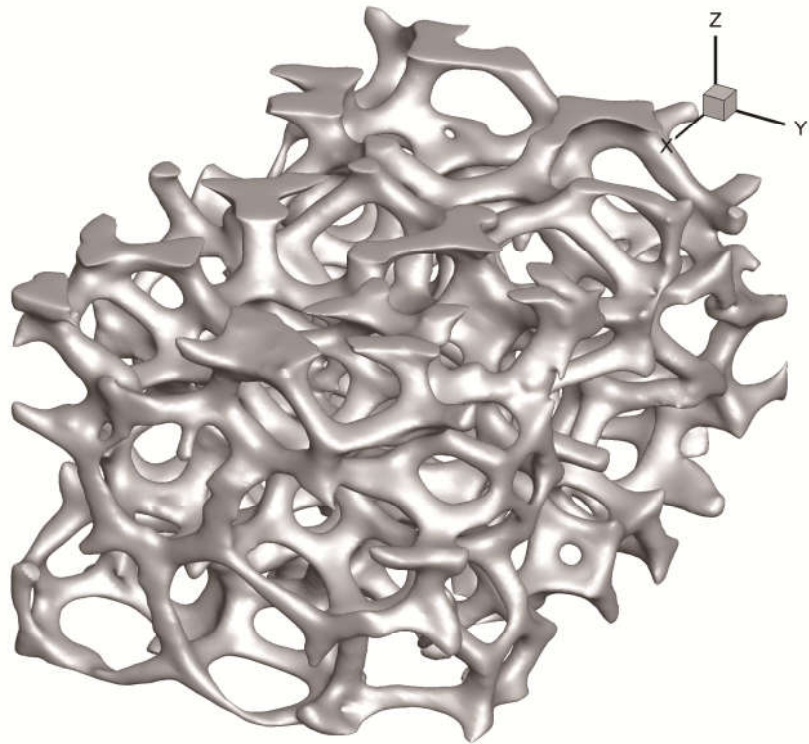

**Fig. 10.** Pore structure reconstructed from CT images. (Reprinted from Du and He [253].)

*5.1.1. Effective thermal conductivity of porous materials*

The effective thermal conductivity is one of the most important parameters that characterize the thermal transport properties of porous media. The effective thermal conductivity of porous media has been studied for over 100 years by using both theoretical and experimental approaches. Owing to the rapid developments in computer and computational techniques in the past three decades, numerical methods have been increasingly employed to predict the effective thermal conductivity of porous media [50,51,258].

In the literature, the first attempt of applying the LB method to predict the effective thermal conductivities of real porous media may be attributed to Qian et al. [47] in 2004. They proposed a D2Q5 thermal LB model to analyze the heat conduction process in porous media, and the effective thermal conductivities of several porous materials were calculated. Qian et al. found that the effective thermal conductivity is strongly dependent upon the porosity and pore structure, but it only has imperceptible dependence on the pore density. Wang et al. [50] built up a full stochastic-statistic-based method for predicting the effective thermal conductivity of random porous media. This method includes an LB model for fluid-solid conjugate heat transfer [184] and a random generation-growth method (the so-called QSGS method) for generating micro-morphology of random porous media. Wang et al.'s method has been successively applied to predict the effective thermal conductivity for both two-phase and three-phase porous materials. Their predictions agree well with a series of experimental data. Subsequently, Wang et al. [51] extended the stochastic-statistic-based method to study the 3D effect on the effective thermal conductivity of porous media. They found that the predicted effective thermal conductivity of 3D porous media varies with the cell number in the third dimension following an exponential relationship.

In 2008, Wang and Pan [52] studied the effective thermal conductivity of open-cell foam materials. They observed that the radiation heat transfer is a non-negligible factor for thermal transports in low-conductivity open-cell foam materials with high porosity. Wang and Pan also found that the effective thermal conductivity of open-cell foam materials is much higher than that of granular materials with the same components and porosity, which means that the inner netlike morphology of the open-cell foam materials can enhance the heat transfer capacity. In the meantime, Jeong et al. [259] studied the thermal and mass diffusivities in a porous medium of complex structure using the LB method. The calculated effective diffusivities were in good agreement with the analytical and numerical results when the inclusions were not overlapped. In particular, it was found that the deviation becomes significant if the thermal diffusivity of the inclusion is larger than that of the fluid in the medium for enhanced thermal conduction. Subsequently, Zhao et al. [256] calculated the effective thermal conductivity for different kinds of 3D porous media. The effects of the porosity and the average particle diameter on the effective thermal conductivity were investigated. They found that the effective thermal conductivity decreases as the porosity increases, while it increases as the average particle diameter increases. Later, Wang et al. [260] studied the effective thermal conductivity of granular porous materials with or without considering the thermal contact resistance. They found that the effective thermal conductivity decreases slightly as the average grain size increases for negligible thermal contact resistance, whereas when the thermal contact resistance is significant, the effective thermal conductivity decreases sharply as the average grain size gets smaller.

In 2012, Yablecki et al. [261] investigated the anisotropic effective thermal conductivity of the gas diffusion layer (GDL) of polymer electrolyte membrane fuel cell. They showed that the anisotropic structure of the GDL leads to an anisotropic thermal conductivity, with a higher value for the in-plane thermal conductivity than the through-plane thermal conductivity. The predicted values of the through-plane effective thermal conductivity from 2D simulations were almost one order of magnitude smaller than those predicted from 3D simulations. Askari et al. [262] have studied the effect of grain shape (elliptical inclusion) on the thermal conduction anisotropy in granular porous media. They demonstrated that the ratio of thermal conductivities of solid to fluid and aspect ratio (the ratio of semi-major to semi-minor axes) are the two most important factors affecting the thermal conduction anisotropy in granular porous media. In addition, using a non-dimensional LB method, Su et al. [263] successfully simulated the 3D thermal diffusion for different types of porous structures (including random isotropic homogenous and shape-constrained anisotropic heterogeneous types). Some results can be found in Fig. 11: it can be seen that the local temperature distributions differ a great deal between different structures. Moreover, it was found that, with equivalent macroscopic volume fractions, structures with higher mesoscopic volume fractions of high conductivity phases have higher effective thermal conductivity due to the greater connectivity of the higher conductive material at mesoscopic scale.

In 2009, Wang et al. [264] investigated the effective thermal conductivity enhancement of carbon fiber composites using a 3D thermal LB model. The predicted thermal conductivity enhancements were in good agreement with the available experimental data for both a low-loading fiber-in-oil suspension and a high-fiber-loading phase-change materials. Later, Zhou and Cheng [265] calculated the effective thermal conductivity of a composite filled with randomly distributed particle. The effects of the thermal conductivity ratio (between the particles and the matrix), the interactions between the particles and matrix, and particle volume fraction on the effective thermal conductivity of the composite were studied. They showed that the consideration of interactions has a great effect on the prediction of the effective thermal conductivity.

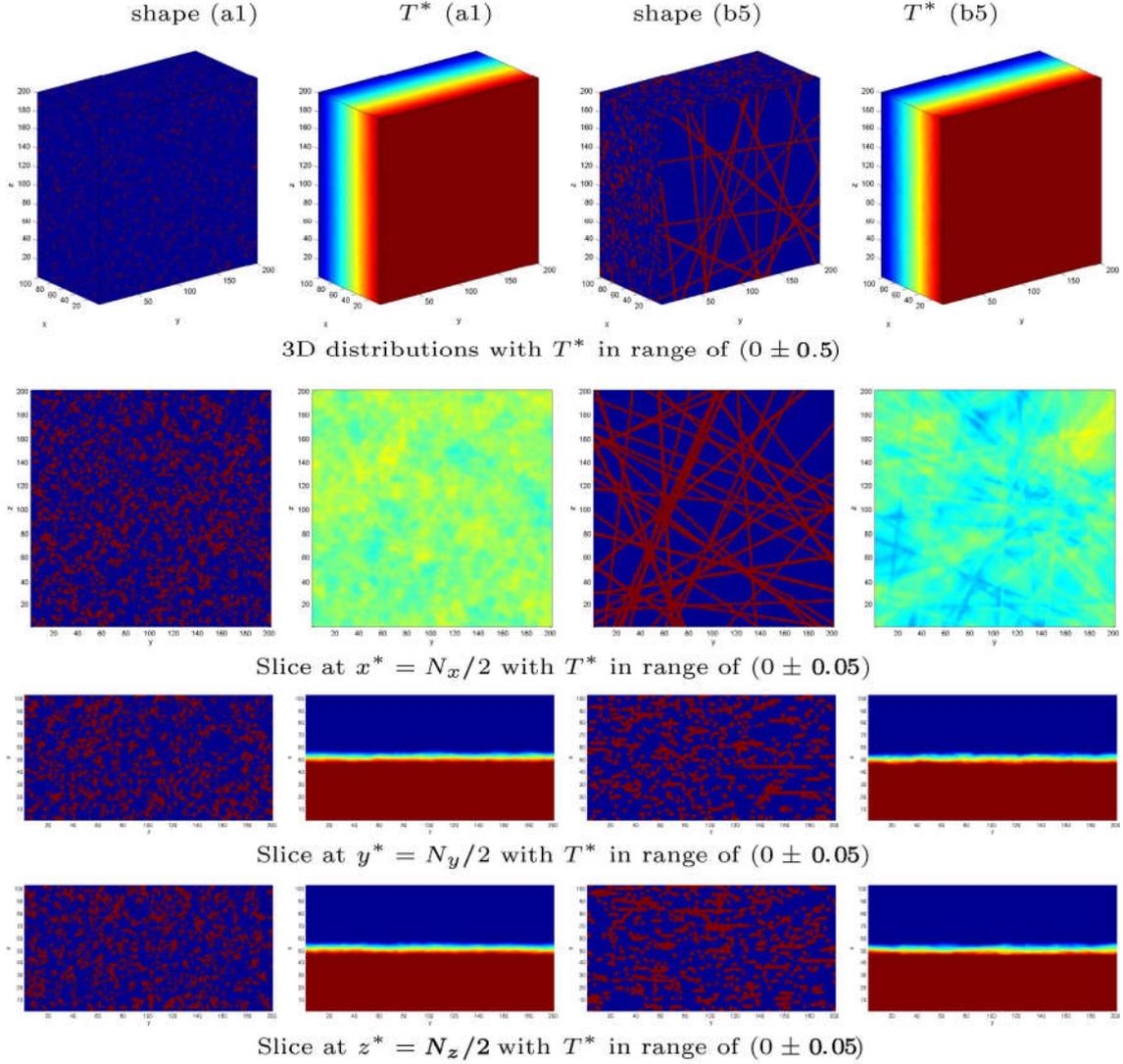

**Fig. 11.** The structures and local temperature distributions of sample (4) for shape (a1) and shape (b5) (the volume fraction of the red phase is 0.2847, and the volume fraction of the blue phase is 0.7153). (Reprinted from Su et al. [263].)

Using an MRT-LB model with off-diagonal elements in the relaxation matrix, Fang et al. [266] predicted the longitudinal and transverse thermal conductivities of needled C/C-SiC composite materials with anisotropic constituents. It was found that the transverse thermal conductivity of the C/C-SiC composite materials is larger than its longitudinal thermal conductivity, and using high thermal conductive fibers can significantly improve the longitudinal thermal conductivity of the needled C/C-SiC composite materials. Li et al. [267] have investigated the effective thermal conductivity of the silicone/phosphor composite and found that the effective thermal conductivity of the composite with larger particles is higher than that with small particles at the same volume fraction. Moreover, Pan and Zhao [268] studied the effective thermal conductivities of the commonly used porous thermochemical materials and the corresponding metal-foam composites. They found that the effective thermal conductivities of the three-phase composites are dominated by the metal phase, and the effect of porosities of metal foams is stronger than that of the operating temperatures.

*5.1.2. Single-phase heat transfer with fluid flow*

In 2005, Yamamoto et al. [48] studied the combustion flow in a 3D porous structure of Ni-Cr metal using a TDF-LB method. They found that the temperature is increased by soot combustion near the wall surface, and the local temperature is different due to the heterogeneous porous structure. This local information is indispensable to improve the thermal duration of diesel particulate filter . Later, Yamamoto and Takada [49] conducted LB simulations on soot combustion in porous media. It was found that the oxygen concentration around the wall surface is decreased by its reaction with soot, and then the maximum temperature is higher than the wall temperature.

In 2009, using Peng et al.'s simplified thermal LB model [143], Cai and Huai [269] conducted numerical simulations to study fluid-solid coupling heat transfer in fractal porous media. The temperature evolution under different ratios (1:1 and 10:1) of thermal conductivity of the solid matrix to that of the fluid was investigated. The results showed that, the temperature increases with a parabola shape along the $x$-direction when the ratio is 1:1, whereas when the ratio is 10:1, the temperature increases with a ladder shape. Later, Li et al. [270] simulated the complex process of the surface catalytic reaction in the catalyst porous media. Their results showed that the increase of the porosity or the decrease of the inlet velocity improves the conversion, and the flow no longer meets the Darcy's law due to the surface catalytic reaction. Moreover, Li et al. [271] have simulated the heat and mass transfer characteristics of the endothermic catalytic reaction of the isopropanol dehydrogenation with buoyancy. The effects of the buoyancy on the distributions of the outlet velocity, concentration and temperature were investigated. It was found that the effects of the buoyancy on the temperature are large. Nevertheless, the lower porosity can restrain the effects of the buoyancy.

Jeong and Choi [272] investigated the thermal dispersion in a porous medium of complex structure. The effects of medium properties (porosity and fluid-to-solid diffusivity ratio) on the dispersivity have been thoroughly examined. For various inclusion shapes and arrangements, new correlations for the longitudinal dispersivities have been proposed. In addition, Dai and Yang [273] have simulated the thermal oscillating flow both in planar planes and in porous media. For the thermal oscillating flow in porous media, they found that the working gas periodically absorbs or releases heat with the solid matrix, and the thermal diffusivity ratio between the solid matrix and fluid has great influence on the heat transfer between the solid matrix and fluid. Nazari et al. [274] have investigated the forced convection heat transfer in a channel partially filled with an anisotropic porous medium. They showed that the topology of the flow and heat transfer depends strongly on the arrangement of blocks, blockage ratios, Reynolds number and porosity. As the Reynolds number increases, the rate of heat transfer increases for both regular and random arrangement of blocks. Subsequently, Nazari et al. [275] studied the power-law fluid flow and heat transfer in a channel partially filled with an anisotropic porous medium for three power-law indices ($n = 0.8$, 1, and 1.2). The local and averaged Nusselt numbers on the channel walls were calculated, and it was found that the pseudoplastic fluids ($n < 0.8$) generate the highest heat transfer rate for all configurations of obstacles.

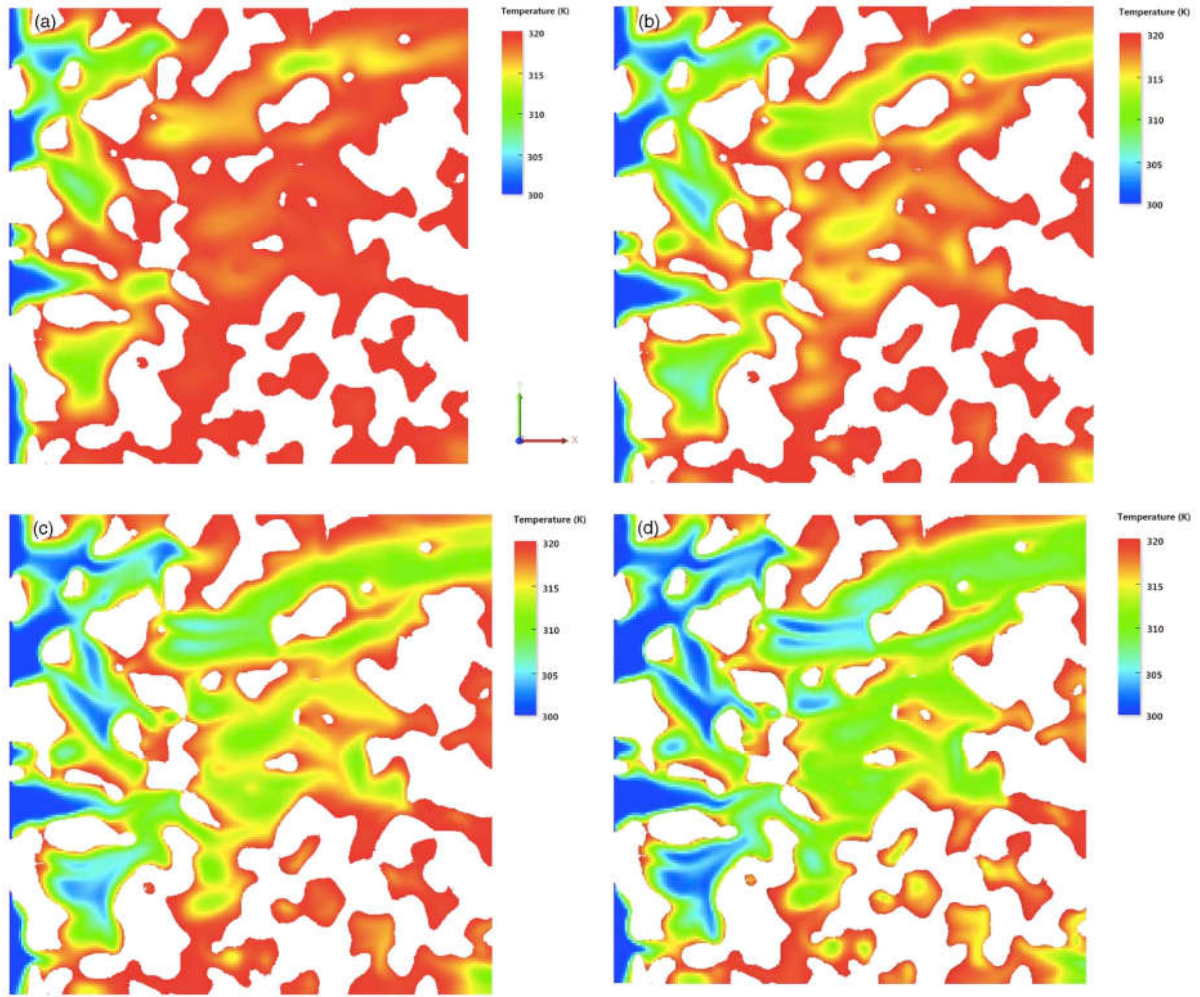

**Fig. 12.** Temperature distributions in the longitudinal section of $z = 2.5$ mm for different inlet pressures. (a) $P_{in}$ = 20 Pa, (b) $P_{in}$ = 40 Pa, (c) $P_{in}$ = 60 Pa, and (d) $P_{in}$ = 80 Pa. (Reprinted from Liu and Wu [53].)

Liu and Wu [53] have investigated the fluid flow and heat transfer in 3D reconstructed porous media at the pore scale. The correlations for fluid flow and heat transfer in the specific porous media were obtained based on the numerical results. The temperature distributions in the longitudinal section of $z = 2.5$ mm for different inlet pressures can be found in Fig. 12. The temperature of the solid matrix is $T_{f,in} = 320$ K, and the temperature of the fluid in the inlet is $T_m = 300$ K. It can be seen that increasing the inlet pressure reduces the fluid temperature in the outlet. For a higher inlet pressure, the flow velocity in the domain increases and complex flow status occurs because of the irregular porous structure, and then the heat will be transferred mainly by convection under such condition. Moreover, an increase in the inlet pressure leads to a higher convective heat transfer coefficient along the flow direction (see Fig. 10 in Ref. [53]).

Yang et al. [276] have investigated the flow and heat transfer in a 2D channel filled with random porous medium composed of cylinders of different diameters. For the random porous medium, the effect of the disordered cylinder arrangement on permeability and Nusselt number were investigated. Heat transfer correlations were proposed and compared with the existing experimental data and empirical correlations. It was found that the Nusselt number increases with the increasing of porosity, and the exponent between the Nusselt number and Reynolds number is 0.786. Yang et al. [277] also studied the heat transfer characteristics in a 3D channel filled with random porous medium composed of cubes or spheres. The effects of randomness of porous structure, particle sizes, and particle shape on the flow and heat transfer characteristics were investigated. The results showed that the randomness has significant effects on the permeability and Nusselt number when the number and sizes of spheres or cubes are fixed. In a recent work carried out by Liu et al. [278], the LB method for transport phenomena was combined with the simulated annealing algorithm for digitized structure generation to study flow and heat transfer in random porous media. Based on the pore-scale numerical results, the effective transport properties at the REV scale were derived with reasonable accuracy through appropriate upscale averaging.

Natural convection in a square cavity filled with solid obstacles has also been studied [192,279,280]. In 2010, Zhao et al. [279] have simulated the 2D natural convection flow in porous media (porous metals). The effects of Rayleigh number, porosity, pore density and metal blocks' shape on the heat transfer process were investigated. Later, Karani and Huber [192] proposed an LB model for conjugate heat transfer in heterogeneous material. They studied the problem of natural convection in a square cavity containing solid obstacles with different thermal properties. In their study, the average Nusselt numbers for different thermal properties were calculated and compared with the FVM results. The results confirmed the reliability of their model in simulating fluid flow and heat transfer in complex geometries. Ren and Chan [280] investigated conjugate heat transfer in a square cavity filled with an array of obstacles. Five cases with different Rayleigh numbers, solid-to-fluid thermal diffusivity ratios and number of solid obstacles were presented. It was demonstrated that the LB method can serve as an effective approach to simulate conjugate heat transfer with solid obstacles. In addition, Zhao et al. [281] investigated the double-diffusive convection heat and mass transfer in a square cavity filled with a porous medium at the pore scale. In Zhao et al.'s study, the Dufour and Soret effects were considered by adding additional source terms into the LB equations of the temperature and concentration fields, respectively. They showed that the heat and mass transfer are enhanced as the Rayleigh number increases. Especially, the temperature gradient has great influence on the mass transfer process.

*5.2. Pore-scale solid-liquid phase-change heat transfer*

In 2006, using Jiaung et al.'s enthalpy-based LB model, Qian et al. [54] studied the heat conduction problem with solid-liquid phase change in porous media (metal foams) at the pore scale. They found that the thermal conductivity of the composite PCM can be significantly enhanced by using the metal foams. Later, Huber et al. [55] investigated pure substance melting with natural convection in a synthetic porous medium (porosity is 0.45) at the pore scale. It was found that the system melts from bottom up until it reaches a critical average height that enables large scale convection, and the melting front is not flat due to the horizontal temperature gradients caused by the inhomogeneous distribution of the solid obstacles. Chen et al. [56] studied the melting behaviors of PCM in metal foams at the pore scale. The effect of metal foam on melting rate of the PCM can be found in Fig. 13 ($V_{fl}$ is the volume of liquid PCM, $V$ is the volume of PCM). As shown in the figure, when $FoSte = 0.2$, the melting rate of the PCM (with metal foam) is 0.89, while the melting rate of the PCM (without metal foam) is 0.50, i.e., the melting rate of the PCM with metal foam is enhanced by 78% compared to that without metal foam.

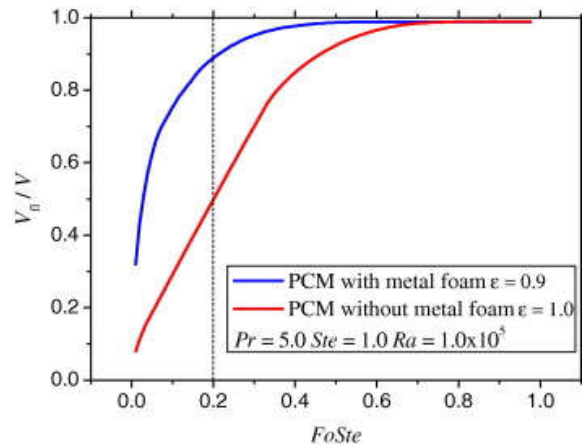

**Fig. 13.** The effect of metal foam on the melting rate of the PCM. (Reprinted from Chen et al. [56].)

In addition, Song et al. [57] simulated the heat and moisture transport phenomena in frozen soil during freezing process. Comparisons between experimental and numerical results of water content distribution, temperature distribution and the position of the freezing front have been made, and the numerical results agree well with the experimental data. Most recently, Ren et al. [58] have studied the PCM melting performance of latent heat thermal energy storage at the pore scale. The effect of metal foam characteristics such as porosity and pore size on the melting performance was studied at different hot wall temperatures and initial subcooled temperatures.

For pore-scale solid-liquid phase change, some results (the evolution processes of the phase field at different times) can be found in Fig. 14. The random porous medium is generated by the QSGS method (the porosity is 0.7, and the blue block represents the medium). A grid size of $N_x \times N_y = 400 \times 80$ is employed for the computational domain. The left and right walls are kept at constant temperatures $T_h$ and $T_c$ ($T_h > T_c$), respectively, while the horizontal walls are adiabatic. Owing to its kinetic nature, the LB method can serve as an accurate and efficient numerical tool for studying solid-liquid phase change in porous media at the pore scale. However, the research of this topic is still in its early stages, and more studies are required along this line.

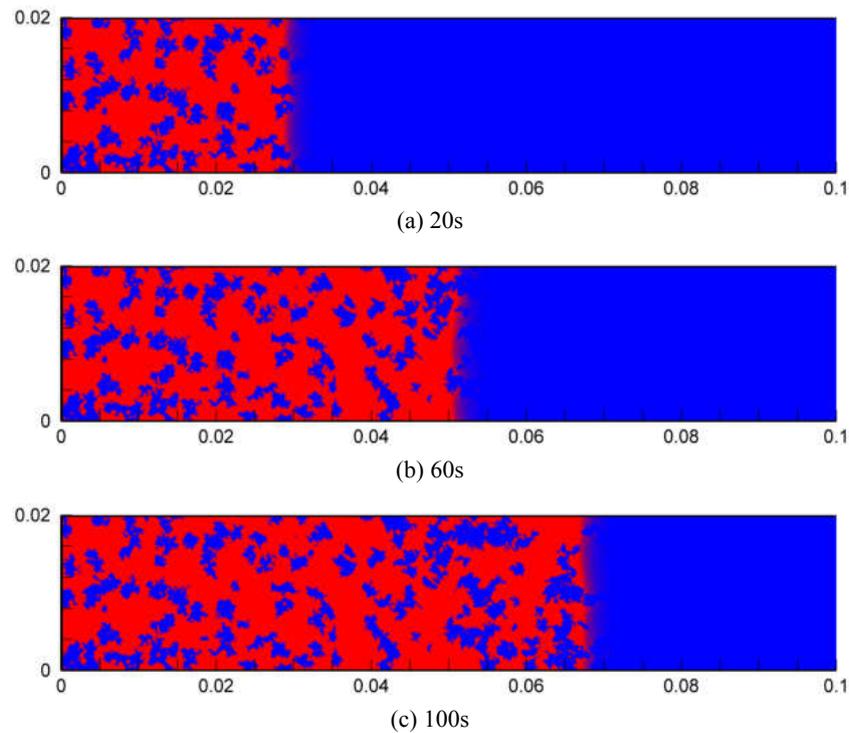

(a) 20s

(b) 60s

(c) 100s

**Fig. 14.** The evolution processes of the phase field at $t$ = 20s, 60s, and 80s. (Computations by He, Liu and Li for this review.)

*5.3. REV-scale single-phase heat transfer*

In 2005, Guo and Zhao [65] studied convection heat transfer in porous media with their temperature-based DDF-BGK model. They showed that their model can serve as a promising numerical tool for studying natural, mixed-, and forced-convection problems in porous media. Furthermore, the variable viscosity effects on the natural convection characteristics in porous media were studied in Guo and Zhao' work [66]. They found that the variation in viscosity has a strong influence on both the flow and heat transfer characteristics of natural convection in porous media, especially at large Darcy-Rayleigh numbers. For natural convection in porous media with variable viscosity, the flow patterns are asymmetric, and the heat transfer is significantly enhanced in comparison with the results for constant viscosity case. Additionally, Rong et al. [282] studied the problem of Bénard convection with temperature-dependent viscosity in a porous cavity. They proposed a new formula of reference temperature and the numerical results showed that the predicted average Nusselt numbers based on the new formula have higher precision than those predicted based on average temperature.

In 2006, Seta et al. [67] employed their internal-energy-based DDF-BGK model to study natural convection in porous media. A comprehensive parametric study was carried out for various values of Rayleigh number, Darcy number, and porosity. Shokouhmand et al. [283] studied the laminar flow and convective heat transfer in conduits filled with porous media. The effects of various parameters (e.g., Darcy number, porous medium thickness) on the conduit thermal performance were investigated. Vishnampet et al. [284] investigated natural convection in a porous cavity at high Rayleigh numbers. They found that for high Rayleigh number, the generalized non-Darcy model predicts up to 20% lower average Nusselt number than the Darcy model. Djebali et al. [285] studied the natural convection flow in a tilted porous cavity under the effect of uniform magnetic field. They showed that the rate of heat transfer inside the porous cavity deceases as the Hartmann number increases. Furthermore, the laminar convection of temperature-sensitive magnetic fluids in a porous square cavity was investigated in Jin and Zhang' work [286].

Table 2 The average Nusselt numbers along the left wall of the cavity. (Adapted from Liu and He [68].)

| Da ($N_x \times N_y$) | Ra | $\phi = 0.4$ | | | $\phi = 0.6$ | | | $\phi = 0.9$ | | |
|---|---|---|---|---|---|---|---|---|---|---|
| | | Ref.[199] | Ref.[65] | Present | Ref.[199] | Ref.[65] | Present | Ref.[199] | Ref.[65] | Present |
| $10^{-2}$ | $10^3$ | 1.010 | 1.008 | 1.007 | 1.015 | 1.012 | 1.012 | 1.023 | - | 1.017 |
| ($120 \times 120$) | $10^4$ | 1.408 | 1.367 | 1.362 | 1.530 | 1.499 | 1.494 | 1.640 | - | 1.628 |
| | $10^5$ | 2.983 | 2.998 | 3.009 | 3.555 | 3.422 | 3.460 | 3.910 | - | 3.939 |
| $10^{-4}$ | $10^5$ | 1.067 | 1.066 | 1.067 | 1.071 | 1.068 | 1.069 | 1.072 | - | 1.073 |
| ($200 \times 200$) | $10^6$ | 2.550 | 2.603 | 2.630 | 2.725 | 2.703 | 2.733 | 2.740 | - | 2.796 |
| | $10^7$ | 7.810 | 7.788 | 7.808 | 8.183 | 8.419 | 8.457 | 9.202 | - | 9.352 |
| $10^{-6}$ | $10^7$ | 1.079 | 1.077 | 1.085 | 1.079 | 1.077 | 1.089 | 1.08 | - | 1.090 |
| ($250 \times 250$) | $10^8$ | 2.970 | 2.955 | 2.949 | 2.997 | 2.962 | 2.957 | 3.00 | - | 3.050 |
| | $10^9$ | 11.460 | 11.395 | 11.610 | 11.790 | 11.594 | 12.092 | 12.01 | - | 12.041 |

Using the temperature-based DDF-MRT model, Liu and He [68] have studied convection heat transfer in porous media. In their work, they simulated mixed-convection in a porous channel, natural convection in a porous cavity, and thermal convection in a porous cavity with internal heat generation. For natural convection in a porous cavity, the average Nusselt numbers along the left wall of the porous cavity are listed in Table 2. Clearly, the predicted results agree well with the numerical results reported in previous studies [65,199]. Furthermore, convection heat transfer in porous media was also studied by Wang et al. [71] with their modified DDF-BGK model. Gao et al. [73] have investigated conjugate heat transfer in a system containing simultaneously a porous medium and other media. In their work, they simulated conjugate natural convection in a cavity partially filled with porous medium and conjugate heat transfer in porous media with a conducting wall. They showed that their model can serve as a feasible tool for conjugate heat transfer problems in porous media. In addition, conjugate heat transfer between fluid-saturated porous media and solid walls was also studied in Chen's work [287].

Furthermore, using the 3D DDF-MRT model, Liu and He [74] have studied natural convection in a cubical cavity filled with fluid-saturated porous media.

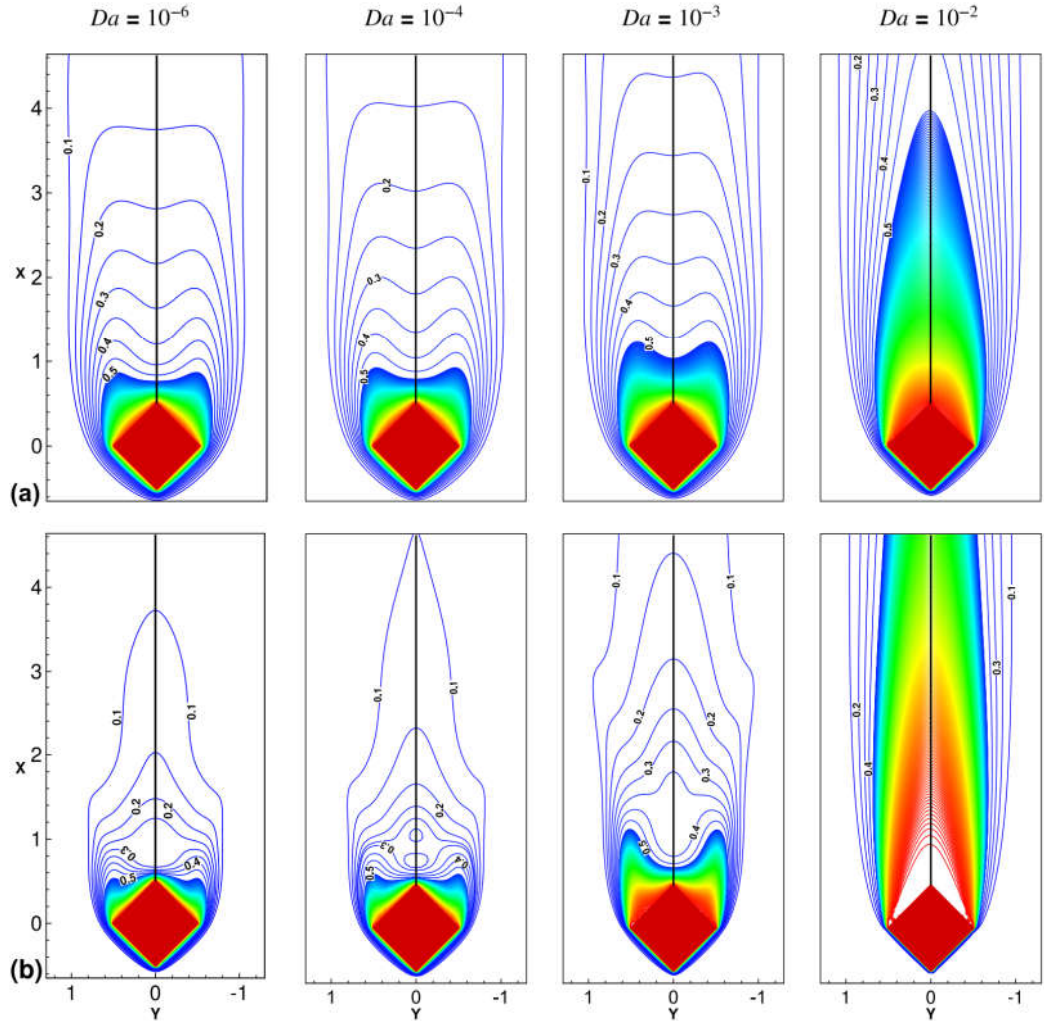

**Fig. 15.** Time-averaged isotherms of mean flow around and through a diamond-shaped porous block at (a) *Re*=50 and (b) *Re*=150 for different values of *Da*. (Reprinted from Vijaybabu et al. [288].)

Most recently, the unsteady flow and heat transfer characteristics of a permeable diamond-shaped block, placed in a uniform flow and maintained at a constant temperature, have been studied in detail by Vijaybabu et al. [288]. The effects of the permeability of the diamond-shaped porous block on the flow and heat transfer behaviors have been investigated for different Darcy numbers ($10^{-6} \leq Da \leq 10^{-2}$) and Reynolds numbers ($50 \leq Re \leq 150$). Some results can be found in Fig. 15, from which the influence of the permeability on the heat transfer characteristics can be clearly observed.

The LB method has also been applied to study axisymmetric thermal flows in porous media [209-211,289]. The first study was conducted by Rong et al. [209] in 2010. In their study, the free convection flow through an annulus packed with porous media and natural convection in a closed cylindrical annulus filled with porous media were modeled. Later, Rong et al. [289] studied heat transfer enhancement in a pipe filled with porous media by using their axisymmetric DDF-LB model based on the graphics processing unit (GPU). As shown in Fig. 16, the heat transfer performance can be significantly improved by inserting porous medium into the pipe center ($Ri = h/R$ is the ratio of porous layer thickness $h$ and pipe radius $R$).

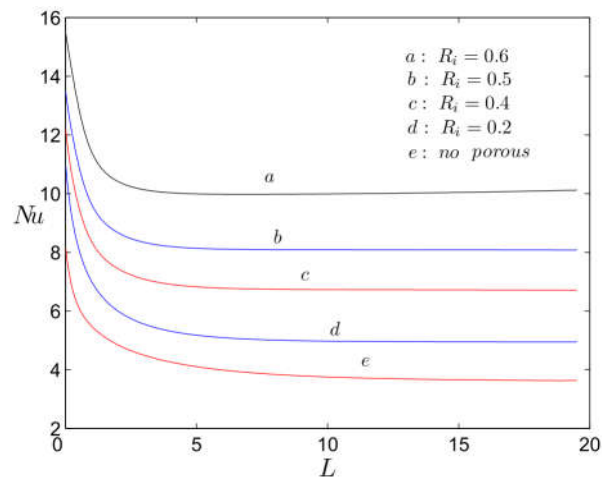

**Fig. 16.** The Nusselt number distributions along the pipe wall for different values of $Ri$. (Reprinted from Rong et al. [289].)

Furthermore, axisymmetric thermal flows in porous media were also studied in Liu and He's work [210]. In this work, the thermally developing flow in a pipe filled with porous media and natural convection in a vertical annulus filled with porous media were simulated, and the numerical results agree well with the data reported in previous studies. Similarly, Grissa et al. [211] have investigated axisymmetric thermal flows in porous media with their axisymmetric DDF-LB model. Their simulation results agree well with the analytical solutions and/or the results of previous studies.

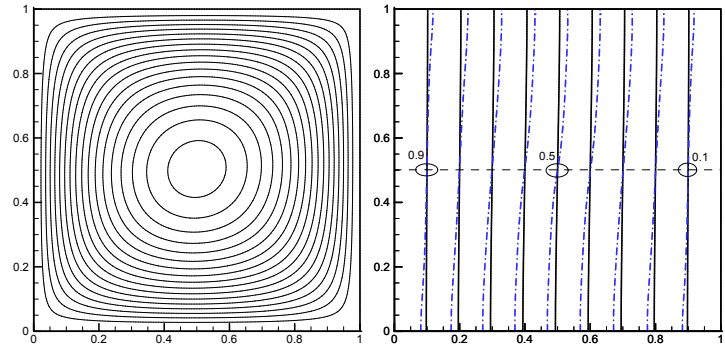

(a) $Ra = 10^6$

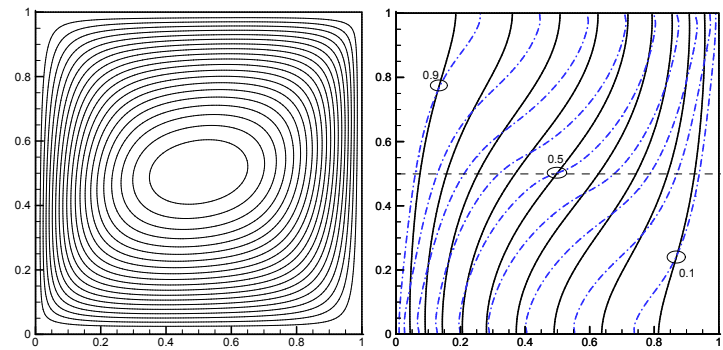

(b) $Ra = 10^8$

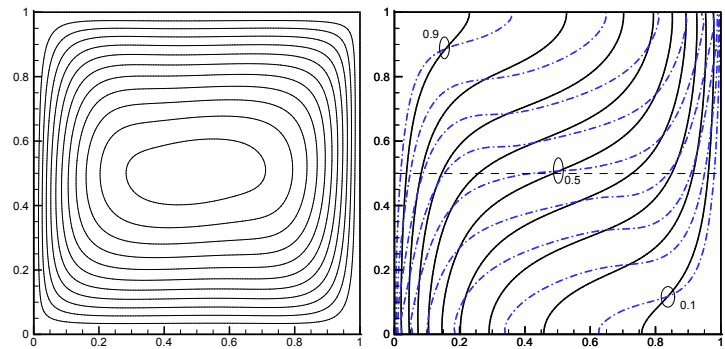

(c) $Ra = 10^9$

**Fig. 17.** Streamlines (left) and isotherms (right) of natural convection in a cavity filled with metallic porous media at the steady state (fluid phase: black lines; solid phase: blue dash-dot lines). (Adapted from Liu and He [215].)

Recently, the LB method has been applied to simulate convection heat transfer in porous media under LTNE conditions [214,215,216]. In 2014, Gao et al. [214] employed a temperature-based TDF-BGK model to simulate natural convection problems in porous media under LTNE conditions. They simulated steady natural convection in a porous cavity with heat-generating solid phase and transient natural convection in a cavity filled with metallic porous media. The results showed that the LTNE assumption should be employed when the ratio of solid-to-fluid thermal conductivities, Darcy number, and Rayleigh number are relatively large and the interstitial Nusselt number is small. Liu and He [215] have simulated natural convection problems in porous media under LTNE conditions with a temperature-based TDF-MRT model. Some results can be found in Fig. 17, from which the thermal non-equilibrium effect can be clearly observed from the isotherms of natural convection in a cavity filled with metallic porous media at the steady state. In addition, Wang et al. [216] studied convection heat transfer problems in porous media under LTNE conditions using a total-energy-based TDF-BGK model. In this work, both the compression work and viscous dissipation were considered, and it was found that the temperature difference between the fluid and solid phases increases as the Eckert number increases.

*5.4. REV-scale solid-liquid phase-change heat transfer*

For solid-liquid phase-change heat transfer in porous media under LTE condition, the first study using LB method was conducted by Gao and Chen [75] in 2011. They simulated melting with natural convection in a square porous cavity and found that, for high Darcy number and high porosity, the Darcy-Rayleigh number $Ra^*$ ($Ra^* = DaRa$) may not be appropriate to correlate the average Nusselt number of the hot wall independently. Shortly afterwards, Jourabian et al. [290] studied melting of ice in $Al_2O_3$ porous matrix. In Jourabian et al.'s study, the effects of the porosity on the temperature contours, flow patterns within the melting domain, complete melting time of the PCM and average Nusselt number were investigated qualitatively and quantitatively. In Fig. 18, the streamlines and temperature contours at different times ($\theta = FoSte$ is the dimensionless time) for $\phi = 0.9$, 0.7, and 0.5 are presented. As shown in the figure, as the porosity decreases, the melting rate increases and better heat transfer performance can be achieved.

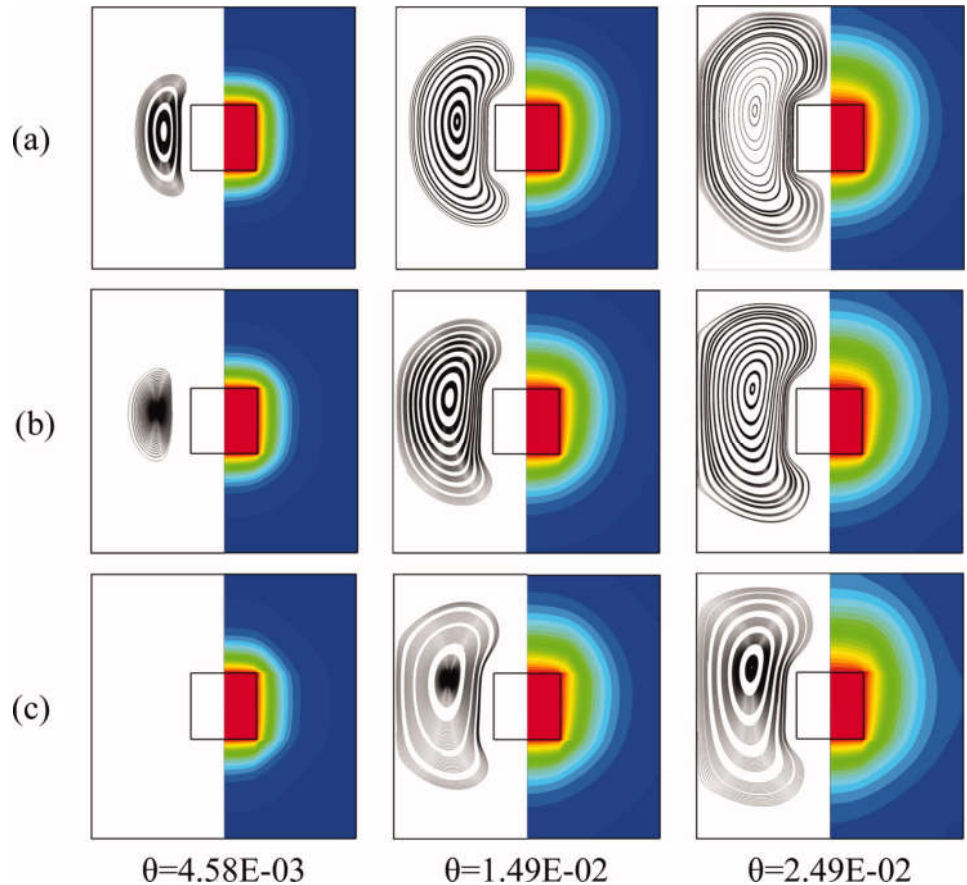

**Fig. 18.** Streamlines (left) and temperature contours (right) at different times for (a) $\phi = 0.9$, (b) $\phi = 0.7$, and (c) $\phi = 0.5$. (Reprinted from Jourabian et al. [290].)

Later, Liu and He [76] studied solid-liquid phase change with natural convection in a square porous cavity using a temperature-based DDF-MRT model. The locations of the solid-liquid interface at different Fourier numbers can be found in Fig. 19 (the experimental and numerical results in the figure are obtained from Ref. [252]). The predicted results agree reasonably well with the experimental and numerical results of the literature, and it can be observed that the melting interface moves faster near the top wall due to the convection effect. Besides, using an enthalpy-based LB model, Wu et al. [77] have simulated melting with natural convection in a square porous cavity.

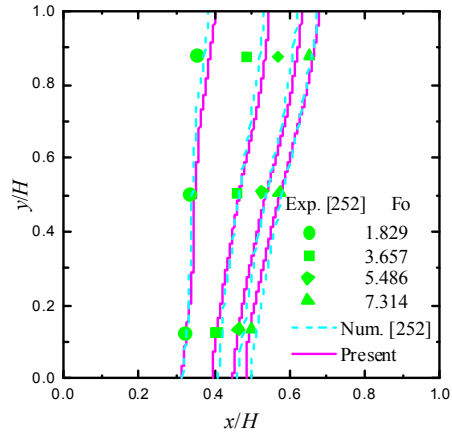

**Fig. 19.** Locations of the phase interface at different Fourier numbers. (Adapted from Liu and He [76].)

Recently, Gao et al. [246] studied convection melting in a square porous cavity with conducting fin using an enthalpy-based LB model. The temperature contours and streamlines at $Fo = 0.00255$ can be found in Fig. 20. It can be observed that adding a conducting fin in the porous cavity can significantly improve the melting heat transfer. By comparing Fig. 20(b) with Fig. 20(c), one can find that for the same thermal conductivity, the fin with low volumetric heat capacity heats more quickly than that with high volumetric heat capacity. This is because in the heat transfer process of melting, a lower volumetric heat capacity leads to a higher thermal diffusivity when the thermal conductivity is fixed. Moreover, it was found that the melting rate increases when the length of the fin becomes larger, and varying the vertical position of the fin has no remarkable impact on the melting rate for the cases considered (see Figs. 12 and 13 in Ref. [246]).

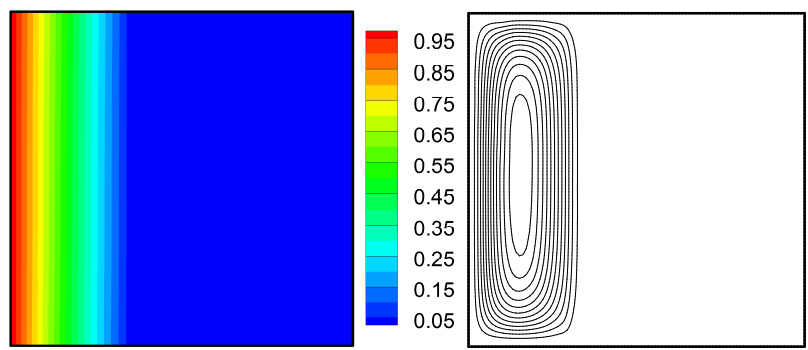

(a) without fin

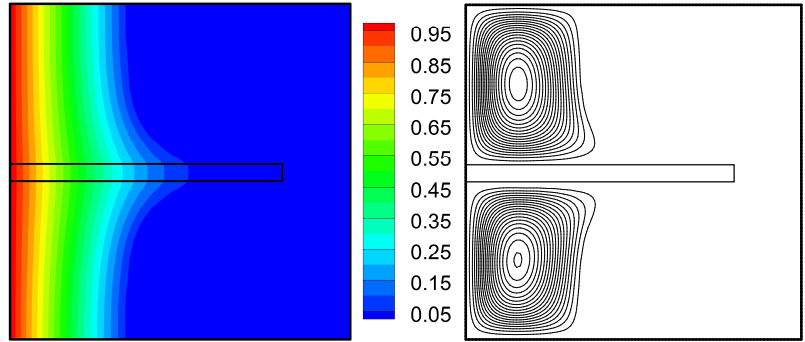

(b) $(\rho c_p)_{\text{fin}}/(\rho c_p)_{\text{f}}=20$, $hf/H=0.5$, $lf/H=0.8$

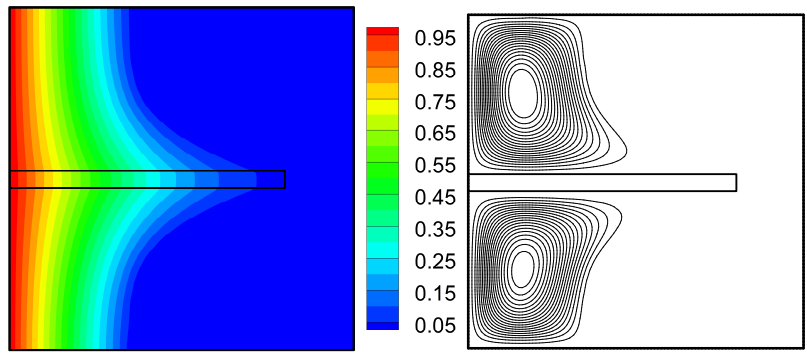

(c) $(\rho c_p)_{\text{fin}}/(\rho c_p)_{\text{f}}=1.05$, $hf/H=0.5$, $lf/H=0.8$

**Fig. 20.** Temperature contours and streamlines for melting in a porous cavity without fin and with a fin at the time $Fo = 0.00255$. (Reprinted from Gao et al. [246].)

Usually, the available PCMs commonly suffer from low thermal conductivities, and embedding PCMs in highly conductive porous materials such as high porosity open-cell metal foams to form composite phase change materials has been widely used in practical applications. Since the thermal conductivity of the metal foam is usually two or three orders of magnitude higher than that of the PCM, the thermal non-equilibrium effects between the PCM and metal foam may play a significant role. Therefore, the LTNE model should be employed for numerical studies. In 2010, Gao et al. [78] successfully simulated convection melting in a square cavity filled with metal-foam-based PCM using a temperature-based TDF-BGK model. They showed that as the porosity decreases, the melting rate increases due to the enhanced effect of heat conduction from metal foam with high heat conductivity. In addition, they found that the pore size has only a limited effect on the melting rate because of the two counteracting effects between conduction and convection heat transfer. Later, Gao et al. [214] studied transient natural convection with solid-liquid phase change in porous media with an improved temperature-based TDF-BGK model. The effect of the interstitial Nusselt number on melting volume fraction was investigated, and it was found that the melting rate increases as the interstitial Nusselt number increases. Tao et al. [79] employed a temperature-based TDF-BGK model to study the latent heat storage performance of copper foams/paraffin composite phase change material. The effects of geometric parameters such as pore density and porosity on PCM melting rate, thermal energy storage capacity and density were investigated.

Using an enthalpy-based TDF-BGK model, Gao et al. [80] simulated transient natural convection in a square cavity filled with metal-foam-based PCM. In Gao et al.'s work, the iteration procedure has been avoided and the relaxation time can be adjusted to reduce the numerical diffusion across the phase interface. Very recently, Liu and He [81] have applied an enthalpy-based TDF-MRT model to simulate transient natural convection in a square cavity filled with metal-foam-based PCM. The iteration procedure in Liu and He's study has also been avoided, and the enthalpy-based TDF-MRT model can be used for large-scale engineering calculations of solid-liquid phase-change heat transfer in metal-foam-based PCMs.

## 6. Summary and outlook

The LB method celebrates its 30th birthday this year. The great Chinese philosopher Confucius said: "At thirty, I had planted my feet firm upon the ground." Over the last 30 years, the LB method has been developed into a powerful and promising numerical tool for computational fluid dynamics and beyond, and it is still undergoing development. It is noticed that the LB method has achieved great success in modeling fluid flow and heat transfer in porous media for its physical representation of microscopic interactions and the capability of handling complex boundaries. The present review has focused on the LB methods for single-phase and solid-liquid phase-change heat transfer in porous media that are widely involved in energy/environmental science and technologies. In the related areas, both the theoretical exploration and new applications of the LB method continue to appear.

There is no doubt that the LB method will play an increasingly important role in studying single-phase and solid-liquid phase-change heat transfer in porous media in the near future. However, it should be noted that, despite its undeniable success, one should be aware of the limits of the thermal LB methods. As pointed out by Succi [103], the available thermal LB models still suffer from the inherent athermal nature of the LB method. In addition, for heat transfer in porous media at the REV scale, the existing thermal LB models are limited to Boussinesq flows, in which the temperature variation is small. Hence, developing numerically stable and thermo-hydrodynamic consistent LB scheme for energy transport in porous media is still required.

As in the conventional numerical methods for solving solid-liquid phase-change problems, an interface-capturing (or interface-tracking) technique is needed in the LB method so as to capture (or track) the phase interface, i.e., the phase interface is treated as a mathematical boundary. In the enthalpy-based LB method, the phase interface is implicitly captured through the liquid fraction, which can be obtained via updating the enthalpy step by step. For solid-liquid phase-change heat transfer with convection effect, the interfacial behaviors predicted by the enthalpy-based LB method may not be consistent with the real physical processes. Microscopically, the interfacial behaviors originate from the interactions among the constituent molecules. Future developments can focus on developing accurate and effective LB schemes for solid-liquid phase change by making full use of the discrete kinetic origin of the LB method. Moreover, there have been only few pore-scale studies on solid-liquid phase change (with convection effect) in porous media. The pore-scale modeling has the potential to provide constitutive parameters needed in the REV-scale modeling through up-scaling of the pore-scale results. With the combination of pore-scale and REV-scale approaches, the capability of the enthalpy-based LB method can be greatly expanded.

**Acknowledgments**

This work was financially supported by the Key Project of National Natural Science Foundation of China (Grant No. 51436007) and the Major Program of the National Natural Science Foundation of China (Grant No. 51590901).